\documentclass[10pt]{article}
\pdfoutput=1

\usepackage{latexsym}
\usepackage{amsmath}
\usepackage{mathtools}
\usepackage{amssymb}
\usepackage{amsfonts}
\usepackage{amsthm}
\usepackage{graphicx}
\usepackage{tikz,tikz-qtree,tikz-qtree-compat}
\usetikzlibrary{shapes,snakes,arrows,calc,fit,topaths,calc,chains,scopes}
\usepackage{subfigure}

\usepackage{fancyheadings}
\usepackage{color,xspace,colortbl}
\usepackage[ruled]{algorithm2e}
\usepackage[noend]{algpseudocode}
\usepackage{enumitem}
\usepackage{balance}
\usepackage{todonotes}
\usepackage{url}

\definecolor{Red}{RGB}{255,204,204}
\definecolor{Green}{RGB}{204,255,204}
\definecolor{Blue}{RGB}{204,204,255}
\definecolor{goodgreen}{rgb}{0.1, 0.5, 0.1}

\colorlet{LightViolet}{violet!40}
\colorlet{LightRed}{red!40}
\colorlet{LightOrange}{orange!40}
\colorlet{LightGreen}{green!40}
\colorlet{LightBlue}{blue!40}
\colorlet{DarkGreen}{green!50!black}
\colorlet{DarkRed}{red!70!black}
\colorlet{DarkCyan}{red!70!black}
\colorlet{DarkBlue}{blue!80!black}
{\definecolor{DarkOrange}{rgb}{1.0, 0.49, 0.0}
\definecolor{Airforceblue}{rgb}{0.36, 0.54, 0.66}

\newcommand{\Bigstar}{\mathop{\bigstar}}

\newcommand{\nop}[1]{}

\newcommand{\vietnam}{\text{\sf vietnam}}
\newcommand{\saigon}{\text{\sf saigon}}
\newcommand{\hanoi}{\text{\sf hanoi}}

\newcommand{\england}{\text{\sf england}}
\newcommand{\usa}{\text{\sf usa}}

\newcommand{\faqcs}{\text{\sf FAQ-SS}}

\newcommand{\faq}{\text{\sf FAQ}}

\newcommand{\functionname}[1]{\text{\sf #1}}

\newcommand{\tw}{\functionname{tw}}

\newcommand{\faqw}{\functionname{faqw}}

\newcommand{\fhtw}{\functionname{fhtw}}

\newcommand{\InsideOut}{\functionname{InsideOut}}

\newcommand{\problemname}[1]{\text{\sf #1}}

\newcommand{\EVO}{\problemname{EVO}}

\newcommand{\fama}{\text{\sf FaMa}}
\newcommand{\pr}{\text{\sf PR}}
\newcommand{\lr}{\text{\sf LR}}

\newcommand{\fcity}{\textsf{city}}
\newcommand{\fcountry}{\textsf{country}}

\newcommand{\calLL}{\mathcal L}

\newcommand{\calE}{\mathcal E}

\newcommand{\calH}{\mathcal H}
\newcommand{\calV}{\mathcal V}
\newcommand{\calU}{\mathcal U}

\newcommand{\R}{\mathbb R} 

\newcommand{\gv}[1]{\ensuremath{\mbox{\boldmath$ #1 $}}} 
\newcommand{\grad}[1]{\gv{\nabla} #1} 
\newcommand{\pd}[2]{\frac{\partial#1}{\partial#2}}

\newcommand{\be}{\begin{enumerate}}
\newcommand{\ee}{\end{enumerate}}
\newcommand{\bi}{\begin{itemize}}
\newcommand{\ei}{\end{itemize}}
\newcommand{\beq}{\begin{equation}}
\newcommand{\eeq}{\end{equation}}

\newcommand{\bp}{\begin{proof}}
\newcommand{\ep}{\end{proof}}
\newcommand{\bcor}{\begin{cor}}
\newcommand{\ecor}{\end{cor}}
\newcommand{\bthm}{\begin{thm}}
\newcommand{\ethm}{\end{thm}}
\newcommand{\blmm}{\begin{lmm}}
\newcommand{\elmm}{\end{lmm}}
\newcommand{\bdefn}{\begin{defn}}
\newcommand{\edefn}{\end{defn}}
\newcommand{\bprop}{\begin{prop}}
\newcommand{\eprop}{\end{prop}}
\newcommand{\bconj}{\begin{conj}}
\newcommand{\econj}{\end{conj}}
\newcommand{\bopm}{\begin{opm}}
\newcommand{\eopm}{\end{opm}}
\newcommand{\brmk}{\begin{rmk}}
\newcommand{\ermk}{\end{rmk}}

\newcommand{\norm}[1]{\left\|#1\right\|}

\newcommand{\suchthat}{\ | \ }
\newcommand{\inner}[1]{\left\langle #1 \right\rangle}

\newcommand{\DIAG}[1]{\mathop{\textnormal{DIAG}} #1}

\newcommand{\mv}[1]{\mathbf{#1}}

\theoremstyle{plain}                   
\newtheorem{thm}{Theorem}[section]
\newtheorem{lemma}[thm]{Lemma}
\newtheorem{prop}[thm]{Proposition}
\newtheorem{cor}[thm]{Corollary}

\theoremstyle{definition}              

\newtheorem{opm}{Open Problem}
\newtheorem{conj}{Conjecture}
\newtheorem{ex}{Example}

\newtheorem{defn}{Definition}
\newtheorem{definition}{Definition}

\newtheorem{rmk}{Remark}

\newcommand{\pluseq}{\mathrel{+}=}
\newcommand{\TAB}{\makebox[2.5ex][r]{}}%
\newcommand{\STAB}{\makebox[1.5ex][r]{}}%
\newcommand{\IF}{\textbf{if}\xspace}%
\newcommand{\FOREACH}{\textbf{foreach}\xspace}%
\newcommand{\DO}{\textbf{do}\xspace}%
\newcommand{\RETURN}{\textbf{return}\xspace}%
\newcommand{\MATCH}{\textbf{switch}\xspace}%
\newcommand{\SUM}{\texttt{SUM}\xspace}%
\newcommand{\COUNT}{\texttt{COUNT}\xspace}%

\newcommand{\aggs}{\mathsf{aggregates}}
\newcommand{\aggreg}{\mathsf{aggregateRegister}}
\newcommand{\rfbox}[1]{{\color{red}\fbox{\color{black}#1}}}

\usepackage{xpatch}
\usepackage{textcase}
\usepackage{amssymb}

\usepackage[left=2.15cm,right=2.15cm,top=2.8cm,bottom=2.8cm]{geometry}

\makeatletter
\xpatchcmd{\@sect}{\uppercase}{\MakeTextUppercase}{}{}
\xpatchcmd{\@sect}{\uppercase}{\MakeTextUppercase}{}{}
\makeatother

\allowdisplaybreaks[1]
\renewcommand{\vec}[1]{\ensuremath\boldsymbol{#1}}

\tikzstyle{path} = [->,double,rounded corners=.1cm]
\tikzstyle{data} = [draw, rectangle, rounded corners = .07cm, align=center, inner sep = .2cm, outer sep = .1 cm]
\tikzstyle{processor} = [draw, ellipse, align=center, inner sep = .1cm, outer sep = .1 cm]
\colorlet{clr_outofDB}{red!80!black}
\colorlet{clr_inDB}{blue!80!black}
\colorlet{clr_inDB_FD}{green!50!black}

\newcommand{\drawtable}[5]
{
   \begin{scope}[shift={(-#1/2,-#2/2)}]
      \pgfmathsetmacro{\cellwidth}{#1/#3}
      \pgfmathsetmacro{\cellheight}{#2/#4}
      \draw[fill=#5!30] (0, #2) rectangle (#1, #2-\cellheight);
      \foreach \i in {1,...,#3}
         \draw[thin, #5!70] (\cellwidth*\i, 0) -- (\cellwidth*\i, #2);
      \foreach \i in {1,...,#4}
         \draw[thin, #5!70] (0, \cellheight*\i) -- (#1, \cellheight*\i);
      \draw[#5] (0, #2-\cellheight) -- (#1, #2-\cellheight);
      \draw[thick, #5] (0, 0) rectangle (#1, #2);
   \end{scope}
}

\newcommand{\drawcylinder}[4]
{
   \begin{scope}[shift={(0,#3/2)}]
      \draw[#4] (0, #2/2) ellipse (#1/2 and #3);
      \draw[#4] (#1/2, -#2/2) arc (0:-180:#1/2 and #3);
      \draw[#4] (-#1/2, #2/2) -- (-#1/2, -#2/2);
      \draw[#4] (#1/2, #2/2) -- (#1/2, -#2/2);
   \end{scope}
}

\newcommand{\drawfilter}[2]
{
   \begin{scope}[scale=#1/7, shift={(-3.5,3)}]
      \draw[#2] (0,0)--(7, 0)--(4, -3)--(4,-5)--(3,-6)--(3,-3)--cycle;
   \end{scope}
}

\newcommand*{\highlight}[1]{%
   \tikz\node[rectangle, fill=yellow, inner sep=0.5mm]{#1};%
}

\title{Learning Models over Relational Data using Sparse Tensors and Functional Dependencies}

\author{
  Mahmoud Abo Khamis\\
  RelationalAI, Inc.
    \and 
   Hung Q. Ngo \\
   RelationalAI, Inc. 
    \and
    XuanLong Nguyen \\
    University of Michigan
    \and
    Dan Olteanu \\
    University of Oxford 
    \and
    Maximilian Schleich\\
    University of Oxford
}

\date{}

\begin{document}

\maketitle

\begin{abstract}
  Integrated solutions for analytics over relational databases
  are of great practical importance as they avoid the costly repeated loop 
  data scientists have to deal with on a daily basis: 
  select features from data residing in relational databases using
  feature extraction queries involving joins, projections, and aggregations;
  export the training dataset defined by such queries; convert this dataset into
  the format of an external learning tool; and train the desired model using
  this tool.  These integrated solutions are also a fertile ground of
  theoretically fundamental and challenging problems at the intersection of
  relational and statistical data models.

  This article introduces a unified framework for training and evaluating a
  class of statistical learning models over relational databases. This class
  includes ridge linear regression, polynomial regression, factorization
  machines, and principal component analysis.  We show that, by synergizing key
  tools from database theory such as schema information, query
  structure, functional dependencies, recent advances in query evaluation
  algorithms, and from linear algebra such as tensor and matrix
  operations, one can formulate relational analytics problems and design
  efficient (query and data) structure-aware algorithms to solve them.

  This theoretical development informed the design and implementation of the
  AC/DC system for structure-aware learning.  We benchmark the performance of
  AC/DC against R, MADlib, libFM, and TensorFlow.  For typical retail
  forecasting and advertisement planning applications, AC/DC can learn
  polynomial regression models and factorization machines with at least the same
  accuracy as its competitors and up to three orders of magnitude faster than
  its competitors whenever they do not run out of memory, exceed 24-hour
  timeout, or encounter internal design limitations.
  
\end{abstract}

\section{Introduction}
\label{sec:introduction}

Although both disciplines of databases and statistics occupy foundational roles for the emerging field of data science, they are largely seen as complementary. Most fundamental contributions made by statisticians and machine learning researchers are abstracted away from the underlying infrastructure for data management. However, there is undoubtedly clear value in tight integration of statistics and database models and techniques. This is 
receiving an increasing interest in both academia and industry~\cite{DBTHEORY:DAGSTUHL:17,Kumar:SIGMOD:Tutorial:17,Polyzotis:SIGMOD:Tutorial:17}. This is motivated by the realization that in many practical cases data used for training resides inside relational databases and bringing the analytics closer to the data saves non-trivial time usually spent on data import/export at the interface between database systems and statistical packages~\cite{MADlib:2012}. A complementary realization is that large chunks of statistical machine learning code can be expressed as relational queries and computed using database techniques~\cite{Bismarck:SIGMOD:2012,KuNaPa15,SOC:SIGMOD:16,SPOOF:CIDR:2017}.

The problem of solving analytics over databases naturally lends itself to a systematic investigation using the toolbox of concepts and techniques developed by the database theorist, 
and by synergizing ideas from both relational and statistical data modeling. One can exploit database schema information, functional dependencies, state-of-the-art query evaluation algorithms, and well-understood complexity analysis.

\subsubsection*{Contributions}

Our {\em conceptual contribution} is the introduction of a framework for training and evaluating a class of statistical learning models over relational databases. This class, commonly used in retail-planning and forecasting applications~\cite{LB15}, includes ridge linear regression, polynomial regression, factorization machines, and principal component analysis. In such applications, the training dataset is the result of a feature extraction query over the database. Typical databases include weekly sales data, promotions, and product descriptions. A retailer would like to compute a parameterized model, which can predict, for instance, the additional demand generated for a given product due to promotion~\cite{Ron2013}. The feature extraction query is commonly a natural join of the database relations, yet it may join additional relations derived from the input ones using aggregation. The features correspond to database attributes, their categorical values, or aggregates over them. As is prevalent in practical machine learning, the models are trained using a first-order optimization algorithm such as batch or stochastic gradient descent, in part because their convergence rates are dimension-free (for well-behaved objectives). This is a crucial property given the high-dimensionality of our problem as elaborated next.

The main {\em computational challenge} posed by analytics over databases is the large
number of records and of features in the training dataset. There are two types
of features: continuous (quantitative) such as price and sales; and
categorical (qualitative) such as colors, cities, and countries.\footnote{Most
of the raw features we observed in datasets for retail applications are categorical. In several domains, such as statistical arbitrage~\cite{StatArb:2016}, it is common to derive many continuous features from categorical features.} While continuous
features allow for aggregation over their domains, categorical features cannot
be aggregated together. To accommodate the latter, the state-of-the-art approach
is to one-hot encode their active domain: each value in the active domain of an
attribute is encoded by an indicator vector whose dimension is the size of the 
domain. For instance, the colors in the domain 
$\{\mbox{red}, \mbox{green}, \mbox{blue}\}$ can be represented by
indicator vectors $[1,0,0]$ for red, $[0,1,0]$ for green, and $[0,0,1]$ for
blue. The one-hot encoding amounts to a relational representation of the
training dataset with one new attribute per distinct category of each categorical 
feature and with wide tuples whose values are mostly 0. This entails huge 
redundancy due to the presence of the many 0 values. The one-hot encoding 
also blurs the usual distinction between schema and data, 
since the schema can become as large as the input database. 

Closely related to the computational challenge is a {\em cultural challenge}:
the feasibility of a tight integration of analytics and databases may be called into question. 
In terms of pure algorithmic performance, why would
such an approach be more efficient than the common approach that decouples the computation of the training dataset from the learning task, given the widely available pletho\-ra of tools and techniques for the latter?

Our answer to these challenges is that, for a large class of feature extraction queries, it is possible to train a model in time {\em sub-linear} in the output size of the feature extraction query! This makes our approach competitive {\em regardless of the learning techniques} used by the mainstream approaches that first materialize the training dataset, including those that use sampling and stochastic gradient descent to only process a subset of the training dataset.

More concretely, our approach entails {\em three database-centric technical contributions}. 

First, we exploit join dependencies and their factorization in the training dataset to asymptotically improve the per-iteration computation time of a gradient descent algorithm. 

Second, we exploit functional dependencies present in the database to reduce the dimensionality of the underlying optimization problem by only optimizing for those parameters that functionally determine the others and by subsequently recovering the functionally determined parameters using their dependencies. 

Third, we address the shortcomings of one-hot encoding by expressing the sum-product aggregates used to compute the gradient and point evaluation as functional aggregate queries (FAQs)~\cite{faq}. The aggregates over continuous features are expressed as FAQs without free  (i.e., group-by) variables and their computation yields scalar values. In contrast, aggregates over categorical features originating from a set $S$ of database attributes are expressed as FAQs with free variables $S$. The tuples in the result of such FAQs are combinations of categorical values that occur in the training dataset. The ensemble of FAQs defining the gradient form a {\em sparse tensor representation and computation solution} with lower space and time complexity than solutions based on one-hot encoding. In particular, the complexity of our end-to-end solution can be arbitrarily smaller than that of materializing the result of the feature extraction query. 

The above three technical contributions led to the design and implementation of
AC/DC, a gradient descent solver for polynomial regression models and
factorization machines over databases. To train such models of up to 66K
features over the natural join of all relations from a real-world dataset of up
to 86M tuples, AC/DC needs up to 15 minutes on eight cores of a commodity
machine. AC/DC is up to 1,031 times faster than its competitors
MadLib~\cite{MADlib:2012}, libFM~\cite{libfm}, and TensorFlow~\cite{tensorflow}
whenever they do not exceed memory limitation, 24-hour timeout, or internal
design limitations.

\begin{figure}[t]
\centering{
      \begin{tikzpicture}[yscale = 0.5, xscale=0.5, every node/.style={transform shape}]
        \node[data,scale=1.7] (FEQ) {Feature Extraction \\ Query};
        
        \begin{scope}[shift={($(FEQ)+(6,0)$)},scale=1,every node/.style={transform shape}]
          \node[inner sep=.3cm,scale=1.7] at (0,0) (DB) {DB};
          \drawcylinder{1.75}{1.25}{.25}{black}
        \end{scope}
        
        \begin{scope}[shift={($(DB)+(4,0)$)}, scale=1, every node/.style={transform shape}]
          \node[inner sep=.9cm] (0,0) (output_table){};
          \drawtable{1.5}{2}{4}{8}{clr_outofDB}
          \node[clr_outofDB, align=center, scale=1.7] at (0,-2) {materialized output\\ $=$ data matrix};
        \end{scope}
        
        \node[data, clr_outofDB, scale=1.7] at($(output_table)+(5,0)$) (ML) {ML Tool};
        \node[data, scale=1.7] at($(ML)+(4,0)$) (theta_opt) {$\vec\theta$};
        
        \node[data, scale=1.7] at($(ML)+(0,-2.5)$) (model) {Model};

        \node[data,clr_inDB,scale=1.7] at($(model)+(-1.5,-2.5)$) (model_reform)
        {Model Reformulation};
        
        \node[data,clr_inDB,scale=1.7] at($(FEQ)+(0,-5)$) (FAQ)
        {Optimized\\ Aggregate Queries};
        
        \node[data, clr_inDB,scale=1.7] at($(FAQ)+(0.6,-4)$) (sigma_c)
        {Factorized Query\\Evaluation};
        
        \node[data, clr_inDB, scale=1.7] at($(model_reform)+(0,-4)$) (GD)
        {Optimization\\\vspace{.1cm}};
        \begin{scope}[shift={($(GD)+(0,-.45)$)},scale=.5,every node/.style={transform shape}]
          \draw[clr_inDB,->,thick] (-1,.1) arc (170:10:1);
          \draw[clr_inDB,->,thick] (1,-.1) arc (-10:-170:1);
        \end{scope}

        \draw[path, clr_outofDB] (FEQ)--(DB);
        \draw[path, clr_outofDB] (DB)--(output_table);
        \draw[path, clr_outofDB] (output_table)--(ML);
        \draw[path, clr_outofDB] (ML)--(theta_opt);
        \draw[path, clr_outofDB] (model)--(ML);

        \draw[path, clr_inDB] (model)--(model_reform);
        \draw[path, clr_inDB, scale=1.7] (model_reform)--(FAQ); 
        \draw[path, clr_inDB] (DB)--(FAQ);
        \draw[path, clr_inDB] (FEQ)--(FAQ);
        \draw[path, clr_inDB] (FAQ)--($(sigma_c.north)-(0.6,0)$);
        \draw[path, clr_inDB] (sigma_c)--(GD);
        \draw[path, clr_inDB, scale=1.7] (model_reform)--(GD); 
        \draw[path, clr_inDB] (GD) -| (theta_opt);
      \end{tikzpicture}
}
\caption{{\color{clr_inDB}\bf Structure-aware} vs. {\color{clr_outofDB}structure-agnostic} learning: High-level diagram.}
\label{fig:high-level-diagram-intro}
\end{figure}

Figure~\ref{fig:high-level-diagram-intro} depicts schematically the workflows of our approach and the mainstream approach for solving optimization problems. The mainstream approach materializes the result of the feature extraction query, exports it out of the database and imports it as the training dataset in the ML tool, where the desired model is learned. We call it {\em structure-agnostic} since it does not exploit the relational structure of the underlying training dataset
to avoid learning over the full materialization of the training dataset.
In contrast, our {\em structure-aware} approach avoids this materialization and has the following steps: (1) it defines a set of aggregates needed to compute the gradient of the objective function for the desired model; (2) it optimizes these aggregates over the feature extraction query and under dependencies holding in the database and join dependencies defined by the feature extraction query; (3) it computes these aggregates in bulk using factorization techniques and exploiting subexpressions common among them; and (5) it uses a gradient descent solver to compute the model parameters based on the computed aggregates.

This article brings together and extends two lines of our prior work: The
theoretical development of model reparameterization under functional
dependencies and of factorized learning~\cite{ANNOS:PODS:18} and a preliminary
report on the design and implementation of AC/DC~\cite{ANNOS:DEEM:18}.  The
extensions concern the treatment of PCA, simplified proof for model
reparameterization under functional dependencies, different experiments, and a
classification of the existing landscape of structure-aware versus
structure-agnostic approaches to analytics.

\nop{
}

\subsubsection*{Organization} 

The structure of the paper follows our
contributions. Section~\ref{sec:preliminaries} introduces preliminary notions
needed throughout the article. Section~\ref{sec:problem} describes our unified
framework for structure-aware analytics. 
Section~\ref{SEC:ALGO} introduces our
sparse tensor representation and computation
approach for square loss problems (learning polynomial regression models and factorization machines) and principal component analysis together with its 
complexity analysis.
Section~\ref{SEC:FDS} shows how to
exploit functional dependencies to reduce the dimensionality of
learned models. 
Section~\ref{sec:acdc} discusses the design and implementation of the AC/DC system for 
learning models over relational databases.
Section~\ref{sec:experiments} presents our experimental findings.  
Section~\ref{sec:relatedwork} overviews several strands of related work. 
Finally, Section~\ref{sec:conclusion} lists promising directions for future work. 
Further preliminaries and proofs of some theorems are deferred to the electronic appendix.


\section{Preliminaries}
\label{sec:preliminaries}

We use the following notational conventions: bold face letters, e.g., $\mv x$, $\vec\theta$, $\mv x_i$, $\vec\theta_j$, denote vectors or matrices, and normal face letters, e.g., $x_i$, $\theta_j$, $\theta^{(j)}_i$, denote scalars. For any positive integer $n$, $[n]$ denotes the set $\{1,\dots,n\}$. For any set $S$ and positive integer $k$, $\binom S k$ denotes the collection of all $k$-subsets of
$S$. 
Let $S$ be a finite set and $\textsf{Dom}$ be any domain, then 
$\mv a_S = (a_j)_{j\in S} \in \textsf{Dom}^{|S|}$
is a {\em tuple} indexed by $S$, whose components are in $\textsf{Dom}$.
If $S$ and $T$ are disjoint, and given tuples
$\mv a_S$ and $\mv a_T$, the tuple $(\mv a_S,\mv a_T)$ is
interpreted naturally as the tuple $\mv a_{S\cup T}$.
\nop{The tuple $\mv e_S$ is the all-$1$ tuple indexed by $S$.}
The tuple $\mv 0_S$ is the all-$0$ tuple indexed by $S$.
If $S \subseteq G$, then the tuple $\mv 1_{S|G}$ is the characteristic
vector of the subset $S$, i.e., $\mv 1_{S|G}(v) = 1$ if $v \in S$, and $0$
if $v \in G-S$.

\subsection{Feature Extraction Query} 

We consider the setting where the
training dataset $D$ used as input to machine learning is the result of a
query $Q$ called {\em feature extraction query}, over a 
relational database $I$. This query is typically the natural join of the relations in the database. It is also common to join in further relations that are derived from the input relations by aggregating some of their columns. These further relations provide derived features, which add to the raw features readily provided by the input relations.

We use standard notation for query hypergraphs. Let $\calH = (\calV, \calE)$
denote the hypergraph of the join query $Q$, where $\calV$ is the set of variables
occurring in $Q$ and $\calE$ is the set of hyperedges with one hyperedge per set
of variables in a relation symbol $R$ in the body of $Q$. We denote by
$V\subseteq \calV$ the subset of variables selected as features, and let $n=|V|$.
The features in $V$ corresponding to qualitative 
attributes are called {\em categorical}, while those corresponding to 
quantitative attributes are {\em continuous}.
Let $N$ be the size of the largest input relation $R$ in $Q$.
Each tuple $(\mv x, y)\in D$ contains a scalar response (regressand) $y$ and a 
tuple $\mv x$ encoding features (regressors).

\begin{ex}\label{ex:feature-extraction-query}
Consider the following natural join query $Q$ that is a simplified version of a feature extraction query:
\begin{align*}
  &Q({\sf sku,store,day,color,quarter,city,country,
    \mathit{unitsSold},\mathit{price},\mathit{size}}) \\ 
  & \hspace*{2em}\leftarrow\hspace*{.5em}
    \textsf{Sales}({\sf sku,store,day},\mathit{unitsSold}),
    \textsf{Items}({\sf sku,color},price),\\
  &\hspace*{4em} \textsf{Quarter}({\sf day,quarter}),
    \textsf{Stores}({\sf store,city},size),
    \textsf{Country}({\sf city,country}).
\end{align*}
Relation {\sf Sales} records the number of units of a given {\tt sku} (stock
keeping unit) sold at a {\tt store} on a particular {\tt day}.  The retailer is
a global business, so it has stores in different cities and countries.  One
objective is to predict the number of blue units to be sold next year in the
Fall quarter in Berlin. The response is the continuous variable
$\mathit{unitsSold}$, $\calV$ is the set of all variables, and
$V = \calV - \{\mathit{unitsSold}, \textsf{day}\}$, all of which are
categorical except $price$ and  $size$.\qed
\end{ex}

\nop{ = $\{ {\sf
sku,store,day,color,quarter,city,country,\mathit{unitsSold}} \}$}

\subsection{Matrix calculus}
\label{sec:matrix:calculus}

We introduce basic concepts of matrix calculus and the following operations: the
Kronecker/tensor product $\otimes$; the Hadamard product $\circ$; the Khatri-Rao
product $\star$; and the Frobenius inner product of two matrices
$\inner{\cdot,\cdot}$, which reduces to the vector inner product when the
matrices have one column each.  We defer further preliminaries on matrix
calculus to Appendix~\ref{app:matrix-calculus} and connection of tensor
computation and the $\faq$ framework~\cite{faq} to
Appendix~\ref{appendix:faq-tensor}.

\subsubsection*{Basics}

We list here common identities we often use in the paper; for more details see
the Matrix Cookbook~\cite{IMM2012-03274}.
We use {\em denominator layout} for differentiation, i.e., the gradient is a column vector.
Let $\mv A$ be a matrix, and $\mv u, \mv v, \mv x,\mv b$ be vectors, 
where $\mv A$ and $\mv b$ are independent of $\mv x$, and $\mv u$ and $\mv v$ are functions of $\mv x$
then
\begin{eqnarray}
   \pd{\inner{\mv b, \mv x}}{\mv x} &=& \mv b \label{eqn:mc:0}\\
   \pd{\mv x^\top \mv A\mv x}{\mv x} &=& (\mv A+\mv A^\top)\mv x\label{eqn:mc:1}\\
   \pd{\norm{\mv A\mv x - \mv b}_2^2}{\mv x} &=& 2\mv A^\top (\mv A\mv x-\mv b)\label{eqn:mc:2}\\
   \pd{\mv u^\top \mv v}{\mv x} &=& \pd{\mv u^\top}{\mv x}\mv v + \pd{\mv v^\top}{\mv x}\mv u\label{eqn:mc:3}\\
   \pd{(\mv B\mv x+\mv b)^\top \mv C (\mv D\mv x+\mv d)}{\mv x} &=&
   \mv B^\top \mv C(\mv D\mv x+\mv d)+\mv D^\top\mv C^\top (\mv B\mv x+\mv b).\label{eqn:mc:4}
\end{eqnarray}

\subsubsection*{The Product Cookbook: Tensor product, Kronecker product, and Khatri-Rao product}

Next, we discuss some identities regarding tensors. We use $\otimes$ to denote
the {\em tensor product}. When taking tensor product of two matrices, this is called
the {\em Kronecker} product, which is {\em not} the same as the outer product for
matrices, even though the two are isomorphic maps. 
If $\mv A=(a_{ij})$ is an $m\times n$
matrix and $\mv B=(b_{k\ell})$ is a $p \times q$ matrix, then the tensor product
$\mv A \otimes \mv B$ is an $mp \times nq$ matrix whose $((i,k), (j,\ell))$
entry is $a_{ij}b_{k\ell}$.
In particular, if $\mv x =(x_i)_{i\in [m]}$ is an $m$-dimensional vector
and $\mv y=(y_j)_{j\in [p]}$ is an $p$-dimensional vector, then $\mv x \otimes \mv y$ is
an $mp$-dimensional vector whose $(i,j)$ entry is $x_iy_j$; this is 
{\em not} an $m \times p$ matrix as in the case
of the outer product. This layout is the correct layout from the definition of
the tensor (Kronecker) product. 
If $\mv A$ is matrix, then $\mv A^{\otimes k}$ denote the tensor power
$\underbrace{\mv A \otimes \cdots \otimes \mv A}_{k \text{ times}}$.

\bdefn[Tensor product]
Let $\mv A$ and $\mv B$ be tensors of order $r$ and $s$ respectively, i.e., functions $\psi_A(X_1,\dots,X_r)$ and $\psi_B(Y_1,\dots,Y_s)$. The
tensor product $\mv A \otimes \mv B$ is the multilinear function
\[\psi(X_1,\dots,X_r,Y_1,\dots,Y_s) = \psi_A(X_1,\dots,X_r) \psi_B(Y_1,\dots,Y_s).\]
A matrix is a tensor of order $2$.
\edefn

\bdefn[Khatri-Rao product]
Let $\mv A$ and $\mv B$ be two matrices each with $n$ columns.
We use $\mv A \star \mv B$ to denote the matrix with $n$ columns, where the $j$th 
column of $\mv A \star \mv B$ is the tensor product of the $j$th column of $\mv A$ 
with the $j$th columns of $\mv B$. The operator $\star$ is a (special case of)
the {\em Khatri-Rao} product~\cite{MR0238416}, 
where we partition the input matrices into blocks 
of one column each. More elaborately, if $\mv A$ has columns $\mv a_1, \dots,
\mv a_n$, and $\mv B$ has columns $\mv b_1, \dots, \mv b_n$,then one can visualize
the $\star$ operator as follows:
\[ \mv A \star \mv B =
\begin{bmatrix}
   \vert&\vert&\cdots&\vert\\
   \mv a_1&\mv a_2&\cdots &\mv a_n\\
   \vert&\vert&\cdots&\vert
\end{bmatrix} \star
\begin{bmatrix}
   \vert&\vert&\cdots&\vert\\
   \mv b_1&\mv b_2&\cdots &\mv b_n\\
   \vert&\vert&\cdots&\vert
\end{bmatrix} =
\begin{bmatrix}
   \vert&\vert&\cdots&\vert\\
   \vert&\vert&\cdots&\vert\\
   \mv a_1 \otimes \mv b_1&\mv a_2 \otimes \mv b_2&\cdots &\mv a_n \otimes \mv
   b_n\\
   \vert&\vert&\cdots&\vert\\
   \vert&\vert&\cdots&\vert
\end{bmatrix}.
\]
(Note $\mv A$ and $\mv B$ do not need to have the same number of rows.)
\edefn

\bdefn[Hadamard product]
Let $\mv A=(a_{ij})$ and $\mv B=(b_{ij})$ be two $m \times n$ matrices, then the Hadamard product
$\mv A \circ \mv B$ is an $m \times n$ matrix, where each $i,j$ element is given
by $(\mv A \circ \mv B)_{ij} = a_{ij}b_{ij}$.
\edefn


\section{Problem formulation}
\label{sec:problem}

This section introduces a general formulation for a range of machine learning tasks and then lays out a versatile mathematical representation suitable for the in-database treatment of these tasks.

\subsection{Continuous features}

We start with a standard formulation in machine learning, where all model features are numerical.

The training dataset $D$, which is defined by a feature extraction query over a relational database, consists of tuples $(\mv x, y)$ of a feature vector $\mv x$ and a response $y$.

In case of continuous features, $\mv x\in \R^n$ is the vector of $n$ raw input features, or equivalently the variables in the feature extraction query. We denote by $\vec\theta=(\theta_1,\dots,\theta_p)\in \R^p$ the vector of $p$ so-called parameters. Let $m \geq n$ be an integer. We define feature and parameter maps as follows.

The feature map $h: \R^n \to \R^m$ transforms the
raw input vector $\mv x$ into an $m$-vector of ``monomial features'' $h(\mv x) = (h_j(\mv x))_{j\in [m]}$.
Each component $h_j$ is a multivariate {\em monomial} designed to capture the \emph{interactions} among dimensions of input $\mv x$. 
In particular, we write $h_j(\mv x) = \prod_{i\in[n]} x_i^{a_j(i)}$, where
degree $a_j(i)$ represents the level of participation of input dimension $i$ in
the $j$-th monomial feature.

The parameters $\vec\theta$ produce the coefficients associated with features $h$
via parameter map $g : \R^p \to \R^m$, $g(\vec\theta) = (g_j(\vec\theta))_{j\in [m]}$. 
Each component $g_j$ is a multivariate {\em polynomial} of $\vec\theta$. 

A large number of machine learning tasks learn a functional quantity of the form $\inner{g(\vec\theta), h(\mv x)}$, where the parameters $\vec\theta$ are obtained by solving $\min_{\vec\theta} J(\vec\theta)$ with
\begin{equation} J(\vec\theta) = \sum_{(\mv x,y) \in D} \calLL 
   \left( \inner{g(\vec\theta), h(\mv x)},y\right) +
\Omega(\vec\theta). \label{eqn:generic:J}\end{equation}
%
%
$\calLL$ is a loss function, e.g., square loss, and $\Omega$ is a regularizer, e.g., $\ell_1$- or $\ell_2$-norm of $\vec\theta$. For square loss and $\ell_2$-regularization, $J(\vec\theta)$ becomes:
\begin{equation}
   J(\vec\theta) = \frac{1}{2|D|} \sum_{(\mv x,y) \in D} (\inner{g(\vec\theta),
   h(\mv x)}-y)^2 + \frac \lambda 2 \norm{\vec\theta}^2_2.
   \label{eqn:square:loss}
\end{equation}

\begin{ex}\label{ex:lr}
The {\em ridge linear regression} ($\lr$) model with response $y$ and 
regressors $x_1,\dots,x_{n}$ has $p=n+1$, parameters $\vec\theta =
(\theta_0,\dots,\theta_{n})$. For convenience, we set $x_0=1$
corresponding to the bias parameter $\theta_0$. Then, $m=n+1$, 
$\mv g(\vec\theta)=\vec\theta$, and $h(\mv x) = \mv x$.
The inner product becomes $\inner{g(\vec\theta), h(\mv x)} = \inner{\vec\theta, \mv x} = \sum_{i=0}^{n}\theta_i
x_i$ and Equation~\eqref{eqn:square:loss} becomes
$J(\vec\theta) = \frac{1}{2|D|}\sum_{(\mv x,y)\in D} \left(\sum_{i=0}^{n}\theta_i
x_i-y\right)^2 + \frac \lambda 2 \norm{\vec\theta}_2^2.$\qed
\end{ex}

\begin{ex}\label{ex:pr}
   The {\em degree-$d$ polynomial regression} ($\pr^d$) mo\-del with response $y$ and 
   regressors $x_0=1,x_1,\dots,x_n$ has 
   $p=m=\binom{n+d}{d}=\sum_{i=0}^d \binom{n+i-1}{i}$ parameters $\vec\theta =
   (\theta_{\mv a})$, where $\mv a = (a_1,\dots,a_{n})$ is a 
   tuple of non-negative integers such that $\norm{\mv a}_1 \leq d$.
   In this case, $g(\vec\theta)=\vec\theta$, while the components of $h$ are given by
   $h_{\mv a}(\mv x) = \prod_{i=1}^{n} x_i^{a_i}$.\qed
\end{ex}

\begin{ex}\label{ex:fama}
In contrast to polynomial regression models, factorization machines~\cite{Rendle13}  factorize the space of model parameters to better capture data correlations. The {\em degree-$2$ rank-$r$ factorization machines}
   ($\fama^2_r$) model with regressors $x_0=1,x_1,\dots,x_n$ and regressand $y$ has
   parameters $\vec\theta$ consisting of
   $\theta_i$ for $i \in \{0,\dots,n\}$ and
   $\theta^{(l)}_i$ for $i \in [n]$ and $l \in [r]$.
   Training $\fama^2_r$ corresponds to minimizing the following $J(\vec\theta)$:
   {\small
   \begin{equation*}        
\hspace*{-0.1cm} \frac{1}{2|D|}\sum_{(\mv x,y) \in D}
      \left( \sum_{i=0}^{n}\theta_ix_i + 
         \!\!\!\! \sum_{\substack{\{i,j\}\in\binom{[n]}{2}\\\ell \in [r]}} \theta_i^{(\ell)}\theta_j^{(\ell)} x_ix_j - y \right)^2
  \!\! +\frac \lambda 2 \norm{\vec\theta}_2^2.
   \end{equation*}
   }

   This loss function follows Equation~\eqref{eqn:square:loss} with $p=1+n+r n$, $m=1+n+ \binom{n}{2}$, and the maps
   \begin{eqnarray*}
      h_{S}(\mv x) &=& \prod_{i \in S} x_i,  \text{ for } S \subseteq [n], |S| \leq 2\\
      g_{S}(\vec\theta) &=&
      \begin{cases}
         \theta_0 & \text{ when } |S|=0\\
         \theta_i & \text{ when } S=\{i\}\\
        \sum_{\ell=1}^r \theta_i^{(\ell)}\theta_j^{(\ell)} & \text{ when } S=\{i,j\}.
      \end{cases}
      \label{eqn:gh:fama}
   \end{eqnarray*}\qed
\end{ex}


\begin{ex}\label{ex:class}
  \emph{Classification methods} such as support vector machines (SVM), logistic
  regression and Adaboost also fall under the same optimization framework, but
  with different choices of loss $\calLL$ and regularizer $\Omega$.  Typically,
  $\Omega(\vec\theta) = \frac{\lambda}{2}\|\vec\theta\|_2^2$.  Restricting to
  binary class labels $y \in \{\pm 1\}$, the loss function $\calLL(\gamma, y)$,
  where $\gamma=\langle g(\vec\theta), h(\mv x) \rangle$, takes the form
  $\calLL(\gamma,y) = \max \{1-y\gamma, 0 \}$ for SVM,
  $\calLL(\gamma,y) = \log (1+ e^{-y\gamma})$ for logistic regression and
  $\calLL(\gamma,y) = e^{-y\gamma}$ for Adaboost.\qed
\end{ex}

\begin{ex}\label{ex:pca} Various unsupervised learning techniques
can be expressed as iterative optimization procedures according to
which each iteration is reduced to an optimization problem of
the generic form given above. For example, the \emph{Principal Component Analysis (PCA)}
requires solving the following optimization problem to obtain a principal
component direction
\[\max_{\|\vec\theta\|_2 = 1} \vec\theta^\top \vec \Sigma \vec\theta 
= \max_{\vec\theta \in \R^p} \min_{\lambda \in \R} 
\vec\theta^\top \vec \Sigma \vec\theta + \lambda (\|\vec\theta\|_2^2-1),\]
where $\vec \Sigma = \frac{1}{|D|}\sum_{\mv x \in D} \mv x \mv x^\top$ 
is the (empirical) correlation matrix of the given data.
Although there is no response/class label $y$, within each iteration
of the above iteration, for a fixed $\lambda$,
there is a loss function $\calLL$ acting on
feature vector $h(\mv x)$ and parameter vector $g(\vec\theta)$, along with 
a regularizer $\Omega$. Specifically,
we have $h(\mv x) = \vec \Sigma \in \R^{p\times p}$, 
$g(\vec\theta) = \vec\theta \otimes \vec\theta\in \R^{p\times p}$, 
$\calLL = \langle g(\vec\theta), h(\mv x) \rangle_F$, where 
the Frobenius inner product is now employed. In addition,
$\Omega(\vec\theta) = \lambda (\|\vec\theta\|_2^2-1)$.\qed
\end{ex}

\subsection{Categorical features}
\label{sec:cat}

The active domain of a categorical feature/variable 
is a set of possible values or categories, e.g., $\vietnam$, $\england$, and $\usa$ are possible
categories of the categorical feature {\sf country}. 
Categorical features constitute the vast majority  of features we 
observed in machine learning applications.

It is common practice to one-hot encode categorical
variables~\cite{OneHot:Book:2012}. Whereas a continuous variable such as
${\textsf{salary}}$ is mapped to a scalar value $x_{\textsf{salary}}$, a
categorical variable such as {\sf country} is mapped to an indicator vector $\mv
x_{\text{\sf country}}$ -- a vector of binary values indicating the 
category that the variable takes on. For example, if the active domain
of {\sf country} consists of $\vietnam$, $\england$, and $\usa$, then 
$\mv x_{\text{\sf country}} = [x_{\vietnam}, x_{\england}, x_{\usa}] \in
\{0,1\}^3$.
If a tuple in the training dataset has {\sf country = ``england''}, then 
$\mv x_{\text{\sf country}} = [0,1,0]$ for that tuple. 

In general, the feature vector $\mv x$ has the form $\mv x = (\mv x_c)_{c\in V}$, where each component $\mv x_c$ is an indicator vector if $c$ is a categorical variable and a scalar otherwise. Similarly, each component of the parameter vector $\vec\theta$ becomes a matrix, or a vector if the matrix has one column.

\subsection{ Tensor product representation}
\label{sec:tensor}

We accommodate both continuous and categorical features
in our problem formulation~\eqref{eqn:square:loss} by 
replacing arithmetic product by tensor product in the 
component functions of the parameter map $g$ and the feature map $h$.
Specifically, monomials $h_j$ now take the form
\begin{equation} h_j(\mv x) = \bigotimes_{f\in V} \mv x_f^{\otimes a_j(f)}
\label{eqn:hk:monomial}
\end{equation}
with degree vector $\mv a_j = (a_j(f))_{f\in V} \in \mathbb N^n$.
For each $j \in [m]$, the set $V_j=\{ f \in V \suchthat a_j(f)>0\}$ consists of
features that participate in the interaction captured by the (hyper-) monomial $h_j$.
Let $C\subseteq V$ denote the set of categorical variables and $C_j = C\cap V_j$ the subset of categorical variables in $V_j$. For $f \in C_j$, $h_j$ represents $\prod_{f\in C_j} |\pi_{f}(D)|$ many monomials, one for each combination of the  categories, where $\pi_f(D)$ denotes the projection of $D$ onto variable $f$.
Due to one-hot encoding, each element in the vector $\mv x_f$ for a categorical 
variable $f$ is either $0$ or $1$, and $\mv x_f^{a_j(f)} = \mv x_f$ for $a_{j}(f) > 0$. 
Hence, $h_j$ can be simplified as follows:
\begin{equation}
   h_j(\mv x) = \prod_{f \in V_j-C_j}x_{f}^{a_{j}(f)} \cdot \bigotimes_{f \in C_j} \mv x_f. 
   \label{eqn:categorical:h}
\end{equation}
Note that we use $x_f$ instead of boldface $\mv x_f$ since each variable $f \in
V_j-C_j$ is continuous.

\begin{ex}\label{ex:categorical:features}
For illustration, consider a query that extracts tuples over schema $(\textsf{country}, a, b,$ $c, \textsf{color})$ from the database, 
where {\sf country} and {\sf color} are categorical variables, while $a,b,c$ are continuous variables. Moreover, there are two countries {\sf vietnam} and {\sf england}, and  three colors {\sf red}, {\sf green}, and {\sf blue} in the training dataset $D$. Consider three of the possible feature functions:
\begin{eqnarray}
   h_1(\mv x) &=& \mv x_{\textsf{country}} \otimes x^2_ax_c\label{eqn:h1}\\
   h_2(\mv x) &=& \mv x_{\textsf{country}} \otimes \mv x_{\textsf{color}}\otimes x_b\label{eqn:h2}\\
   h_3(\mv x) &=& x_bx_c.\label{eqn:h3}
\end{eqnarray}
Under the one-hot encoding, the schema of the tuples becomes:
$$(\textsf{vietnam}, \textsf{england}, a, b, c, \textsf{red}, \textsf{green},
\textsf{blue}).$$ 

Equation~\eqref{eqn:categorical:h} says that the functions $h_1$ and $h_2$ are actually encoding $8$ functions:
\begin{eqnarray*}
   h_{1,\textsf{vietnam}}(\mv x) &=& x_{\textsf{vietnam}} x_a^2x_c\\
   h_{1,\textsf{england}}(\mv x) &=& x_{\textsf{england}} x_a^2x_c\\
   h_{2,\textsf{vietnam},\textsf{red}}(\mv x) &=& x_{\textsf{vietnam}} x_{\textsf{red}}x_b\\
   h_{2,\textsf{vietnam},\textsf{green}}(\mv x) &=& x_{\textsf{vietnam}} x_{\textsf{green}}x_b\\
   h_{2,\textsf{vietnam},\textsf{blue}}(\mv x) &=& x_{\textsf{vietnam}} x_{\textsf{blue}}x_b\\
   h_{2,\textsf{england},\textsf{red}}(\mv x) &=& x_{\textsf{england}} x_{\textsf{red}}x_b\\
   h_{2,\textsf{england},\textsf{green}}(\mv x) &=& x_{\textsf{england}} x_{\textsf{green}}x_b\\
   h_{2,\textsf{england},\textsf{blue}}(\mv x) &=& x_{\textsf{england}}
   x_{\textsf{blue}}x_b.
\end{eqnarray*}\qed
\end{ex}

We elaborate the tensor product representation for the considered learning models.

\begin{ex}
   In linear regression, parameter $\vec\theta$ is a vector of vectors:
   $\vec\theta = [\vec\theta_0,\dots,\vec\theta_n]$. Since our inner product is
   Frobenius, when computing $\inner{\vec\theta,\mv x}$ we should be multiplying,
   for example, $\theta_{\text{\sf usa}}$ with $x_{\text{\sf usa}}$
   correspondingly. \qed
\end{ex}

\begin{ex}
   In polynomial regression, the parameter $\vec\theta$ is a vector of tensors
   (i.e., high-dimensional matrices).
   Consider for instance the second order term $\theta_{ij}x_ix_j$. When 
   both $i$ and $j$ are continuous, $\theta_{ij}$ is just a scalar.
   Now, suppose $i$ is {\sf country} and $j$ is {\sf color}. Then, the model
   has terms 
   $\theta_{\vietnam,\textsf{red}}x_{\vietnam}x_{\textsf{red}}$,
   $\theta_{\usa,\textsf{green}}x_\usa x_{\textsf{green}}$, and so on.
   {\bf All} these terms are captured by the Frobenius inner product
   $\inner{\vec\theta_{ij},\mv x_i \otimes \mv x_j}$.
   The component $\vec\theta_{ij}$ is a matrix whose number of entries is the
   number of pairs {\sf (country, color)} that appear together in some
   tuple in the training dataset. This number can be much less than the product of the 
   numbers of countries and of colors in the input database.\qed
\end{ex}

\begin{ex}
   Consider the $\fama^2_r$ model from Example~\eqref{ex:fama}, but now with
   categorical variables. From the previous examples, we already know how to
   interpret the linear part $\sum_{i=0}^n \theta_ix_i$ of the model when
   features are categorical. Consider a term in the quadratic part such as
   $\sum_{\ell\in[r]} \theta_i^{(\ell)}\theta_j^{(\ell)} x_ix_j$. 
   When $i$ and $j$ are categorical, the term becomes
   $\inner{\sum_{\ell\in[r]} \vec\theta_i^{(\ell)}\otimes\vec\theta_j^{(\ell)}, \mv x_i
      \otimes \mv x_j}.$\qed
\end{ex}

\section{Database-centric problem reformulation}
\label{SEC:ALGO}

In this section, we show how we reformulate the square loss optimization problems (learning polynomial regression and factorization machine models) and PCA to encode their data-intensive components as FAQs. 
The ensemble of these FAQs form a sparse tensor representation and computation solution with lower space and time complexity than solutions based on one-hot encoding.

\subsection{Solution for square loss problems}
\label{sec:solution-squareloss}

\begin{algorithm}[t]
   \caption{BGD with Armijo line search.}
   \label{algo:bgd}
   $\vec\theta \gets $ a random point\;
   \While{not converged yet}{
      $\alpha \gets $ next step size\;
      $\mv d \gets \grad J(\vec\theta)$\;
      \While{$\left(J(\vec\theta-\alpha \mv d) \geq J(\vec\theta)-\frac
      \alpha 2 \norm{\mv d}_2^2\right)$}{
         $\alpha \gets \alpha/2$ \tcp*[f]{line search}\;
      }
      $\vec\theta \gets \vec\theta - \alpha \mv d$\;
   }
\end{algorithm}

We introduce our approach to learning statistical models for the
setting of square loss function $J(\vec\theta)$ and $\ell_2$-norm as
in~\eqref{eqn:square:loss}. We use a gradient-based optimization algorithm that
employs the first-order gradient information to optimize the loss function
$J(\vec\theta)$. It repeatedly updates the parameters $\vec\theta$ by some step
size $\alpha$ in the direction of the gradient $\grad J(\vec\theta)$ until
convergence. To guarantee convergence, it uses backtracking line search to
ensure that $\alpha$ is sufficiently small to decrease the loss for each
step. Each update step requires two computations: (1) {\em Point evaluation}:
Given $\vec\theta$, compute the scalar $J(\vec\theta)$; and (2) {\em Gradient
  computation}: Given $\vec\theta$, compute the vector $\grad J(\vec\theta)$. In
particular, we use the batch gradient descent (BGD) algorithm with the Armijo
line search condition and the Barzilai-Borwein step size
adjustment~\cite{MR967848,MR2144378}, as depicted in
Algorithm~\ref{algo:bgd}. Quasi-Newton optimization algorithms (e.g., L-BFGS)
and other common line search conditions are also applicable in our framework. We
refer the reader to the excellent review
article~\cite{DBLP:journals/corr/GoldsteinSB14} for additional details on fast
implementations of gradient-descent optimization methods.

\subsubsection*{Continuous features}

We first consider the case without categorical
features. We rewrite the square-loss function~\eqref{eqn:square:loss} to factor
out the data-dependent part of the point evaluation and gradient computation.
Recall that, for $j\in [m]$, $h_j$ denotes the $j$th component function of the
vector-valued function $h$, and $h_j$ is a multivariate monomial in $\mv x$.

\bthm
Let $J(\vec\theta)$ be the function in~\eqref{eqn:square:loss}.
Define the matrix $\vec\Sigma = (\sigma_{ij})_{i,j\in [m]}$, 
the vector $\mv c = (c_i)_{i \in [m]}$, and the scalar $s_Y$ by
\begin{align}
   \vec\Sigma &= \frac{1}{|D|} \sum_{(\mv x, y)\in D} h(\mv x)h(\mv x)^\top
   \label{eqn:sigma}\\
   \vec c &= \frac{1}{|D|} \sum_{(\mv x,y)\in D} y\cdot h(\mv x)\label{eqn:c}\\
   s_Y &= \frac{1}{|D|}\sum_{(\mv x,y)\in D} y^2.
\end{align}
Then,
{\small
\begin{align}
   J(\vec\theta) &= \frac 1 2 g(\vec\theta)^\top \vec\Sigma g(\vec\theta)
   - \inner{g(\vec\theta), \mv c} + \frac{s_Y}{2}
   +\frac \lambda 2 \norm{\vec\theta}^2\label{eqn:point:eval}\\
   \grad J(\vec\theta) &= \pd{g(\vec\theta)^\top}{\vec\theta}\vec\Sigma
   g(\vec\theta)-\pd{g(\vec\theta)^\top}{\vec\theta} \mv c+\lambda\vec\theta.
\label{eqn:gradient}
\end{align}
}
\label{thm:Sigma:h}
\ethm

Note that $\pd{g(\vec\theta)^\top}{\vec\theta}$ is a $p \times m$ matrix, and
$\vec\Sigma$ is an $m \times m$ matrix.  Statistically, $\vec\Sigma$ is related
to the covariance matrix, $\mv c$ to the correlation between the response and
the regressors, and $s_Y$ to the empirical second moment of the response
variable.  Theorem~\ref{thm:Sigma:h} allows us to compute the two key steps of
BGD {\em without} scanning through the data again, because the quantities
$(\vec\Sigma, \mv c, s_Y)$ can be computed efficiently in a preprocessing step
{\em inside the database} as aggregates over the feature extraction query $Q$.
\begin{ex}\label{ex:sigma:ij:cont}
  Consider the query $Q$ in Example~\ref{ex:feature-extraction-query}, where the
  set of features is {\sf \{sku, store, day, color, quarter, city, country,
    $price$, $size$\}} and \textit{unitsSold} is the response variable. In this
  query $n=9$, and thus for a $\pr_2$ model we have $m=1+9+\binom{9}{2}=46$
  parameters. Consider two indices $i$ and $j$ to the component functions of $g$
  and $h$, where $i = (price)$ and $j = (size)$. Then we can compute the entry
  $\sigma_{ij} \in \vec\Sigma$ with the following SQL query:
  \begin{align*}
    &\texttt{SELECT SUM($price$ * $size$) FROM D;}
  \end{align*}

  \vspace*{-1em} \qed
\end{ex}

When $g$ is the identity function, i.e., the model is linear,
as is the case in $\pr$ (and thus $\lr$) model,  
\eqref{eqn:point:eval}
and~\eqref{eqn:gradient} become particularly simple:

\bcor
In a linear model (i.e., $g(\vec\theta)=\vec\theta$), 
\begin{align}
   J(\vec\theta) &= \frac 1 2 \vec\theta^\top \vec\Sigma \vec\theta
   - \inner{\vec\theta, \mv c} + \frac{s_Y}{2}
   +\frac \lambda 2 \norm{\vec\theta}_2^2\label{eqn:point:eval:pr}\\
   \grad J(\vec\theta) &= \vec\Sigma \vec\theta+
   \lambda\vec\theta-\mv c.
\label{eqn:gradient:pr}
\end{align}
Let $\mv d=\grad J(\vec\theta)$. 
Then,
\begin{equation}
   \grad J(\vec\theta - \alpha \mv d) = (1-\alpha)\mv d - \alpha \vec\Sigma \mv
   d.
   \label{eqn:next:grad}
\end{equation}
The Armijo condition
$J(\vec\theta-\alpha\mv d)\geq J(\vec\theta)-\frac \alpha
2\norm{\mv d}_2^2$ becomes: 
\begin{equation}
   \alpha \vec\theta^\top\vec\Sigma\mv d 
   -\frac{\alpha^2}{2}\mv d^\top\vec\Sigma\mv d
      -\alpha \inner{\mv c,\mv d}
   + \lambda \alpha \inner{\vec\theta,\mv d}
   \leq \frac \alpha 2(\lambda\alpha+1)\norm{\mv d}_2^2.
   \label{eqn:alternative:armijo}
\end{equation}
\label{cor:Sigma:h}
\ecor
The significance of~\eqref{eqn:alternative:armijo} is as follows.
In a typical iteration of BGD, we have to backtrack a few times (say $t$
times) for each value of $\alpha$. If we were to recompute
$J(\vec\theta-\alpha\mv d)$ using~\eqref{eqn:point:eval:pr} each time, then
the runtime of Armijo backtracking search is $O(tm^2)$, even after we have
already computed $\mv d$ and $J(\vec\theta)$.
Now, using~\eqref{eqn:alternative:armijo}, we can compute in advance 
the following quantities (in this order): 
$\mv d$, 
$\norm{\vec\theta}_2^2$,
$\vec\Sigma\mv d$, 
$\inner{\mv c, \mv d}$, 
$\inner{\vec\theta, \mv d}$, 
$\mv d^\top\vec\Sigma\mv d$,
$\vec\theta^\top\vec\Sigma\mv d$.
Then, each check for
inequality~\eqref{eqn:alternative:armijo} can be done in $O(1)$-time, for a
total of $O(m^2+t)$-times.
Once we have determined the step size $\alpha$, \eqref{eqn:next:grad}
allows us to compute the next gradient (i.e., the next $\mv d$) in $O(m)$, because we have already computed $\vec\Sigma \mv d$ for line search.

To implement BGD, we need to compute four quantities efficiently:
the $\vec\Sigma$ matrix in~\eqref{eqn:sigma},
the vector $\mv c$ in~\eqref{eqn:c},
point evaluation in~\eqref{eqn:point:eval},
and the gradient in~\eqref{eqn:gradient}.
The covariance matrix and the correlation vector only have to be computed once
in a pre-processing step. The gradient is computed at every iteration,
which includes several point evaluations as we perform line search.\footnote{In
our implementation, each iteration typically involves $1$-$4$ backtracking
steps.}
We do not need to compute the second moment $s_Y$
because optimizing $J(\vec\theta)$ is the same as optimizing
$J(\vec\theta)-s_Y$. 
Before describing how those four quantities can be computed efficiently, we
discuss how we deal with categorical features.

\subsubsection*{Categorical features via sparse tensors}

The more interesting, more common, and also considerably challenging situation is in the presence of categorical features. We next explain how we accommodate categorical features in the computation of $\vec\Sigma$ and $\mv c$.

\begin{ex}
  In Example~\ref{ex:categorical:features}, the matrix $\vec\Sigma$ is of size
  $8\times 8$ instead of $3\times 3$ after one-hot encoding. However, many of
  those entries are $0$, for instance ($\forall (\mv x,y) \in D$):
\begin{align*}
   h_{1,\textsf{vietnam}}(\mv x)  h_{1,\textsf{england}}(\mv x) &=0\\
   h_{1,\textsf{england}}(\mv x)  h_{2,\textsf{vietnam},\textsf{blue}}(\mv x) &= 0\\
   h_{2,\textsf{vietnam},\textsf{blue}}(\mv x)
   h_{2,\textsf{england},\textsf{blue}}(\mv x) &= 0\\
   h_{2,\textsf{vietnam},\textsf{blue}}(\mv x)
   h_{2,\textsf{vietnam},\textsf{red}}(\mv x) &= 0.
\end{align*}
The reason is that the indicator variables $x_{\textsf{blue}}$ and
$x_{\textsf{england}}$ act like selection clauses
$x_{\textsf{color}} = \textsf{blue}$ and
$x_{\textsf{country}} = \textsf{england}$. More concretely, we can rewrite the
entry $\sigma_{ij}$ as an aggregate over a more selective query. For instance,
the entry that corresponds to the product of functions
$h_{1,\textsf{vietnam}}(\mv x)$ and $h_{2,\textsf{vietnam},\textsf{red}}(\mv x)$
from Example~\ref{ex:categorical:features} can be rewritten as follows:
\begin{align*}
  &\sum_{(\mv x,y) \in D} h_{1,\textsf{vietnam}}(\mv x) 
  h_{2,\textsf{vietnam},\textsf{red}}(\mv x) = \sum_{\phi} x_a^2x_cx_b,
\end{align*}
where
$\phi = ((\mv x,y) \in D \wedge x_{\textsf{color}} = \textsf{red}\wedge
x_{\textsf{country}} = \textsf{vietnam})$.\qed
\end{ex}
Extrapolating straightforwardly, if we were to write $\vec\Sigma$ down in the
one-hot encoded feature space, then the entries $\sigma_{ij}$ under one-hot
encoding got unrolled into many entries.  Let $C_i$ and $C_j$ be the set of
categorical variables for $h_i$ and $h_j$ as defined in
Section~\ref{sec:tensor}. Then, $\sigma_{ij}$ is in fact a tensor
$\vec\sigma_{ij}$ of dimension
$\prod_{f\in C_i } |\pi_f(D)| \times \prod_{f\in C_j} |\pi_f(D)|$, because
\begin{equation}
   \vec\sigma_{ij} 
   = \frac{1}{|D|} \sum_{(\mv x, y)\in D} h_i(\mv x) h_j(\mv x)^\top.
\label{eqn:sigma:ij:categorical}
\end{equation}
Similarly, each component $c_j$ of $\mv c$ defined in \eqref{eqn:c} is a 
tensor $\mv c_j$ of dimension $\prod_{f\in C_j } |\pi_f(D)|$, because
$h_j(\mv x)$ is a tensor in the categorical case. The following follows
immediately.
\bthm
Theorem~\ref{thm:Sigma:h} remains valid even when some features are categorical.
\ethm

Note that the outer product in~\eqref{eqn:sigma:ij:categorical} specifies 
the matrix layout of $\vec\sigma_{ij}$, and so $\vec\Sigma$ is a block matrix,
each of whose blocks is $\vec\sigma_{ij}$.
Furthermore, if we were to layout the tensor $\vec\sigma_{ij}$ as a vector, 
we can also write it as
\begin{equation} \vec\sigma_{ij} = \frac{1}{|D|} \sum_{(\mv x, y)\in D} h_i(\mv x) \otimes
h_j(\mv x).
   \label{eqn:sigma:ij:tensor}
\end{equation}
The previous example demonstrates that the dimensionalities of 
$\vec\sigma_{ij}$ and $\mv c_j$ can be very large.
Fortunately, the tensors are very sparse, and a sparse representation 
of them can be computed with functional aggregate queries (in the
$\faq$-framework~\cite{faq}) as shown in 
Proposition~\ref{prop:precomputation:time} below. We next illustrate the
sparsity.

\begin{ex}\label{ex:store:city}
  We extend the Example~\ref{ex:sigma:ij:cont} for entries in Sigma with
  categorical variables. Consider two indices $i$ and $j$ to the component
  functions of $g$ and $h$, where $i = (\textsf{store, city})$ and
  $j = (\textsf{city})$.  Suppose the query result states that the retailer has
  $N_s$ stores in $N_c$ countries. Then, the full dimensionality of the tensor
  $\vec\sigma_{ij}$ is $N_s \times N_c^2$, because by definition it was defined
  to be
   \begin{equation} \vec\sigma_{ij} = \frac{1}{|D|}\sum_{(\mv x,y)\in D} 
        \underbrace{\mv x_{\textsf{store}} \otimes 
        \mv x_{\textsf{city}}}_{h_i(\mv x)} \otimes 
        \underbrace{\mv x_{\textsf{city}}}_{h_j(\mv x)}.
        \label{eqn:store:city}
   \end{equation}
   Recall that $\mv x_{\textsf{store}}$ and $\mv x_{\textsf{city}}$ are both
   indicator vectors. The above tensor has the following
   straightforward interpretation: for every triple $(store,city_1,city_2)$, where $s$
   is a store and $c_1$ and $c_2$ are cities, this triple entry of the
   tensor counts the number of data points $(\mv x, y) \in D$ for this particular
   combination of store and cities (divided by $1/|D|$).  
   Most of these $(s,c_1,c_2)$-entries are $0$. For example, if 
   $c_1\neq c_2$ then the count is zero. Thus, we can concentrate on computing
   entries of the form $(s,c,c)$:
   \begin{align*}
      &\texttt{SELECT store, city, COUNT(*) FROM D GROUP BY store, city;}
   \end{align*}
   Better yet, since {\sf store} functionally determines {\sf city}, the number
   of entries in the query output is bounded by $N_s$. 
   Using relations to represent sparse tensor results 
   in massive space saving.

   We can also succinctly represent entries in $\vec\Sigma$ that are composed of
   continuous and categorical variables. Consider the entry that corresponds to
   dimensions $i = (\textsf{store, city})$ and $j = ({\sf city}, price)$.  We
   can compute this entry with the following SQL query:
   \begin{align*}
     &\texttt{SELECT store, city, SUM(price) FROM D GROUP BY store, city;}
   \end{align*}

   \vspace*{-1em} \qed
\end{ex}

\subsection{Solution for Principal Component Analysis}
\label{sec:solution-pca}

We next consider principal component analysis (PCA) over the training dataset
defined by a feature extraction query. We focus on the problem of computing the
top-$K$ principal components, which correspond to the eigenvectors of the
covariance matrix. Once computed, the principal components are then used to
transform the data to a lower dimensional space. We show that the solution to
this problem requires similar computations as our solution for square loss
problems in Section~\ref{sec:solution-squareloss}.

\subsubsection*{Continuous features}

We first consider the case with continuous features  only. Let
$\vec\mu = \frac{1}{|D|}\sum_{\mv x \in D} \mv x$ be the vector of means for
each variable in the feature extraction query, and
$\vec \Sigma_1 = \frac{1}{|D|}\sum_{\mv x \in D} (\mv x-\vec\mu) (\mv x -
\vec\mu)^\top$ the centered covariance matrix. The top-$K$ eigenvectors
$\vec\theta = (\vec\theta_1,\ldots,\vec\theta_K)$ and the corresponding
eigenvalues $\vec\lambda = (\lambda_1,\ldots,\lambda_K)$ can be computed one at
a time using the min-max theorem based on the Rayleigh quotient~\cite{strang:la}:
\begin{align}
  \max_{\vec\theta_j \in \R^p} \min_{\lambda_j \in \R} \vec\theta_j^\top \vec \Sigma_j
  \vec\theta_j + \lambda_j (\|\vec\theta_j\|^2-1)
\end{align}

We compute the optimal solution for the top eigenvector $\vec\theta_1$ using a
gradient-based optimization algorithm, which optimizes the following loss
function by alternating between performing gradient ascent with respect to
$\vec\theta_1$ and gradient descent with respect to $\lambda_1$ until
convergence:
\begin{align}\label{eqn:point:PCA}
  J (\vec\theta_1, \lambda_1)  = \vec\theta_1^\top \vec \Sigma_1
  \vec\theta_1 + \lambda_1 (\|\vec\theta_1\|^2-1)
\end{align}
The gradient optimization steps can then be done with Algorithm~\ref{algo:bgd},
where the gradient of $J(\vec\theta_1, \lambda_1)$ for the two subproblems is given
by:
\begin{align}
  \label{eqn:grad:PCA}
  \grad_{\vec\theta_1} J(\vec\theta_1, \lambda_1) &= \vec\Sigma_1\vec\theta_1 -
                                                    2\lambda_1 \vec\theta_1 \\
  \grad_{\lambda_1}  J(\vec\theta_1, \lambda_1) &= \norm{\vec\theta_1}^2 - 1
\end{align}

The subsequent eigenvectors are computed with the same optimization procedure
but over an updated covariance matrix that subtracts all previously computed
principal components. The iteration step assumes we already computed the
covariance matrix $\vec\Sigma_l$, the eigenvector $\vec\theta_l$, and the
eigenvalue $\lambda_l$ for $l \in [K-1]$. The step then computes the eigenvector
$\vec\theta_{l+1}$ and the eigenvalue $\lambda_{l+1}$ over the covariance matrix
\begin{align}
  \vec\Sigma_{l+1} = \vec\Sigma_{l} - \lambda_{l}\vec\theta_{l}\vec\theta_{l}^\top.
  \label{eqn:update:sigma}
\end{align}

Once we computed the top-$K$ eigenvectors $\vec\theta$, the projection of a
training sample $\mv x \in D$ onto the lower $K$-dimensional space is given by
the inner product $\mv x^\top \vec\theta$.

As for the square-loss problems, we can compute $\vec\Sigma_1$ once, and then
compute the eigenvectors without scanning the data again. If the data is
centered in a preprocessing step, then the computation of $\vec\Sigma_1$ for PCA
is identical to \eqref{eqn:sigma} for the case of linear regression. If the data
is not centered, we can compute the covariance matrix with the following
reformulation:
\begin{align}
  \vec\Sigma_1
  &= \frac{1}{|D|} \sum_{\mv x \in D} (\mv x-\vec\mu)(\mv x - \vec\mu)^\top
   \;=\; \frac{1}{|D|} \sum_{\mv x \in D} \mv x\mv x^\top -
    \frac{2\vec\mu}{|D|} \sum_{\mv x \in D}\mv x^\top +
    \frac{1}{|D|} \sum_{\mv x \in D}\vec\mu\vec\mu^\top\nonumber\\
  &= \frac{1}{|D|} \sum_{\mv x \in D} \mv x \mv x^\top
    - 2\vec\mu\vec\mu^\top + \vec\mu\vec\mu^\top 
   \;=\; \frac{1}{|D|} \sum_{\mv x \in D} \mv x \mv x^\top - \vec\mu\vec\mu^\top
  \label{eq:center:sigma}
\end{align}
Thus, we can compute the covariance matrix by first computing the matrix
from~\eqref{eqn:sigma} where $h(\mv x) = \mv x$ and then subtracting
$\vec\mu\vec\mu^\top$ to center the data.

The gradient with respect to $\vec\theta$ for PCA requires the same computation over
$\vec\Sigma_1$ as the gradient for linear regression models
(c.f. \eqref{eqn:gradient:pr}).

\subsubsection*{Categorical features via sparse tensors}

PCA is based on the analysis of variance between variables, and therefore it
cannot be computed directly over categorical data. It can however be meaningful
to compute PCA over one-hot encoded categorical data, which would provide
insights into the variance of the frequency of the co-occurrence of categories for 
different categorical variables. We can compute PCA over one-hot
encoded categorical variables efficiently by computing it over the sparse
representation of the covariance matrix, which is a variant of the sparse tensor
representation of the $\vec\Sigma$ matrix that we introduced for the case of
square-loss problems.

One difference between the square loss problems and PCA is that PCA requires its
features to be linearly independent. This property is not satisfied by one-hot
encoding, because it is possible to derive the indicator value for one category
based on a linear combination of the indicator values for all other
categories. For this reason, it is common practice to do one-hot encoding of the
categorical variables for all but one category. In our problem formulation, this
means that for a categorical variable $c$, we encode the corresponding component
$\mv x_c$ as an indicator vector whose size is the number of its categories
minus one, so that we one-hot encode over all but the last category of $c$. This
encoding is often referred to as {\em dummy encoding} in many data science
tools.

Another difference is the requirement to center the data. For categorical data,
it is not desirable to center the data in a preprocessing step, as this would
require a one-hot encoding of the input relations. To avoid the materialization of the
one-hot encoding, we compute the non-centered matrix first, and then
subtract $\vec\mu\vec\mu^\top$, as shown in~\eqref{eq:center:sigma}. 
The sparse representation of the covariance matrix is then a block
matrix, where each entry $\vec\sigma_{ij} \in \vec\Sigma_1$ is defined as:
\begin{align}
  \vec\sigma_{ij} = \frac{1}{|D|} \sum_{\mv x \in D} \mv x_i \mv x_j^\top  -
  \vec\mu_i\vec\mu_j^\top
  \label{eqn:covar:ij:categorical}
\end{align}
The vector of means $\vec\mu$ has the same dimension as $\mv x$, where for each
categorical variable $c \in V$ the component $\vec\mu_c \in \vec\mu$ is the
vector of frequencies for all but one category in the domain of $c$:
\begin{align}
  \vec\mu_c = \frac{1}{|D|} \sum_{\mv x \in D} \mv x_c.
  \label{eqn:mu:categorical}
\end{align}
The vector $\vec\mu_c$ can be computed efficiently as a SQL count query with 
group-by variable $c$. We drop the group with the
lowest count and divide the count for each other group by $|D|$.

The resulting matrix $\vec\Sigma_1$ has the same structure as the sparse tensor that is
computed for linear regression problems. In fact, the quantity $\vec\sigma_{ij}$
in~\eqref{eqn:covar:ij:categorical} is simply the centered variant of the
expression in~\eqref{eqn:sigma:ij:categorical} for the case where
$h(\mv x) = \mv x$. The centering of the $\vec\Sigma_1$ as well as updating the
matrix for subsequent eigenvectors can be expressed as group-by aggregate queries
and computed without materializing the quantities $\vec\mu\vec\mu^\top$ and
$\vec\theta\vec\theta^\top$.


\begin{ex}\label{ex:store:city:pca}
  Consider the entry $\vec\sigma_{ij} \in \vec\Sigma_1$ where
  $i = (\textsf{store})$ and $j = (\textsf{city})$. We can compute the centered
  entry in the covariance matrix based on the non-centered entry
  $\vec\sigma_{ij}$ and the frequency vectors for {\sf store} and {\sf
    city}. The non-centered entry is computed with the SQL query in
  Example~\ref{ex:store:city},

  Let $\vec\mu_s({\sf store,val})$ and $\vec\mu_c({\sf city,val})$ be the
  relational encoding of the frequency vectors for $\textsf{store}$ and
  respectively $\textsf{city}$. The relations store tuples that give for each
  city and respectively store the corresponding frequency that is denoted by
  {\sf val}. We can then compute the $(i,j)$ entry in the centered covariance
  matrix without materializing the product of $\vec\mu_s$ and $\vec\mu_c^\top$
  with the following SQL query:
   \begin{align*}
     &\texttt{SELECT store, city, SUM($\vec\sigma_{ij}$.val -
       $\vec\mu_c$.val * $\vec\mu_s$.val)}\\
     &\texttt{FROM $\vec\sigma_{ij}$,$\vec\mu_c$,$\vec\mu_s$ WHERE
       $\vec\sigma_{ij}$.city = $\vec\mu_c$.city AND
       $\vec\sigma_{ij}$.store = $\vec\mu_s$.store}\\
     &\texttt{GROUP BY store, city;}
   \end{align*}

   Let $\vec\theta_s({\sf store,val})$ and $\vec\theta_c({\sf city,val})$ be the
   relational encodings of the components in $\vec\theta_1$ that correspond to
   {\sf store} and respectively {\sf city}. We can compute the updated entry
   $\vec\sigma_{ij} \in \vec\Sigma_2$ based on~\eqref{eqn:update:sigma} without
   materializing the product of $\vec\theta_s$ and $\vec\theta_c^\top$ with the
   following query:
   \begin{align*}
     &\texttt{SELECT store, city, SUM($\vec\sigma_{ij}$.val -
       $\lambda_1$ * $\vec\theta_c$.val * $\vec\theta_s$.val)}\\
     &\texttt{FROM $\vec\sigma_{ij}$,$\vec\theta_c$,$\vec\theta_s$ WHERE
       $\vec\sigma_{ij}$.city = $\vec\theta_c$.city AND
       $\vec\sigma_{ij}$.store = $\vec\theta_s$.store;}\\
     &\texttt{GROUP BY store, city;}
   \end{align*}
   
   \vspace*{-1em} \qed
\end{ex}

The eigenvectors and eigenvalues can be computed on top of the sparse
representation of the covariance matrix without touching the input database, and
the gradient with respect to the eigenvectors requires similar computation as
the gradient for square loss problems. We next show how we can compute the
sparse tensor representation.

\subsection{Efficient computation of the sparse tensor representation}
\label{sec:evaluation}

We consider the problem of computing the sparse tensor representation for a
given optimization problem. For square-loss problems the sparse tensor captures
the quantities $\vec\Sigma$ and $\mv c$, and for PCA we compute the
non-centered covariance matrix (referred to as $\vec\Sigma$ for uniformity),
which is then centered in a subsequent step as shown in
Example~\ref{ex:store:city:pca}.

An immediate approach to computing this representation is to first materialize
the result of the feature extraction query $Q$ using an efficient query engine,
e.g., a worst-case optimal join algorithm, and then compute the entries in the
representation as aggregates over the query result.  This approach, however, is
suboptimal, since the listing representation of the query result is highly
redundant and not necessary for the computation of the aggregates.

We employ two orthogonal observations to avoid this redundancy.

First, we use the FAQ~\cite{faq} and FDB~\cite{OS:SIGREC:2016} frameworks for
factorized computation of aggregates over joins. In a nutshell, factorized
aggregate computation unifies three powerful ideas: worst-case optimal join
processing, query plans defined by fractional hypertree decompositions of join
queries, and pushing aggregates past joins.

Second, we exploit the observation that in the computation of $\vec\Sigma$
many {\em distinct} tensors $\vec\sigma_{ij}$ have {\em identical} sparse
representations. For instance, the tensor $\vec\sigma_{ij}$ from
Example~\ref{ex:store:city} corresponding to $i = (\textsf{store, city})$ and
$j = (\textsf{city})$ has the same sparse representation as any of the following
tensors: $(i,j) \in \{ (({\sf city, city}), {\sf store}),$
$(({\sf store, store}), {\sf city}),$
$(({\sf store, city}), {\sf store}),\ldots\}$. This is because \textsf{store}
and \textsf{city} are categorical features and taking any power of the binary
values in their indicator vectors does not change these values.  Furthermore,
any of the two features can be in $i$ and/or $j$.

The time complexity of computing the representation can be lower than that of
materializing the result of the feature extraction query $Q$. Let
$|\vec\sigma_{ij}|$ denote the {\em size} (i.e., number of tuples) of the sparse
representation of the $\vec\sigma_{ij}$ tensor. Let $\faqw(i,j)$ denote the {\em
  $\faq$-width} of the $\faq$-query\footnote{We show in the proof of
  Proposition~\ref{prop:precomputation:time} in Appendix~\ref{app:sec:algo} how
  to express $\vec\sigma_{ij}$ and $\mv c_j$ as $\faq$-queries.} that expresses
the aggregate $\vec\sigma_{ij}$ over $Q$; $\fhtw$ be the fractional hypertree
width of $Q$; and $\rho^*$ be the fractional edge cover number \footnote{Due to
  space limitation, these width notions are defined in
  Appendix~\ref{app:subsec:faqw}.} of $Q$. Let $I$ be the input database and
$D=Q(I)$.  Let $N$ be the size of the largest input relation in $Q$, which means
that $|D| = O(N^{\rho^*})$. Recall that $\calV$ is the set of query variables in
$Q$, $\calE$ is the set of relations in $Q$, and $m$ is the number of
features. The time to compute the sparse tensor representation can be bounded as
follows.

\bprop
The tensors $\vec\sigma_{ij}$ and $\mv c_j$ can be sparsely represented by 
$\faq$-queries with group-by variables $C_i \cup C_j$ and $C_j$, respectively.
They can be computed in time   
\[ 
O\left( |\calV|^2\cdot |\calE| \cdot \sum_{i,j\in [m]} ( N^{\faqw(i,j)} + |\vec\sigma_{ij}| )\cdot\log N \right).
\]
\label{prop:precomputation:time}
\eprop

In case all features in $D$ are continuous, i.e., $C_j=\emptyset$ for all
$j \in [m]$, then $\faqw(i,j) = \fhtw$~\cite{faq} and the overall runtime
becomes $O(|\calV|^2\cdot|\calE|\cdot m^2 \cdot N^{\fhtw}\cdot\log N)$.  When
some features are categorical, we can also bound the width $\faqw(i,j)$ and
tensor size.

\bprop Let $c = \max_{i,j}|C_i\cup C_j|$ be the maximum number of categorical
variables for any $\vec\sigma_{ij}$. Then, $\faqw(i,j) \leq \fhtw+c-1$ and
$|\vec\sigma_{ij}|\leq N^{\min\{\rho^*,c\}}\}$, $\forall i,j\in [m]$.

For any query $Q$ with $\rho^* > \fhtw+c-1$, there are infinitely many database
instances for which
\begin{equation}
   \lim_{N\to\infty}
\frac{N^{\rho^*}}{\sum_{i,j\in [m]}(N^{\faqw(i,j)}+N^{\min\{\rho^*,c\}})\log N} =
\infty.
   \label{eqn:unbounded:gap}
\end{equation}
\label{prop:output:size}
\eprop

Our precomputation step takes strictly sub-output-size runtime for infinitely
many queries and database instances. If we were to compute $\vec\sigma_{ij}$ on
a training dataset with categorical variables one-hot encoded, then the
complexity would raise to
$O(|\calV|^2\cdot|\calE|\cdot m^2 \cdot N^{\fhtw+2d}\log N)$, where $d$ is the
degree of the polynomial regression model or factorization machine.

\subsection{Point evaluation and gradient computation}
\label{sec:gradientcomputation}

We introduce two ideas for efficient point evaluation and gradient
computation. 

First, we employ a sparse representation of tensors in the {\em
  parameter space}.  We need to evaluate the component functions of $g$, which
are polynomial.  In the $\fama^2_r$ example, for instance, we evaluate
expressions of the form
\begin{equation}
   g_{\textsf{store, city}}(\vec\theta) = 
   \sum_{\ell=1}^r 
   \vec\theta^{(\ell)}_{\textsf{store}} \otimes
   \vec\theta^{(\ell)}_{\textsf{city}}.
   \label{eqn:g:store:city}
\end{equation}
The result is a $2$-way tensor whose CP-decomposition (a sum of rank-$1$ tensors) is already given by \eqref{eqn:g:store:city}! There is no point in materializing the result of $g_{\textsf{store, city}}(\vec\theta)$ and we instead keep it as is. Assuming
$N_c$ distinct cities and $N_s$ distinct stores in the training dataset $D$, if we were to materialize the tensor, then we would end up with an $\Omega(N_cN_s)$-sized result for absolutely no gain in computational  and space complexity, while the space complexity of the CP-decomposition is only $O(N_c+N_s)$. This is a prime example of factorization of the parameter space.

Second, we explain how to evaluate \eqref{eqn:point:eval}
and~\eqref{eqn:gradient} with our sparse tensor representation. The same
techniques can also be applied to evaluate \eqref{eqn:point:PCA} and
\eqref{eqn:grad:PCA} for PCA. There are two aspects of our solution worth
spelling out: (1) how to multiply two tensors, e.g., $\vec\sigma_{ij}$ and
$g_j(\vec\theta)$, and (2) how to exploit that some tensors have the same
representation to speed up the point evaluation and gradient computation.

To answer question (1), we need to know the intrinsic dimension of the tensor
$\vec\sigma_{ij}$. In order to compute $\vec\Sigma g(\vec\theta)$ in
Example~\ref{ex:store:city}, we need to multiply $\vec\sigma_{ij}$ with
$g_j(\vec\theta)$ for $i = (\textsf{store, city})$ and $j = (\textsf{city})$. In
a linear model, $g_j(\vec\theta) = \vec\theta_j = \vec\theta_{\textsf{city}}$.
In this case, when computing $\vec\sigma_{ij} \vec\theta_{\textsf{city}}$ we
marginalize away one {\sf city} dimension of the tensor, while keeping the other two dimensions {\sf store, city}. This is captured by the following query:
\begin{align*}
&\texttt{SELECT store, city, SUM}(\vec \sigma_{i,j}.\texttt{val} * \vec \theta_j.\texttt{val})\\
&\texttt{FROM } \vec \sigma_{i,j}, \vec \theta_j \texttt{ WHERE } \vec
   \sigma_{i,j}.\texttt{city} = \vec \theta_j.\texttt{city}\\
&\texttt{GROUP BY store, city;}
\end{align*}
where the tensors $\vec \sigma_{i,j}$ and $\vec \theta_j$ map $(\textsf{store, city})$ and respectively $(\textsf{city})$ to aggregate values. In words, $\vec\sigma_{ij} g_j(\vec\theta)$ is computed by a group-by aggregate query where the group-by variables are precisely the variables in $C_i$.

For question (2), we use the CP-decomposition of the parameter space as discussed earlier. Suppose now we are looking at the $\vec\sigma_{ij}$ tensor where  $i = (\textsf{city})$ and $j = (\textsf{store, city})$. Note that this tensor has the identical representation as the above tensor, but it is a {\em different} tensor. In a $\fama^2_r$ model, we would want to multiply this tensor with the component function $g_j(\vec\theta)$ defined in~\eqref{eqn:g:store:city} above. We do so by multiplying it with each of the terms $\vec\theta^{(\ell)}_{\textsf{store}} \otimes \vec\theta^{(\ell)}_{\textsf{city}}$, one by one for $\ell = 1, \dots, r$, and then add up the result. Multiplying the tensor $\vec\sigma_{ij}$ with the first term $\vec\theta^{(1)}_{\textsf{store}} \otimes \vec\theta^{(1)}_{\textsf{city}}$ corresponds precisely to the following query:
\begin{align*}
&\texttt{SELECT city, SUM}(\vec\sigma_{i,j}.\texttt{val} * \vec\theta^{(1)}_{\texttt{store}}.\texttt{val} * \vec\theta^{(1)}_{\texttt{city}}.\texttt{val})\\
&\texttt{FROM } \vec\sigma_{i,j}, \vec\theta^{(1)}_{\texttt{store}}, \vec\theta^{(1)}_{\texttt{city}}
\texttt{WHERE } \vec\sigma_{i,j}.\texttt{city} =  \vec\theta^{(1)}_{\texttt{city}}.\texttt{city} \texttt{ AND } \vec\sigma_{i,j}.\texttt{store} =  \vec\theta^{(1)}_{\texttt{store}}.\texttt{store}\\
&\texttt{GROUP BY city;}
\end{align*}
where the tensors $\vec\sigma_{i,j}$, $\vec\theta^{(1)}_{\textsf{city}}$, and
$\vec\theta^{(1)}_{\textsf{store}}$ map $(\textsf{store, city})$, $(\textsf{city})$, and respectively 
$(\textsf{store})$ to aggregate values.
Finally, to answer question (2), note that for the same column $j$ (i.e., the
same component function $g_j(\vec\theta)$), there can be multiple tensors
$\vec\sigma_{ij}$ which have identical sparse representations. (This holds
especially in models of degree $>1$.)

In such cases, we have queries for point evaluation and gradient computation
with identical from-where blocks but different select-group-by clauses, because
the tensors have different group-by variables. Nevertheless, all such queries
can share computation as we can compute the from-where clause once for all of
them and then scan this result to compute each specific tensor. This analysis
gives rise to the following straightforward (and conservative) estimates.

For each $j \in [m]$, let $d_j$ denote the degree and $t_j$ denote the number of
terms in the polynomial $g_j$ (a component function of $g$). Recall that $p$ is the number of parameters.

\bprop
Point evaluation~\eqref{eqn:point:eval} and gradient computation~\eqref{eqn:gradient}
can be computed in
time $O(\sum_{i,j\in [m]}$ $t_it_jd_id_j|\vec\sigma_{ij}|)$,
and respectively $O(p\sum_{i,j\in [m]}t_it_jd_id_j|\vec\sigma_{ij}|)$.
\label{prop:per:iteration:time}
\eprop

The times for point evaluation and  gradient computation are:  
$O(d^2\sum_{i,j\in [m]}|\vec\sigma_{ij}|)$ and respectively
$O(n^d\sum_{i,j\in [m]}|\vec\sigma_{ij}|)$ for the $\pr^d$ model;
$O(r^2d^2\sum_{i,j\in [m]}|\vec\sigma_{ij}|)$ and respectively 
$O(nr^3d^2\sum_{i,j\in [m]}|\vec\sigma_{ij}|)$ for the $\fama^d_r$ model; and
$O(\sum_{i,j\in [m]}|\vec\sigma_{ij}|)$ and respectively 
$O(n\sum_{i,j\in [m]}|\vec\sigma_{ij}|)$ for PCA.
Recall that the case for PCA is similar to that of $\lr$, or equivalently $\pr_1$.

Overall, there are a couple of remarkable facts regarding the overall runtime of
our approach. Without loss of generality, suppose the number of iterations of
BGD is bounded. (This bound is typically dimension-free, dependent on the
Lipschitz constant of $J$.)  Then, from Proposition~\ref{prop:output:size}, there
are infinitely many queries for which the overall runtime of BGD is unboundedly
better than the output size.  First, our approach is faster than even the
data-export step of the mainstream approach that uses an external tool to train the model.
Second, it is often well-agreed upon that SGD is faster than BGD. However, a
single iteration of SGD requires iterating through all data tuples, which takes
time at least the output size. In particular, by training the model using BGD in
the factorized form, BGD can be unboundedly faster than a single iteration of
SGD.

\subsection{Diagram of our structure-aware approach revisited}
\label{sec:low-level-diagram}

Figure~\ref{fig:low-level-diagram} refines Figure~\ref{fig:high-level-diagram-intro} and depicts key ideas behind the performance improvements of our structure-aware framework over structure-agnostic learning approaches.

\begin{figure}[ht!]
   \centering{
      \begin{tikzpicture}[scale=.7, every node/.style={transform shape}]
      \node[data, scale=.8] (FEQ) {Feature Extraction\\ Query\\$R_1\Join\ldots\Join R_k$};
      
      \begin{scope}[shift={($(FEQ)+(4,0)$)}, scale=1, every node/.style={transform shape}]
         \node[inner sep=.75cm, scale=2, align=center] at (0,0) (DB) {~~~~~};
         \node[inner sep=.3cm, align=center] at(0,.85) () {DB};
         \drawcylinder{3}{1.5}{.25}{black}
         \begin{scope}[shift={(-1,-.1)}]
            \drawtable{.5}{.8}{3}{4}{black}
         \end{scope}
         \begin{scope}[shift={(0,-.1)}]
            \drawtable{.5}{.8}{3}{4}{black}
         \end{scope}
         \begin{scope}[shift={(1,-.1)}]
            \drawtable{.5}{.8}{3}{4}{black}
         \end{scope}
         \node at (-.5,-.2) {$\Join$};
         \node at (+.5,-.2) {$\Join$};
      \end{scope}
      
      \begin{scope}[shift={($(DB)+(3.5,0)$)}, scale=1, every node/.style={transform shape}]
         \node[inner sep=1.1cm] (0,0) (output_table){};
         \pgfmathsetmacro{\width}{2}
         \pgfmathsetmacro{\height}{3}
         \pgfmathsetmacro{\numCols}{5}
         \pgfmathsetmacro{\numRows}{10}
         \pgfmathsetmacro{\cellwidth}{\width/\numCols}
         \pgfmathsetmacro{\cellheight}{\height/\numRows}
         \drawtable{\width}{\height}{\numCols}{\numRows}{clr_outofDB}
         \draw [clr_outofDB, decorate,decoration={brace,amplitude=5pt, raise=2pt}]
            (-\width/2, \height/2)--({-\width/2+\cellwidth*(\numCols-1)},\height/2)
            node [midway, above, yshift=10pt] (output_x) {$\mv x$};
         \draw [clr_outofDB, decorate,decoration={brace,amplitude=2pt, raise=2pt}]
            ({-\width/2+\cellwidth*(\numCols-1)},\height/2)--(\width/2, \height/2)
            node [midway, above, yshift=3pt] (output_y) {$y$};
         \draw [clr_outofDB, <->] (\width/2+\cellwidth/2, \height/2-\cellheight) -- (\width/2+\cellwidth/2, -\height/2)
            node [midway, right] (Z){\highlight{$|D|$}};
      \end{scope}
      
      \node[processor, clr_outofDB] at($(output_table)+(3.5,0)$) (ML) {ML Tool};
      
      \node[data] at($(ML)+(3,0)$) (theta_opt) {$\vec\theta^*$};
      
      \node[data] at($(ML)+(0,-3.25)$) (model) {Model};
      
      \node[processor, clr_inDB, scale=.8] at($(model)+(-3,0)$) (model_reform) {Model\\ Reformulation};
      
      \node[data, clr_inDB, inner sep = .2cm] at($(DB)+(0, -5)$) (FAQ-queries) 
         {\underline{\underline{~FAQ Queries:~}}\\\vspace{3.3cm}};
      \node[clr_inDB] at($(FAQ-queries)+(0,1.25)$) (FAQ-queries-1) {$\vec\sigma_{11}=\varphi^{(1,1)}$};
      \node[clr_inDB] at($(FAQ-queries-1)+(0,-.6)$) (FAQ-queries-2) {$\vdots~~$};
      \node[clr_inDB] at($(FAQ-queries-2)+(0,-.6)$) (FAQ-queries-3) {$\vec\sigma_{i,j}=\varphi^{(i,j)}$};
      \node[clr_inDB] at($(FAQ-queries-3)+(0,-.6)$) (FAQ-queries-4) {$\underline{\hspace{.8cm}\vdots~~\hspace{.8cm}}$};
      \node[clr_inDB] at($(FAQ-queries-4)+(0,-.6)$) (FAQ-queries-5) {$~~\mv c_1=\varphi^{(1)}~~$};
      \node[clr_inDB] at($(FAQ-queries-5)+(0,-.6)$) (FAQ-queries-6) {$\vdots~~$};
      
      \node[processor, clr_inDB] at($(FAQ-queries-3)+(-4, 0)$) (FAQ-optimizer) {Query\\Optimizer};
      
      \newcommand{\TDnode}[5]
      {
         \begin{scope}[shift={(#1,#2)}]
            \node[clr_inDB] at(0,.9) {#4};
            \begin{scope}[shift={(-.1,-.2)}]
               \pgfmathsetmacro{\width}{1.5}
               \pgfmathsetmacro{\height}{1.5}
               \pgfmathsetmacro{\numCols}{4}
               \pgfmathsetmacro{\numRows}{5}
               \pgfmathsetmacro{\cellwidth}{\width/\numCols}
               \pgfmathsetmacro{\cellheight}{\height/\numRows}
               \drawtable{\width}{\height}{\numCols}{\numRows}{clr_outofDB}
               \drawtable{\width}{\height}{\numCols}{\numRows}{clr_inDB};
               \draw [clr_inDB, <->] (\width/2+\cellwidth/2, \height/2-\cellheight) -- (\width/2+\cellwidth/2, -\height/2)
               node [midway, right] (#5){};
            \end{scope}
            \node[clr_inDB, draw, ellipse, inner sep = .95cm, outer sep = 0] at (0, 0) (#3) {};
         \end{scope}
      }
      
      \begin{scope}[shift={($(FAQ-optimizer)+(3,-6.5)$)}, scale=1, every node/.style={transform shape}]
         \node[clr_inDB] at(0, 1.75) {\underline{\underline{Subqueries for Query $\varphi^{(i, j)}$:}}};
         \node[data, clr_inDB] at (0, -1.7) (TD) {\hphantom{0}\hspace{9.5cm}\hphantom{0}\\\vspace{7cm}};
         \TDnode{0}{0}{TDnode_a}{$\varphi_{a}^{(i,j)}$}{TDnode_a_Z}
         \TDnode{-2.5}{-3}{TDnode_b}{$\varphi_{b}^{(i,j)}$}{TDnode_b_Z}
         \TDnode{2.5}{-3}{TDnode_c}{$\varphi_{c}^{(i,j)}$}{TDnode_c_Z}
         \node[clr_inDB] at($(TDnode_b)+(-1.7,-2)$) (TDnode_b1) {\vdots};
         \node[clr_inDB] at($(TDnode_b)+(+1.7,-2)$) (TDnode_b2) {\vdots};
         \node[clr_inDB] at($(TDnode_c)+(-1.7,-2)$) (TDnode_c1) {\vdots};
         \node[clr_inDB] at($(TDnode_c)+(+1.7,-2)$) (TDnode_c2) {\vdots};
         \draw[clr_inDB] (TDnode_a)--(TDnode_b);
         \draw[clr_inDB] (TDnode_a)--(TDnode_c);
         \draw[clr_inDB] (TDnode_b)--(TDnode_b1);
         \draw[clr_inDB] (TDnode_b)--(TDnode_b2);
         \draw[clr_inDB] (TDnode_c)--(TDnode_c1);
         \draw[clr_inDB] (TDnode_c)--(TDnode_c2);
         \node[clr_inDB] at(3.7, 0) (faqw) {\highlight{$\leq N^{\faqw} \ll |D|$}};
         \draw[clr_inDB, dotted] (TDnode_a_Z) -- (faqw.west);
         \draw[clr_inDB, dotted] (TDnode_b_Z) -- (faqw.west);
         \draw[clr_inDB, dotted] (TDnode_c_Z) -- (faqw.west);
      \end{scope}
      
      \node[data, clr_inDB] at($(FAQ-queries)+(0, -3)$) (sigma_c) {$\vec\Sigma, \mv c$};
      
      \node[data, clr_inDB] at($(model_reform)+(0,-2.5)$) (theta) {$\vec\theta$};
      
      \node[data, clr_inDB] at($(theta)+(6,-2)$) (gradJ) {$J(\vec\theta)$\\$\grad J(\vec\theta)$};
      
      \node[processor, clr_inDB] at($(gradJ)+(0,2)$) (converge) {converged?};
      
      \coordinate (GD_corner1) at ($(theta)+(-1 ,+1)$);
      \coordinate (GD_corner2) at ($(gradJ)+(+1.5 ,-1)$);
      \draw[processor, opacity=.1, fill=clr_inDB, rounded corners = .5cm] (GD_corner1) rectangle (GD_corner2);
      \node[clr_inDB, scale=1.3, opacity=.5] at ($(GD_corner1)!.5!(GD_corner2)$) {\emph{Gradient Descent}};
      
      \draw[path, clr_outofDB] (FEQ)--(DB);
      \draw[path, clr_outofDB] (DB)--(output_table);
      \draw[path, clr_outofDB] (output_x) -| ([xshift=+.5cm]ML);
      \draw[path, clr_outofDB] (output_y) -| ([xshift=-.5cm]ML);
      \draw[path, clr_outofDB] (ML)--(theta_opt);
      \draw[path, clr_outofDB] (model)--(ML);
      
      \draw[path, clr_inDB, thick] (model)--(model_reform);
      \draw[path, clr_inDB, thick] (model_reform) -- (model_reform -| FAQ-queries.east) node [midway, above]{$h$};
      \draw[path, clr_inDB, thick] (DB)--(FAQ-queries);
      \draw[path, clr_inDB, thick] (FEQ)--(FAQ-queries);
      \draw[path, clr_inDB, thick, dashed, ->, double=none] (FAQ-queries-1)--(FAQ-optimizer);
      \draw[path, clr_inDB, thick] (FAQ-queries-3)--(FAQ-optimizer);
      \draw[path, clr_inDB, thick, dashed, ->, double=none] (FAQ-queries-5)--(FAQ-optimizer);
      \draw[path, clr_inDB, thick] (FAQ-optimizer) -- (FAQ-optimizer |- TD.north);
      \draw[path, clr_inDB,<-, thick] (sigma_c) -- (sigma_c |- TD.north);
      \draw[path, clr_inDB, thick] (model_reform)--(theta) node [midway, right]{$g$};
      \draw[path, clr_inDB, thick] (theta) --node[midway, right]{$g(\vec\theta)$} (theta |- gradJ) -- (gradJ);
      \draw[path, clr_inDB, thick] (sigma_c) -- (sigma_c -| gradJ.west);
      \draw[path, clr_inDB, thick] (gradJ) -- (converge);
      \draw[path, clr_inDB, thick] (converge) -- (theta) node [at start, pos = .2, above]{No};
      \draw[path, clr_inDB, thick] (converge) -- (theta_opt) node [at start, pos = .05, right]{Yes};
      \end{tikzpicture}
   }
   \caption{{\color{clr_inDB}\bf Structure-aware} vs. {\color{clr_outofDB}structure-agnostic} learning: Low-level diagram.}
   \label{fig:low-level-diagram}
\end{figure}

In structure-agnostic learning, a query engine takes the input relations (of size $\leq N$) and joins them into a potentially much larger output relation of size $|D|$, which might in turn get blown up even more inside the machine learning tool.
In our structure-aware framework, the input query, the input relations, and function $h$ are first translated into $\faq$~\cite{faq}, which is a language that is suitable for aggregate query specification and optimization. In particular, each entry $\vec\sigma_{i,j}$ (and $\vec c_j$) of our target matrix $\vec\Sigma$ (and vector $\vec c$) is expressed as the answer to an $\faq$ query $\varphi^{(i, j)}$ (or $\varphi^{(j)}$). All those queries are fed into an $\faq$ query optimizer. The optimizer factorizes each query $\varphi^{(i,j)}$ into small sub-queries $\varphi^{(i,j)}_a, \varphi^{(i,j)}_b, \varphi^{(i,j)}_c, \ldots$ and solves them individually. Each sub-query results in a relation of size $\leq N^{\faqw}$, which can be much smaller than the size of the data matrix $D$.
By solving the $\faq$ queries $\varphi^{(i, j)}$, we obtain $\vec\Sigma$ and $\vec c$, which are all that is needed as input for the convergence step in a batch gradient descent solver.


\section{FD-Aware Optimization}
\label{SEC:FDS}

In this section, we show how to exploit functional dependencies among variables
to reduce the dimensionality of the optimization problem by eliminating
functionally determined variables and re-parameterizing the model. We compute
the quantities ($\mv \Sigma$, $\mv c$) on the subset of features that are not
functionally determined, and then solve the lower-dimensional optimization
problem. Finally, we recover the parameters in the original space in
closed form. Exploiting functional dependencies drastically reduces the
computation time for ($\mv \Sigma$, $\mv c$) and the gradient.

\nop{
%
}

\subsection{Introduction to the main ideas}
\label{sec:intro-fd}

Consider a query $Q$ with categorical variables {\sf country} and {\sf city}.
For simplicity, assume that there are only two countries ``vietnam'' and
``england'', and $5$ cities ``saigon'', ``hanoi'', ``oxford'',
``leeds'', and ``bristol''.
Under one-hot encoding, the corresponding features are encoded as indicators
$x_{\vietnam}$,
$x_{\text{\sf england}}$,
$x_{\saigon}$,
$x_{\hanoi}$,
$x_{\textsf{oxford}}$,
$x_{\textsf{leeds}}$,
$x_{\textsf{bristol}}$.
Since {\sf city} $\to$ {\sf country} is a functional dependency (FD), 
for a given tuple $\mv x$ in the training dataset, the following hold:
\begin{align}
   x_{\vietnam} &= x_{\saigon} + x_{\hanoi}\label{eqn:x:vietnam}\\
   x_{\england} &=
   x_{\textsf{oxford}}+x_{\textsf{leeds}}+x_{\textsf{bristol}}\label{eqn:x:england}.
\end{align}
The first identity states that if a tuple has ``$\vietnam$'' as the
value for {\sf country} ($x_{\vietnam}=1$), then its value for $\fcity$ can only be either ``$\saigon$'' or ``$\hanoi$'', i.e., $[x_{\saigon}, x_{\hanoi}]$ is either $[1,0]$ or $[0,1]$, respectively. The second identity is explained similarly.

How do we express the identities such as~\eqref{eqn:x:vietnam}
and~\eqref{eqn:x:england} in a formal manner
in terms of the input vectors 
$\mv x_{\textsf{city}}$ and $\mv x_{\textsf{country}}$?  
We can extract in a preprocessing step from the database a relation of the 
form $R(\fcity, \fcountry)$ with $\fcity$ as primary key. 
Let $N_\fcity$ and $N_\fcountry$ be the number of cities and countries,
respectively.  The predicate $R(\fcity, \fcountry)$ is the sparse representation
of a matrix $\mv R$ of size $N_\fcountry \times N_\fcity$, such that 
if $\mv x_\fcity$ is an indicator vector for {\sf saigon},
then $\mv R\mv x_\fcity$ is an indicator for {\sf vietnam}. 
In this language,
identities~\eqref{eqn:x:vietnam} and~\eqref{eqn:x:england} can written simply
as $\mv x_\fcountry = \mv R\mv x_\fcity$.
For example, in the above particular example $N_\fcity = 5$, $N_\fcountry=2$,
and
\begin{align}\label{eqn:R:city:country}
  \mv R = \begin{matrix}
   & \text{saigon}&\text{hanoi}&\text{oxford}&\text{leeds}&\text{bristol}\\
   &1&1&0&0&0&\text{vietnam}\\
   &0&0&1&1&1&\text{england}
\end{matrix}
\end{align}
This relationship suggests a natural idea: 
replace any occurrence of statistics $\mv x_\fcountry$ by its functionally 
determining quantity $\mv x_\fcity$. Since these quantities are present only in
the loss function $\calLL$ via inner products
$\langle g(\mv x), h(\vec\theta) \rangle$, such replacements result in
a (typically) linear reparameterization of the loss. What happens next is less obvious,
due to the presence of the nonlinear penalty function $\Omega$.
Depending on the specific structure of FDs
and the choice of $\Omega$,  
many parameters associated with redundant statistics, which do not affect
the loss $\calLL$, can be optimized out directly with respect to the transformed 
$\Omega$ penalty. 

The remainder of this subsection is a gentle introduction of our idea in the
presence of {\em one} simple FD in the $\lr$ model. 
Consider a query $Q$ in which
$\fcity$ and $\fcountry$ are two of the categorical features and functionally
determine one another via a matrix $\mv R$ such that $\mv R \mv x_\fcity = \mv
x_\fcountry$ for all $\mv x = (\cdots, \mv x_\fcity,\mv x_\fcountry, \cdots) \in D$. 
We exploit this fact to ``eliminate'' $\mv x_\fcountry$ as follows.
\begin{align*}
   \inner{g(\vec\theta),h(\mv x)} = \inner{\vec\theta,\mv x}
   &= \sum_{j\notin \{\fcity,\fcountry\}}\inner{\vec\theta_j,\mv x_j}+
   \inner{\vec\theta_\fcity,\mv x_\fcity}
   + \inner{\vec\theta_\fcountry,\mv x_\fcountry}\\
   &= \sum_{j\notin \{\fcity,\fcountry\}}\inner{\vec\theta_j,\mv x_j}+
   \inner{\vec\theta_\fcity,\mv x_\fcity} + \inner{\vec\theta_\fcountry, \mv R \mv x_\fcity}\\
   &= \sum_{j\notin \{\fcity,\fcountry\}}\inner{\vec\theta_j,\mv x_j}+
   \inner{\underbrace{\vec\theta_\fcity + \mv R^\top
   \vec\theta_\fcountry}_{\vec\gamma_\fcity},\mv x_\fcity}.
\end{align*}
Define a new parameter vector $\vec\gamma = (\vec\gamma_j)_{j \in
V-\{\fcountry\}}$
(note that there is no $\vec\gamma_\fcountry$), 
and two functions $\overline g : \R^{n-1} \to \R^{n-1}$,
$\overline h : \R^n \to \R^{n-1}$:
\begin{align}
   \vec\gamma_j &= 
   \begin{cases}
   \vec\theta_j & j \neq \fcity\\
   \vec\theta_\fcity +\mv R^\top \vec\theta_\fcountry &  j = \fcity.
   \end{cases}\\
   \overline g(\vec\gamma) &= \vec\gamma\\
   \overline h_j(\mv x) &= \mv x_j, \ j \neq \fcity.
\label{eqn:var:change:1}
\end{align}
Then, we can reparameterize $J(\vec\theta)$ in terms of $\vec\gamma$ by
\begin{align*} 
   J(\vec\theta) &=  \frac{1}{2|D|}\sum_{(\mv x,y)\in D}
   (\inner{g(\vec\theta),h(\mv x)}  - y)^2 
   + \frac\lambda 2 \norm{\vec\theta}_2^2\\
   &= \frac{1}{2|D|}\sum_{(\mv x,y) \in D} \left(\inner{\overline g(\vec\gamma),
   \overline h(\mv x)} - y\right)^2 +
   \frac\lambda 2 \left( \sum_{j\neq \fcity} \norm{\vec\gamma_j}_2^2 
   + \norm{\vec\gamma_\fcity - \mv R^\top\vec\theta_\fcountry}_2^2
   + \norm{\vec\theta_\fcountry}_2^2
   \right).
\end{align*}
Note how $\vec\theta_\fcountry$ has disappeared from the loss term, but it still
remains in the penalty term. We now ``optimize out'' $\vec\theta_\fcountry$
by observing that
\begin{equation}
   \frac{1}{\lambda} \pd{J}{\vec\theta_\fcountry} 
   = \mv R (\mv R^\top \vec\theta_\fcountry- \vec\gamma_\fcity) +
   \vec\theta_\fcountry \label{eqn:pd:10}
\end{equation}
By setting~\eqref{eqn:pd:10} to $0$ we obtain
$\vec\theta_\fcountry$ in terms of $\vec\gamma_\fcity$:
\begin{align}
    \vec\theta_\fcountry 
    &= {(\mv I_{\fcountry} + \mv R \mv R^\top)^{-1}\mv R} \vec\gamma_{\fcity}
    = \mv R(\mv I_{\fcity} + \mv R^\top \mv R)^{-1} \vec\gamma_\fcity, 
    \label{eqn:55}
\end{align}
where $\mv I_{\fcountry}$ is the order-$N_{\fcountry}$ identity
matrix and similarly for $\mv I_{\fcity}$.
We can thus express $J$ and its gradient completely in terms of 
$\vec\gamma$:
\begin{align}
   J(\vec\theta) &=
   \frac{1}{2|D|}\sum_{(\mv x,y) \in D} \left(\inner{\overline g(\vec\gamma),
   \overline h(\mv x)}  - y\right)^2 + 
   \frac\lambda 2 \left( \sum_{j\neq \fcity} \norm{\vec\gamma_j}_2^2 
   + \inner{(\mv I_{\fcity} + \mv R^\top \mv R)^{-1}\vec\gamma_\fcity,\vec \gamma_\fcity}
   \right) \label{eqn:56}\\
   \frac 1 2 \pd{\norm{\vec\theta}_2^2}{\vec\gamma_j} 
   &=
   \begin{cases}
      \vec\gamma_j & j \neq \fcity\\
      \left( \mv I_{\fcity} + \mv R^\top \mv R \right)^{-1}
      \vec\gamma_\fcity & j = \fcity.
   \end{cases}
\end{align}
(Appendix~\ref{app:subsec:intro-fd} contains formal proofs of the above claims.)
The gradient of the loss term is computed using the matrix $\overline{\vec\Sigma}$
and the vector $\overline{\mv c}$ with respect to the pair $(\overline g,\overline h)$ of reduced dimensionality.
The matrix $(\mv I_{\fcity} + \mv R^\top\mv R)$ is a rank-$N_\fcountry$ update to the identity matrix $\mv I_{\fcity}$, strictly positive definite and thus invertible. The inverse can be obtained using database aggregate queries; for numerical
stability, one may compute its Cholesky decomposition which can {\em also} be expressed
by aggregate queries. 
These ``linear algebra via aggregate queries'' co\-mputations are possible
because our matrices admit a database interpretation, cf.\@
Section~\ref{subsec:linear:algebra}.


\subsection{Functional dependencies (FDs)}
\label{sec:fd-def}

Composite FDs lead to more complex identities. For
instance, the FD {\sf (guest, hotel, date)} $\to$ {\sf room} leads to the
identity $x_{\textsf{room}} = \sum
x_{\textsf{guest}}x_{\textsf{hotel}}x_{\textsf{date}}$.
Let $R$ be a relation on attributes \textsf{guest}, \textsf{hotel},
\textsf{date}, and \textsf{room}, encoding this dependency, 
i.e., $R$ has a compound key
$(\textsf{guest},\textsf{hotel},\textsf{date})$. Then, corresponding to $R$ 
there is a matrix $\mv R$ of dimension
$
N_{\textsf{room}}
\times
N_{\textsf{guest}}\cdot 
N_{\textsf{hotel}}\cdot 
N_{\textsf{date}} 
$
for which
$\mv x_{\textsf{room}} = \mv R(\mv x_{\textsf{guest}}\otimes \mv x_{\textsf{hotel}} \otimes \mv
x_{\textsf{date}})$. 
Our results can be extended to the case of composite FDs, yet with a great
notational burden; for the sake of clarity, we only state the results for {\em simple} FDs. 

\bdefn

An FD is {\em simple} if its left-hand side is one variable.

Let a query $Q$ in which there are
$k$ disjoint groups $G_1,\dots,G_k$ of features, among other features. 
The $i$th group is $G_i = \{f_i\} \cup S_i$, where $f_i$ is a feature,
$S_i$ a set of features, and $f_i \to S_i$ is an FD. 
We shall refer to these as 
{\em groups of simple FDs}.
\label{defn:groups:of:FDs}
\edefn
\begin{ex}
In a typical feature extraction query for retailer customers, 
we have $k=3$ groups
(in addition to other features): the first group contains {\sf week}
$\to$ {\sf month} $\to$ {\sf quarter} $\to$ {\sf year}, and thus $f_1$ = {\sf
week} and $S_1$ = $\{$ {\sf month, quarter, year} $\}$. In the second
group, $f_2$ = {\sf sku} and $S_2 = \{$ {\sf type, color, size,~...}$\}$ (a
rather large group). In the third group $f_3 = $ {\sf store} and
$S_3 = \{$ {\sf city, country, region} $\}$. \qed
\label{ex:groups:of:FDs}
\end{ex}

For each feature $c \in S_i$, let $\mv R_c$ denote
the matrix for which $\mv x_c = \mv R_c \mv x_{f_i}$.
For the sake of brevity, we also define a matrix $\mv
R_{f_i} = \mv I_{f_i}$ (the identity matrix of dimension equal to the active
domain size of attribute $f_i$), so the equality 
$\mv R_c\mv x_{f_i} = \mv x_c$ holds for every $c \in G_i$.

The linear relationship holds even if the variables are not categorical. 
For example, consider the FD {\sf sku} $\to$ {\sf price} (assuming
every stock-keeping unit has a fixed sale-price). The relationship is modeled
with a $1 \times N_{\sf sku}$ matrix $\mv R$, where the entry corresponding
to a {\sf sku} is its {\sf price}. 
Then, $\mv R\mv x_{\textsf{sku}} = x_{\textsf{price}}$
for any indicator vector $\mv x_{\textsf{sku}}$. 

\bdefn[FD-reduced pairs of functions]
\label{defn:FD-reduced}
Given a pair of functions $g$ and $h$ in our problem setting.
Recall that $C_j$'s are defined in Section~\ref{sec:tensor}, while
$S_k$'s are given in Definition~\ref{defn:groups:of:FDs}.
Define
\[ K = \{ j \in [m] \suchthat C_j \cap (S_1 \cup \dots \cup S_k) \neq \emptyset
  \} \]
($K$ is the set of component functions of $h$ containing at least one
functionally determined variable.)

The group of simple FDs induces an {\em FD-reduced pair} of functions 
$\overline g : \R^{p-|K|} \to \R^{m-|K|}$ and 
$\overline h : \R^{n} \to \R^{m-|K|}$ as follows:
The component functions of $\overline h$ are obtained from 
the component functions of $h$ by removing all component functions $h_j$
for $j \in K$. Similarly, 
$\overline g$ is obtained from $g$ by removing all component functions 
$g_j$ for which $j \in K$. 
Naturally, define the covariance matrix $\overline{\vec\Sigma}$ and the correlation vector
$\overline{\mv c}$ as in~\eqref{eqn:sigma} and~\eqref{eqn:c}, but with respect to $\overline h$.
\edefn

We next generalize the above technique to
speedup the training of $\pr^d$ and $\fama$ under an arbitrary collection of simple FDs. 

\subsection{Polynomial regression under FDs}
\label{subsec:pr:fd}

Recall the $\pr^d$-model formulated in Example~\ref{ex:pr}. 
Consider the set $A_V$
of all tuples $\mv a_V = (a_w)_{w\in V} \in \mathbb N^V$ of non-negative
integers such that $\norm{\mv a_V}_1 \leq d$. 
For any $(\mv x,y) \in D$ and $\mv a \in A_V$, define
$\mv x^{\otimes \mv a} = \bigotimes_{v \in V} \mv x_v^{\otimes a_v}.$
In the $\pr^d$ model we have $\vec\theta = (\vec\theta_{\mv a})_{\norm{\mv a}_1
\leq d}$, $g(\vec\theta) = \vec\theta$, and $h_{\mv a}(\mv x) = \mv x^{\otimes \mv
a}$. If a feature, say $v \in V$, is non-categorical, then $\mv
x_v^{\otimes a_v} = x_v^{a_v}$.
If we knew $x_v \in \{0,1\}$, then
$x_v^{a_v}=x_v$ and thus there is no need to have terms for which $a_v>1$.
A similar situation  occurs when $v$ is a categorical variable. To see this,
let us consider a simple query where $V=\{b,c,w,t\}$, and all four variables
are categorical. Suppose the $\pr^d$ model has a term corresponding to 
$\mv a = (a_b,a_c,a_w,a_t) = (0,2,0,1)$. The term of $\inner{\vec\theta,h(\mv
x)}$ indexed by tuple $\mv a$ is of the form 
\[ \inner{\vec\theta_{\mv a}, \mv x_c^{\otimes 2} \otimes \mv x_t}
= \inner{\vec\theta_{\mv a}, \mv x_c \otimes \mv x_c \otimes \mv x_t}.
\]
For the dimensionality to match up, $\vec\theta_{\mv a}$ is a 
3rd-order tensor, say indexed by $(i,j,k)$. The above expression can be simplified as
\begin{align*} \sum_i\sum_j\sum_k \vec\theta_{\mv a}(i,j,k) \cdot \mv x_c(i)\cdot \mv
x_c(j)\cdot \mv x_t(k)
=
 \sum_j\sum_k \vec\theta_{\mv a}(j,j,k)\mv x_c(j)\mv x_t(k),
\end{align*}
where the equality holds due to the fact that $\mv x_c(j)$ is idempotent.
In particular, we only need the entries indexed by $(j,j,k)$ of $\vec\theta_{\mv
a}$. Equivalently, we write:
\[ \inner{\vec\theta_{\mv a}, \mv x_c \otimes \mv x_c \otimes \mv x_t}
=
 \inner{((\mv I_c \star \mv I_c)^\top \otimes \mv I_t)\vec\theta_{\mv a}, \mv x_c \otimes \mv x_t}.
\]
Multiplying on the left by the matrix $(\mv I_c \star \mv I_c)^\top \otimes \mv
I_t$ has precisely the same effect as selecting out only entries
$\vec\theta_{\mv a}(j,j,k)$ from the tensor $\vec\theta_{\mv a}$. 
More generally, in the $\pr^d$ model we can assume that all the indices 
$\mv a_V = (a_v)_{v\in V}$ satisfy the condition that $a_v \in \{0,1\}$
whenever $v$ is categorical. (This is in addition to the degree requirement that
$\norm{\mv a_V}_1\leq d$.)

Given $k$ groups of FDs represented by $G_1,\dots,G_k$, let
$G = \bigcup_{i=1}^kG_i$, $S = \bigcup_{i=1}^k S_i$, $\overline G = V-G$, 
$\overline S = V-S$, and $F = \{f_1,\dots,f_k\}$.
For every non-empty subset $T \subseteq [k]$, define $F_T = \{f_i \suchthat i \in T\}$.
Given a natural number $q < d$, 
and a non-empty set $T \subseteq [k]$ with size $|T| \leq d-q$, define the collection
\begin{align}
   \calU(T,q) = \{ U \suchthat U \subseteq G 
                    \wedge U \cap G_i \neq \emptyset, \forall i \in T
                    \wedge   U\cap G_i = \emptyset, \forall i \notin T 
                    \wedge  |U| \leq d-q \}.
  \label{eqn:calU:T:h}
\end{align}
For every tuple $\mv a_{\overline G} \in \mathbb N^{\overline G}$ with
$\norm{\mv a_{\overline G}}_1=q<d$ and 
every $U \in \calU(T,q)$, define the following matrices, which
play the same role as $\mv I_{\fcity} + \mv R^\top \mv R$ in
Section~\ref{sec:intro-fd}:

\begin{align}
  \mv R_{\mv a_{\overline G}, U} 
   &= 
      \bigotimes_{\substack{w\in \overline G\\ a_w>0}}  \mv I_w \otimes \bigotimes_{i\in T} \Bigstar_{c \in U \cap G_i} \mv R_c.
      \label{eqn:R:a:U} \\
   \mv B_{\mv a_{\overline G},T} &=
            \sum_{W \in \calU(T, \norm{\mv a_{\overline G}}_1)}
   \mv R_{\mv a_{\overline G}, W}^\top
   \mv R_{\mv a_{\overline G}, W} 
   \label{eqn:B:T:h}
\end{align}
Note that the matrices $\mv B_{\mv a_{\overline G},T}$ can be further factorized
as each of its terms is a tensor product, but we 
refrain from doing so here to avoid heavy notational complexity in the proofs and the theorem
statement. See~\eqref{eqn:B:0:ij} and~\eqref{eqn:B:w:i} in the appendix for examples of what
we mean by factorization of these matrices.
The following theorem reparameterizes $J(\vec\theta)$ for $\pr^d$ ($d\geq 1$)
to become $\overline
J(\vec\gamma)$. While $\vec\theta = (\vec\theta_{\mv a})$ is a vector indexed by
tuples $\mv a = \mv a_V \in \mathbb N^V$, 
the new parameters $\vec\gamma = (\vec\gamma_{\mv b})$ are indexed 
by integer tuples $\mv b = \mv b_{\bar S} \in \mathbb N^{\bar S}$.

\bthm
Let the $\pr^d$-model with parameters
$\vec\theta=(\vec\theta_{\mv a_V})_{\norm{\mv a_V}_1\leq d}$, and
$k$ groups of simple FDs $G_i = \{f_i\}\cup S_i$, $i \in [k]$. 
Define the reparameterization:
\begin{equation*}
   \vec\gamma_{\mv b_{\overline S}} =
   \begin{cases}
       \vec\theta_{(\mv b_{\overline G}, \mv 0_{G})} & \norm{\mv b_{G}}_1 = 0\\
      \sum_{U \in \calU(T,q)} \mv R_{\mv b_{\overline G}, U}^\top
      \vec\theta_{(\mv b_{\overline G},\mv 1_{U|G})} & 
      \norm{\mv b_{G}}_1 > 0,
      T := \{j \suchthat j \in F, b_{f_j}=1\}, 
   \end{cases}
\end{equation*}
Then, minimizing $J(\vec\theta)$ is equivalent to minimizing the function
\begin{equation}
   \overline J(\vec\gamma) = \frac 1 2 \vec\gamma^\top
\overline{\vec\Sigma}\vec\gamma - \inner{\vec\gamma, \overline c} +
\frac \lambda 2 \Omega(\vec\gamma),
   \label{eqn:overline:J}
\end{equation}
where
\begin{align}
   \Omega(\vec\gamma)=
   \sum_{\substack{\norm{\mv b_{\overline S}}_1\leq d\\ \norm{\mv b_F}_1=0}}
     \norm{\vec\gamma_{\mv b_{\overline S}}}_2^2
   + \sum_{\substack{\norm{\mv b_{\overline G}}_1 = q\\ q< d}} 
     \sum_{\substack{T \subseteq [k]\\ 0<|T|\leq d-q}}
   \inner{
       \mv B_{\mv b_{\overline G},T}^{-1}
   \vec\gamma_{(\mv b_{\overline G},\mv 1_{F_T|F})},
   \vec\gamma_{(\mv b_{\overline G},\mv 1_{F_T|F})}}.
   \label{eqn:Omega:gamma}
\end{align}
(Recall $\overline{\vec\Sigma}$ and $\overline c$ from Definition~\ref{defn:FD-reduced}.)
\label{thm:pr:d:fd}
\ethm

The proof of this theorem
(Appendix~\ref{sec:monsterproof})
is technically involved.
$\overline J$ is defined above with respect to the FD-reduced pair of
functions $\overline{g}, \overline{h}$ and a reduced parameter
space of $\vec\gamma$. Its gradient is simple to compute, since
\begin{align}
   \frac 1 2 \pd{\Omega(\vec\gamma)}{\vec\gamma_{\mv b_{\overline S}}} &= 
   \begin{cases}
      \vec\gamma_{\mv b_{\overline S}}, & \mv b_F = \mv 0_F, \\
       \mv B_{\mv b_{\overline G},T}^{-1}
       \vec\gamma_{(\mv b_{\overline G},\mv 1_{F_T|F})}, 
       & T = \{j \suchthat j \in F, b_j=1\}.
   \end{cases}
   \label{eqn:main:derivative}
\end{align}
Moreover, once a minimizer $\vec\gamma$ of $\overline J$ is obtained,
we can compute a minimizer $\vec\theta$ of $J$ by setting 
\begin{equation}
   \vec\theta_{\mv a_{V}} =
   \begin{cases}
      \vec\gamma_{\mv a_{\overline S}}, & \norm{\mv a_G}_1=0 \\
   \mv R_{\mv a_{\overline G}, U} 
   \mv B_{\mv a_{\overline G},T}^{-1}
   \vec\gamma_{(\mv a_{\overline G},\mv 1_{F_T|F})}
      & \norm{\mv a_G}_1>0, T := \{i \suchthat \exists c \in G_i, a_c>0\}
   \end{cases}
   \label{eqn:main:gamma:to:theta}
\end{equation}

Theorem~\ref{thm:pr:d:fd} might be a bit difficult to grasp at first glance due
to its generality. To give the reader a sense of how the theorem is applied in
specific instances, Appendix~\ref{subsec:lr:special:case}
and~\ref{subsec:pr:2:special:case} present two specializations of the theorem
for (ridge) linear regression ($\pr^1$), and degree-2 polynomial regression
($\pr^2$).

\subsection{Factorization machines under FDs}
\label{subsec:fama:fd}

We now turn our attention to $\fama^2_r$.

\bthm
Consider the $\fama$ model of degree $2$, rank $r$,
parameters 
$\vec\theta=(\vec\theta_i, (\vec\theta_i^{(\ell)})_{\ell\in [r]})_{i\in V}$ 
and $k$ groups of simple FDs $G_i = \{f_i\}\cup S_i$, $i \in [k]$. 
Let $G = \cup_{i\in [k]}G_i$, 
\begin{align}
\vec\beta_{f_i} &= \sum_{\ell=1}^r \sum_{\{c,t\} \in \binom{G_i}{2}} \mv
   R^\top_c\vec\theta_c^{(\ell)}\circ \mv R_t^\top \vec\theta_t^{(\ell)}, \ i
   \in [k]
\end{align}
and the following reparameterization:
\begin{align*}
   \vec\gamma_w &= {\begin{cases}
      \vec\theta_w, & w\notin \bigcup_{i=1}^kG_i\\
      \displaystyle{ \vec\theta_{f_i} + \sum_{c\in S_i}\mv R_c^\top\vec\theta_c + \vec\beta_{f_i}}, & w = f_i, i \in [k].
   \end{cases}}\\
   \vec\gamma^{(\ell)}_w &=
   \begin{cases}
      \vec\theta^{(\ell)}_w, & w \notin F \\
      \vec\theta^{(\ell)}_{f_i} + \sum_{c\in S_i}\mv R_c^\top \vec\theta^{(\ell)}_c, & w=f_i, i \in [k].
   \end{cases}
\end{align*}
Then, minimizing $J(\vec\theta)$ is equivalent to minimizing the
function 
$\overline J(\vec\gamma) = \frac 1 2 \overline g(\vec\gamma)^\top \overline{\vec\Sigma}
\overline g(\vec\gamma)
- \inner{\overline g(\vec\gamma),\overline{\mv c}} + \frac\lambda 2 \Omega(\vec\gamma),$
where 
\begin{multline}
   \Omega(\vec\gamma) = 
   \sum_{w\notin G}\norm{\vec\gamma_w}_2^2
+ \sum_{i=1}^k \inner{\mv B_i^{-1}(\vec\gamma_{f_i} - \vec\beta_{f_i}),
(\vec\gamma_{f_i} - \vec\beta_{f_i})}
   + \sum_{\substack{\ell\in [r]\\w \notin  F}} \norm{\vec\gamma^{(\ell)}_w}_2^2
   + \sum_{\substack{i\in [k]\\ \ell \in [r]}} \norm{\vec\gamma^{(\ell)}_{f_i} -
   \sum_{c\in S_i}\mv R^\top_c \vec\gamma^{(\ell)}_c}_2^2.
   \label{eqn:omega:fama2}
\end{multline}
(Recall $\overline g$, $\overline{\vec\Sigma}$ and $\overline c$ from Definition~\ref{defn:FD-reduced}.)
\label{thm:fd:fama2}
\ethm

In order to optimize $\overline J$ with respect to $\vec\gamma$,
the following proposition provides a closed form formulae for the relevant
gradient.

\bprop
The gradient of $\Omega(\vec\gamma)$ defined in~\eqref{eqn:omega:fama2} can
be computed using
$\vec\delta^{(\ell)}_{i} = \sum_{c\in S_i}\mv R_c^\top
\vec\gamma^{(\ell)}_c$, and
  \[ \vec\beta_{f_i} = \sum_{\ell=1}^r \left[ 
   \left(\vec\gamma^{(\ell)}_{f_i} - \frac 1 2 \vec\delta^{(\ell)}_{i} \right)\circ \vec\delta^{(\ell)}_{i}
   - \frac 1 2 \sum_{t \in S_i} \mv R^\top_t (\vec\gamma^{(\ell)}_t \circ \vec\gamma^{(\ell)}_t)
  \right]
  \]
Then,
{\small
\begin{align}
   \hspace{-5pt}   \frac 1 2 \pd{\Omega(\vec\gamma)}{\vec\gamma_w} &=
   \begin{cases}
      \vec\gamma_w, & w \notin G\\
      \mv B_i^{-1}(\vec\gamma_{f_i}-\vec\beta_{f_i}), & w = f_i, i \in [k].
   \end{cases}\\
   \hspace{-5pt}   \frac 1 2 \pd{\Omega(\vec\gamma)}{\vec\gamma^{(\ell)}_w}&=
   \begin{cases}
      \vec\gamma^{(\ell)}_w, & w \notin G, \ell\in [r]\\
      \displaystyle{\vec\gamma^{(\ell)}_{f_i} - \vec\delta^{(\ell)}_{i} -
      \frac 1 2 \vec\delta^{(\ell)}_{i} \circ \pd{\Omega(\vec\gamma)}{\vec\gamma_{f_i}}
      }, & w={f_i}, \ell\in [r]\\
      \vec\gamma^{(\ell)}_w - \mv R_w \left[ \vec\gamma_{f_i}^{(\ell)} \circ
      \frac 1 2 \pd{\Omega(\vec\gamma)}{\vec\gamma_{f_i}} + \frac 1 2
      \pd{\Omega(\vec\gamma)}{\vec\gamma_{f_i}^{(\ell)}}\right], &
       w \in S_i, \ell\in [r].
   \end{cases}
   \label{eqn:second:degree:fm:pd}
\end{align}
}
\label{prop:omega:fama2:gradient}
\eprop

Suppose that the minimizer $\vec\gamma$ of $\overline J$
has been obtained, then a minimizer $\vec\theta$ of $J$ is available in closed form:
\begin{align*}
   \vec\theta_w &= 
   \begin{cases}
      \vec\gamma_w, & w \in V\setminus G\\
      \mv R_{t} \mv B_i^{-1} (\vec\gamma_{f_i} - \vec\beta_{f_i}), & \forall t \in G_i, i \in [k].
   \end{cases}\\
   \vec\theta_{w}^{(\ell)} &= 
    \begin{cases}
       \vec\gamma_w^{(\ell)}, & \forall w \notin F, \ell \in [r].\\
       \vec\gamma_w^{(\ell)} - \delta^{(\ell)}_i, & w = f_i, \ell \in [r].\\
    \end{cases}
\end{align*}

This section shows that our technique applies to a non-linear model too. It
should be obvious that a similar reparameterization works for $\fama^d_r$ for
any $d\geq 1$.
There is some asymmetry in the
reparameterization of $1$st-order parameters $\vec\theta_i$ and $2$nd-order
parameters $\vec\theta^{(\ell)}_i$ in Theorem~\ref{thm:fd:fama2}, 
because we can solve a system of linear equation with matrix inverses, but we
don't have closed form solutions for quadratic equations.



\subsection{Principal Component Analysis under FDs}

In this section, we show how to exploit functional dependencies to reduce the
number of dimensions of the input to PCA by computing the top-$K$ eigenvectors 
and eigenvalues over the lower dimensional covariance
matrix without the functionally determined features.
We show that the eigenvalues of the lower dimensional problem are identical to those
of the original problem, while the original eigenvectors can be
derived from the solution to the lower dimensional problem.

Recall the  functional dependencies of the form 
$f_i\to S_i$, the sets $S$ of functionally determined variables and  
$\overline{V} = V - S$ of all other variables, as in
Section~\ref{sec:fd-def}. Also, recall that each $\mv x \in D$ is an $n$-dimensional
vector, and for each categorical variable $c$, the component $\mv x_c$ is an
indicator vector.

We define $\overline{\mv x}$ to be the vector of size $q = |\overline{V}|$,
which is obtained by removing all components from $\mv x$ that correspond to
functionally determined variables (i.e., all $\mv x_c$ for which $c \in
S$). Similar to~\eqref{eq:center:sigma} and~\eqref{eqn:mu:categorical}, we can
express the $q$-dimensional vector of means and the $q \times q$ covariance
matrix over $\overline{\mv x}$:
\begin{align*}
&\overline{\vec\mu} = \frac{1}{|D|} \sum_{\mv x \in D} \overline{\mv x}\\
&\overline{\vec \Sigma}_1 = \frac{1}{|D|} \sum_{\mv x \in D} \overline{\mv
  x}\,\overline{\mv x}^\top - \overline{\vec\mu}\, \overline{\vec\mu}^\top.
\end{align*}
The covariance matrix $\overline{\vec\Sigma}_1$ can be computed directly over the input
database as in Example~\ref{ex:store:city:pca}.  
The effect of computing a covariance matrix
over $\overline{\mv x}$ instead of $\mv x$ is that its sparse tensor
representation does not require the computation of any aggregate over
 a functionally determined variable.

 For each functionally determined variable $c \in S_i$, the FD $f_i \to c$
 induces a mapping from a component in $\overline{\mv x}$ to a component in
 $\mv x$. We define $\mv U$ to be the rank-$q$ matrix of all such mappings, so
 that $\mv x = \mv U \overline{\mv x}$ and each index $\mv u_{kl} \in \mv U$
 maps $\overline{\mv x}_l$ to $\mv x_k$. For a variable $c_k$, let $N_k$ be its
 domain size if $c_k$ is categorical or one otherwise. Take two such variables
 $c_k$ and $c_l$.  Then, if $c_k = c_l$ the entry $\mv u_{kl}$ is the identity
 matrix $I_{N_k}$. If $c_k \neq c_l$ and there is no functional dependency
 between them, then the entry $\mv u_{kl}$ is the $N_k\times N_l$ matrix of
 zeros. In case there is a functional dependency $c_l\to c_k$, the entry
 $\mv u_{kl}$ is the $N_k\times N_l$ matrix that encodes this functional
 dependency.  For instance, in case $l=\textsf{city}$ and $k=\textsf{country}$,
 the entry $\mv u_{kl}$ is the $N_{\sf country} \times N_{\sf city}$ matrix
 whose entries $(m,n)$ are one if the $n$-th city is located in
 the $m$-th country, or zero otherwise (as exemplified
 in~\eqref{eqn:R:city:country} in Section~\ref{sec:intro-fd}, and generalized as
 $\mv R_c$ matrices in Section~\ref{sec:fd-def}). We can compute a sparse
 representation of the matrix $\mv U$ as a collection of group-by queries over
 the input relations, and without materializing the result of the feature
 extraction query.

The following lemma shows that the eigenvalues of the covariance matrix are preserved under FDs
while the eigenvectors are subject to a simple transformation.

\begin{lemma} For some $K \leq q$, let $\lambda_1,\ldots,\lambda_K > 0$ be the
  top-$K$ (positive-valued) eigenvalues of $q\times q$ matrix
  $\mv U^\top \mv U \overline{\vec\Sigma}$ and
  $\vec\eta_1,\ldots,\vec \eta_K \in \mathbb{R}^q$ be the corresponding
  eigenvectors. Then $\lambda_1,\ldots,\lambda_K$ are also the top-$K$ eigenvalues
  of $\Sigma_1$. Moreover, the eigenvectors of $\Sigma_1$ are
  \[\forall j \in [K] \; : \; \vec \theta_j = \frac{1}{\lambda_j} \mv U
    \overline{\vec\Sigma}_1\vec \eta_j\]
\end{lemma}

\begin{proof}
  First note that
  $\vec \mu = \frac{1}{|D|}\sum_{\mv x \in D} \mv x = \frac{1}{|D|}\sum_{\mv x
    \in D} \mv U \overline{\mv x} = \mv U \overline{\vec\mu}$ and
  $\vec\Sigma_1 = \frac{1}{|D|}\sum_{\mv x \in D} \mv x \mv x^\top-
  \vec\mu\vec\mu^\top = \frac{1}{|D|}\sum_{\mv x \in D}\mv U \overline{\mv x}\,
  \overline{\mv x}^\top \mv U^\top - \mv
  U\overline{\vec\mu}\,\overline{\vec\mu}^\top\mv U^\top =
  \mv U (\frac{1}{|D|}\sum_{\mv x \in D}\overline{\mv x}\, \overline{\mv x}^\top
  - \overline{\vec\mu}\,\overline{\vec\mu}^\top)\mv U^\top= \mv U
  \overline{\vec\Sigma}_1 \mv U^\top$. For any eigen-pair
  $(\lambda,\vec \theta)$ of $\vec\Sigma_1$, it holds
  $\vec\Sigma_1 \vec \theta = \lambda \vec \theta$ by definition. Thus,
  $\mv U \overline{\vec\Sigma}_1 \mv U^\top \vec \theta = \lambda \vec \theta$.
  Multiplying both sides by $\mv U^\top$ to the left, we obtain
  $\mv U^\top \mv U \overline{\vec\Sigma}_1 \mv U^\top \vec \theta = \lambda \mv
  U^\top \vec \theta$. Hence, $(\lambda, \mv U^\top \vec \theta)$ is an
  eigen-pair of $\mv U^\top \mv U \overline{\vec\Sigma}_1$.  Since
  $\mv U^\top \mv U$ is full-ranked, the set of positive eigenvalues of
  $\vec\Sigma_1$ is identical to that of
  $\mv U^\top \mv U \overline{\vec\Sigma}_1$. Moreover, let
  $(\lambda,\vec \eta)$ be any of the eigen-pairs of
  $\mv U^\top \mv U \overline{\vec\Sigma}_1$ in which $\lambda > 0$, then the
  corresponding eigenvector of $\vec\Sigma_1$ may be obtained by the identity:
  $\mv U \overline{\vec\Sigma}_1 \mv U^\top \vec \theta = \lambda \vec \theta$,
  which yields $\mv U \overline{\vec\Sigma}_1 \vec \eta = \lambda \vec
  \theta$. This gives
  $\vec \theta = (1/\lambda) \mv U \overline{\vec\Sigma}_1 \vec \eta$ to
  conclude the proof.
\end{proof}


\nop{

}

\subsection{Linear algebra with database queries}
\label{subsec:linear:algebra}

To apply the above results, we need to solve several computational primitives.
The first primitive is to compute the matrix inverse $\mv B_{T,q}^{-1}$ and its
product with another vector. This task can be done by either explicitly
computing the inverse, or computing the Cholesky decomposition of the matrix
$\mv B_{T,q}$.  We next explain how both of these tasks can be done using
database queries.

\paragraph*{Maintaining the matrix inverse with rank-$1$ updates}

Using Sher\-man-Morrison-Woodbury formula
\cite{MR997457},
we can incrementally compute the inverse of the matrix $\mv I + \sum_{c \in G_i}\mv R_c^\top\mv R_c$ as follows.
Let $S \subset G_i$ be some subset and suppose we have already computed the 
inverse for $\mv M_S = \mv I + \sum_{s \in S} \mv R_s^\top \mv R_s$. We now explain how
to compute the inverse for $\mv M_{S\cup \{c\}}=\mv I + \sum_{s \in S \cup \{c\}}\mv R_s^\top\mv
R_s$. For concreteness, let the matrix $\mv R_c$ map {\sf city} to {\sf country}. 
For each country $\fcountry$, let $\mv e_\fcountry$ denote the $01$-vector where 
there is a $1$ for each city the country has. For example, 
$\mv e_{\text{cuba}} = [1 \ 1 \ 0 \ 0 \ 0]^\top.$
Then, 
$\mv R_c^\top \mv R_c = \sum_{\fcountry} \mv e_\fcountry \mv e^\top_\fcountry.$
And thus, starting with $\mv M_S$, we
apply the Sherman-Morrison-Woodbury formula for each country, such as:
\begin{equation} (\mv M + \mv e_{\textsf{cuba}} \mv e_{\textsf{cuba}}^\top)^{-1} = \mv M^{-1} - 
   \frac{\mv M^{-1} \mv e_{\textsf{cuba}} \mv e_{\textsf{cuba}}^\top \mv M^{-1}}
               {1 + \mv e_{\textsf{cuba}}^\top \mv M^{-1} \mv e_{\textsf{cuba}}}.
   \label{eqn:inverse:update}
\end{equation}
This update can be done with database aggregate queries,
because $\mv e_{\textsf{cuba}}^\top \mv M^{-1} \mv e_{\textsf{cuba}}$ 
is a sum of entries $(i,j)$ in $\mv M^{-1}$ where both $i$ and $j$ are
cities in {\sf cuba};
$\mv v = \mv M^{-1}\mv e_{\textsf{cuba}}$ is the sum of
columns of $\mv M^{-1}$ corresponding to {\sf cuba};
and 
$\mv M^{-1} \mv e_{\textsf{cuba}} \mv e_{\textsf{cuba}}^\top \mv M^{-1}$
is exactly $\mv v \mv v^\top$.

Overall, each update~\eqref{eqn:inverse:update} can be done in $O(N_{\fcity}^2)$-time,
for an overall runtime of $O(N_{\fcity}^2N_{\fcountry})$.
This runtime should be contrasted with Gaussian-elimination-based inverse
computation time, which is $O(N_{\fcity}^3)$.
When the FDs form a chain, the blocks are nested
inside one another, and thus each update is even cheaper as we do not have to
access all $N_{\fcity}^2$ entries.

\paragraph*{Maintaining a Cholesky decomposition with rank-$k$ update}

Maintaining a matrix inverse can be numerically unstable. It would be
best to compute a Cholesky decomposition of the matrix, since this
strategy is numerically more stable.
There are known rank-$1$ update algorithms~\cite{MR0343558,MR1824053}, using strategies
similar to the inverse rank-$1$ update above.
A further common computational primitive is to multiply a tensor
product with a vector, such as in $(\mv B_i^{-1}\otimes \mv
B_j^{-1})\vec\gamma_{f_if_j}$ (also expressible as aggregate queries).

\subsection{Discussion}

\subsubsection*{Diagram of our structure-aware learning in the presence of FDs}

Figure~\ref{fig:low-level-diagram-FD} depicts the enhancements that we introduce to our framework in order to take advantage of FDs in the input database instance and reduce our previous runtime even further (but still compute \emph{the same} $\vec\theta^*$ as before).

As explained earlier, computing each entry of the matrix $\vec\Sigma$ and of the vector $\vec c$ requires solving an $\faq$ query. However, by utilizing FDs we can filter out many of those entries as unneeded for later stages, thus significantly reducing the number of $\faq$ queries that we have to solve. After the filtering process, $\vec\Sigma$ and $\vec c$
shrink down to  $\overline{\vec\Sigma}$ and $\overline{\vec c}$, which we compute and feed to gradient-descent (GD).
We also filter the function $g$ down to $\bar g$ and feed the latter to GD.
Now, we run GD in the space of $\vec\gamma$ (instead of the original higher-dimensional space of $\vec\theta$).
During each iteration of GD, in order to compute the objective function $\overline J(\vec\gamma)$ and its gradient $\grad\overline J(\vec\gamma)$, we need to use the matrices $\mv R$ that represent the functional dependencies.
And after GD finishes, we have to convert the resulting optimal solution $\vec\gamma^*$ back into the original space to get $\vec\theta^*$. Such conversion also requires the FD-matrices $\mv R$.

\colorlet{clr_outofDB}{clr_outofDB!50!black!50!white}
\colorlet{clr_inDB}{clr_inDB!50!black!50!white}

\begin{figure}[ht!]
   \centering{
      \begin{tikzpicture}[scale=.7, every node/.style={transform shape}]
      \node[data, scale=.8] (FEQ) {Feature Extraction\\ Query\\$R_1\Join\ldots\Join R_k$};
      
      \begin{scope}[shift={($(FEQ)+(4,0)$)}, scale=1, every node/.style={transform shape}]
         \node[inner sep=.75cm, scale=2, align=center] at (0,0) (DB) {~~~~~};
         \node[inner sep=.3cm, align=center] at(0,.85) () {DB};
         \drawcylinder{3}{1.5}{.25}{black}
         \begin{scope}[shift={(-1,-.1)}]
            \drawtable{.5}{.8}{3}{4}{black}
         \end{scope}
         \begin{scope}[shift={(0,-.1)}]
            \drawtable{.5}{.8}{3}{4}{black}
         \end{scope}
         \begin{scope}[shift={(1,-.1)}]
            \drawtable{.5}{.8}{3}{4}{black}
         \end{scope}
         \node at (-.5,-.2) {$\Join$};
         \node at (+.5,-.2) {$\Join$};
      \end{scope}
      
      \begin{scope}[shift={($(DB)+(3.5,0)$)}, scale=1, every node/.style={transform shape}]
         \node[inner sep=1.1cm] (0,0) (output_table){};
         \pgfmathsetmacro{\width}{2}
         \pgfmathsetmacro{\height}{3}
         \pgfmathsetmacro{\numCols}{5}
         \pgfmathsetmacro{\numRows}{10}
         \pgfmathsetmacro{\cellwidth}{\width/\numCols}
         \pgfmathsetmacro{\cellheight}{\height/\numRows}
         \drawtable{\width}{\height}{\numCols}{\numRows}{clr_outofDB}
         \draw [clr_outofDB, decorate,decoration={brace,amplitude=5pt, raise=2pt}]
            (-\width/2, \height/2)--({-\width/2+\cellwidth*(\numCols-1)},\height/2)
            node [midway, above, yshift=10pt] (output_x) {$\mv x$};
         \draw [clr_outofDB, decorate,decoration={brace,amplitude=2pt, raise=2pt}]
            ({-\width/2+\cellwidth*(\numCols-1)},\height/2)--(\width/2, \height/2)
            node [midway, above, yshift=3pt] (output_y) {$y$};
      \end{scope}
      
      \node[processor, clr_outofDB] at($(output_table)+(4.5,0)$) (ML) {ML Tool};
      
      \node[data] at($(ML)+(2.75,0)$) (theta_opt) {$\vec\theta^*$};
      
      \node[data] at($(ML)+(0,-3.25)$) (model) {Model};
      
      \node[processor, clr_inDB_FD, scale=.8] at($(model)+(-4.5,0)$) (model_reform) {Model\\ Reformulation};
      
      \node[data, clr_inDB_FD, inner sep = .2cm] at($(DB)+(0, -5)$) (FAQ-queries) 
         {\underline{\underline{~FAQ Queries:~}}\\\vspace{3.3cm}};
      \node[clr_inDB_FD] at($(FAQ-queries)+(0,1.25)$) (FAQ-queries-1) {$\vec\sigma_{11}=\varphi^{(1,1)}$};
      \node[clr_inDB_FD] at($(FAQ-queries-1)+(0,-.6)$) (FAQ-queries-2) {$\vdots~~$};
      \node[clr_inDB_FD] at($(FAQ-queries-2)+(0,-.6)$) (FAQ-queries-3) {$\vec\sigma_{i,j}=\varphi^{(i,j)}$};
      \node[clr_inDB_FD] at($(FAQ-queries-3)+(0,-.6)$) (FAQ-queries-4) {$\underline{\hspace{.8cm}\vdots~~\hspace{.8cm}}$};
      \node[clr_inDB_FD] at($(FAQ-queries-4)+(0,-.6)$) (FAQ-queries-5) {$~~\mv c_1=\varphi^{(1)}~~$};
      \node[clr_inDB_FD] at($(FAQ-queries-5)+(0,-.6)$) (FAQ-queries-6) {$\vdots~~$};
      
      \begin{scope}[shift={($(FAQ-queries)+(0, -4)$)}]
         \drawfilter{2.25}{clr_inDB_FD}
         \node[data, clr_inDB_FD, draw=none, inner sep =2.5] at(0,.05) (FAQ-filter) {FD-filter\\~\vspace{.9cm}};
      \end{scope}
      
      \node[data, clr_inDB_FD, inner sep = .2cm] at($(FAQ-filter)+(0, -3.5)$) (FAQ-queries-FD) 
      {\underline{\underline{Selected queries:}}\\\vspace{2.1cm}};
      \node[clr_inDB_FD] at($(FAQ-queries-FD)+(0,.7)$) (FAQ-queries-2-FD) {$\vdots~~$};
      \node[clr_inDB_FD] at($(FAQ-queries-2-FD)+(0,-.6)$) (FAQ-queries-3-FD) {$\vec\sigma_{i',j'}=\varphi^{(i',j')}$};
      \node[clr_inDB_FD] at($(FAQ-queries-3-FD)+(0,-.6)$) (FAQ-queries-4-FD) {$\underline{\hspace{.8cm}\vdots~~\hspace{.8cm}}$};
      \node[clr_inDB_FD] at($(FAQ-queries-4-FD)+(0,-.6)$) (FAQ-queries-6-FD) {$\vdots~~$};
      
      \node[processor, clr_inDB_FD] at($(FAQ-queries-3)+(-4, 0)$) (FAQ-optimizer) {FAQ Query\\Optimizer};
      
      \newcommand{\TDnode}[5]
      {
         \begin{scope}[shift={(#1,#2)}]
            \node[clr_inDB_FD] at(0,.9) {#4};
            \begin{scope}[shift={(-.1,-.2)}]
               \pgfmathsetmacro{\width}{1.5}
               \pgfmathsetmacro{\height}{1.5}
               \pgfmathsetmacro{\numCols}{4}
               \pgfmathsetmacro{\numRows}{5}
               \pgfmathsetmacro{\cellwidth}{\width/\numCols}
               \pgfmathsetmacro{\cellheight}{\height/\numRows}
               \drawtable{\width}{\height}{\numCols}{\numRows}{clr_outofDB}
               \drawtable{\width}{\height}{\numCols}{\numRows}{clr_inDB_FD};
               \draw [clr_inDB_FD, <->] (\width/2+\cellwidth/2, \height/2-\cellheight) -- (\width/2+\cellwidth/2, -\height/2)
               node [midway, right] (#5){};
            \end{scope}
            \node[clr_inDB_FD, draw, ellipse, inner sep = .95cm, outer sep = 0] at (0, 0) (#3) {};
         \end{scope}
      }
      
      \node[data, clr_inDB, scale=1.4] at($(FAQ-queries)+(0, -14.25)$) (sigma_c) {$\vec\Sigma, \vec c$};
      
      \node[data, clr_inDB] at($(model_reform)+(0,-13.75)$) (theta) {$\vec\theta$};
      
      \node[data, clr_inDB] at($(theta)+(7.25,-2)$) (gradJ) {$J(\vec\theta)$\\$\grad J(\vec\theta)$};
      
      \node[processor, clr_inDB] at($(gradJ)+(0,2)$) (converge) {converged?};
      
      \coordinate (GD_corner1) at ($(theta)+(-1 ,+1)$);
      \coordinate (GD_corner2) at ($(gradJ)+(+1.5 ,-1)$);
      \draw[processor, opacity=.1, fill=clr_inDB, rounded corners = .5cm] (GD_corner1) rectangle (GD_corner2);
      \node[clr_inDB, scale=1.3, opacity=.5] at ($(GD_corner1)!.5!(GD_corner2)$) {\emph{Gradient Descent}};

      \node[data, clr_inDB_FD] at($(FAQ-queries-FD)+(0, -2.5)$) (sigma_c_bar) {$\overline{\vec\Sigma}, \overline{\mv c}$};
      
      \begin{scope}[shift={($(sigma_c_bar)+(4.125,2)$)}, scale=.85]
         \node[data, clr_inDB_FD] at($(0, 0)$) (theta-FD) {$\vec\gamma$};
         
         \node[data, clr_inDB_FD] at($(theta-FD)+(6,-2)$) (gradJ-FD) {$\overline J(\vec\gamma)$\\$\grad \overline J(\vec\gamma)$};
         
         \node[processor, clr_inDB_FD] at($(gradJ-FD)+(0,2)$) (converge-FD) {converged?};
         
         \coordinate (GD_corner1-FD) at ($(theta-FD)+(-.5 ,+1)$);
         \coordinate (GD_corner2-FD) at ($(gradJ-FD)+(+1.5 ,-.7)$);
         \draw[processor, opacity=.1, fill=clr_inDB_FD, rounded corners = .5cm] (GD_corner1-FD) rectangle (GD_corner2-FD);
         \node[clr_inDB_FD, scale=1.3, opacity=.5] at ($(GD_corner1-FD)!.5!(GD_corner2-FD)$) {\emph{Gradient Descent}};
      \end{scope}
      
      \draw[path, clr_inDB, dashed, ->, double=none] (FAQ-queries-1)--(FAQ-optimizer);
      \draw[path, clr_inDB] (FAQ-queries-3)--(FAQ-optimizer);
      \draw[path, clr_inDB, dashed, ->, double=none] (FAQ-queries-5)--(FAQ-optimizer);
      \draw[path, clr_inDB] ([xshift=-6cm]FAQ-optimizer) |- (sigma_c);
      \draw[path, clr_inDB] (model_reform)--(theta) node [at start, pos = .58, left]{$g$};
      \draw[path, clr_inDB] (theta) --node[midway, right]{$g(\vec\theta)$} (theta |- gradJ) -- (gradJ);
      \draw[path, clr_inDB] (sigma_c) -- (sigma_c -| gradJ.west);
      \draw[path, clr_inDB] (gradJ) -- (converge);
      \draw[path, clr_inDB] (converge) -- (theta) node [at start, pos = .2, above]{No};
      \draw[path, clr_inDB] (converge) -- (theta_opt) node [at start, pos = .02, right]{Yes};
      
      \begin{scope}[shift={($(FAQ-filter)+(4.125, 0)$)}]
         \drawfilter{2.25}{clr_inDB_FD}
         \node[data, clr_inDB_FD, draw=none, inner sep =2.5] at(0,.05) (g-filter) {FD-filter\\~\vspace{.9cm}};
      \end{scope}
      
      \node[processor, clr_inDB_FD, scale = 1.2, outer sep =0cm] at($(g-filter)+(6.5,0)$) (FD-unit) {matrix\\multiplier/inverter};
      
      \node[data, clr_inDB_FD] at($(gradJ-FD)+(3,0)$) (FD-matrices) {FD\\Matrices};
      
      \draw[path, clr_outofDB] (FEQ)--(DB);
      \draw[path, clr_outofDB] (DB)--(output_table);
      \draw[path, clr_outofDB] (output_x) -| ([xshift=+.5cm]ML);
      \draw[path, clr_outofDB] (output_y) -| ([xshift=-.5cm]ML);
      \draw[path, clr_outofDB] (ML)--(theta_opt);
      \draw[path, clr_outofDB] (model)--(ML);
      
      \draw[path, clr_inDB_FD, thick] (model)--(model_reform);
      \draw[path, clr_inDB_FD, thick] (model_reform) -- (model_reform -| FAQ-queries.east) node [midway, above]{$h$};
      \draw[path, clr_inDB_FD, thick] (DB)--(FAQ-queries);
      \draw[path, clr_inDB_FD, thick] (FEQ)--(FAQ-queries);
      
      \draw[path, clr_inDB_FD, thick] (FAQ-queries)--(FAQ-filter);
      \draw[path, clr_inDB_FD, thick] (FAQ-filter)--(FAQ-queries-FD);
      \draw[path, clr_inDB_FD, thick] (FAQ-queries-3-FD.west)--(FAQ-optimizer);
      \draw[path, clr_inDB_FD, thick] ([xshift=1]FAQ-optimizer) |- (sigma_c_bar);
      
      \draw[path, clr_inDB_FD, thick] (theta-FD) --node[midway, right]{$\overline g(\vec\gamma)$} (theta-FD |- gradJ-FD) -- (gradJ-FD);
      \draw[path, clr_inDB_FD, thick] (sigma_c_bar) -- (sigma_c_bar -| gradJ-FD.west);
      \draw[path, clr_inDB_FD, thick] (gradJ-FD) -- (converge-FD);
      \draw[path, clr_inDB_FD, thick] (converge-FD) -- (theta-FD) node [at start, pos = .2, above]{No};
      \draw[path, clr_inDB_FD, thick] (model_reform)--(model_reform |- g-filter.north) node [midway, right]{$g$};
      \draw[path, clr_inDB_FD, thick,<-] (theta-FD)--(theta-FD |- g-filter.south) node [midway, right]{$\overline g$};
      \draw[path, clr_inDB_FD, thick] (converge-FD)--(converge-FD |- FD-unit.south) node [at start, pos = .02, right]{Yes} node[midway, left] {$\vec\gamma^*$};
      \draw[path, clr_inDB_FD, thick, <-] (theta_opt)--(theta_opt |- FD-unit.north) node [at start, pos = .7, left]{$\vec\theta^*$};
      \draw[path, clr_inDB_FD, thick] (FD-matrices)--(FD-matrices |- FD-unit.south);
      \draw[path, clr_inDB_FD, thick] (FD-matrices)--(FD-matrices -| gradJ-FD.east);
      \end{tikzpicture}
   }
   \caption{Structure-aware learning: {\color{clr_inDB_FD}with FDs} vs. {\color{clr_inDB} without FDs}.}
   \label{fig:low-level-diagram-FD}
   \vspace*{-1em}
\end{figure}

\subsubsection*{Impact of FDs on model complexity}

The prevalence of FDs presents new challenges from both computational and statistical viewpoints. On the one hand, a reasonable and well-worn rule of thumb in statistics dictates that one should always eliminate features that are functionally dependent on others, because this helps reduce both computation and model's complexity, which in turn leads to reduced generalization  error (as also noted in~\cite{Kumar:SIGMOD:16}). On the other hand, the statistical effectiveness of such a rule is difficult to gauge when the nature of dependence goes beyond linearity. In such scenarios, it might be desirable to keep some redundant variables, but only if they help construct simpler forms of regression/classification functions, leading to improved approximation ability for the model class.

It is, however, difficult to know a priori which redundant features lead to simple functions. Therefore, the problem of dimensionality reduction cannot be divorced from the model class under consideration. While this remains unsolved in general, in this work we restricted ourselves to specific classes of learning models, the complexity of which may still be varied through regularization via (non-linear) penalties. Within a regularized parametric model class, we introduced dimensionality reduction techniques (variable elimination and re-parameterization) that may not fundamentally change the model's capacity. The reduction in the number of parameters may still help reduce the variance of parameter estimates, leading to improved generalization error guarantees. 

\subsubsection*{Impact of FDs on computational complexity}

Model reparameterization under FDs does not lower the {\em data complexity} 
from Proposition~\ref{prop:precomputation:time} for the computation
of the sparse tensor representation. Under a simple FD $A\rightarrow B$,
the number of categories of the functionally determined categorical variable $B$ 
cannot exceed that of the functionally determining categorical variable $A$. 
This means that by avoiding the computation of aggregates involving $B$, 
the data complexity for the computation of the sparse tensor representation with both $A$ and $B$ 
is the same as with $A$ only. 

Computing less aggregates means however a reduction in the {\em query complexity}.
In case only $q<n$ variables functionally determine the entire set of $n$ variables,
the dimensionality of $\overline{\vec\Sigma}$ for $\pr^d$ is $\Theta(q^d) \times \Theta(q^d)$, which is much
smaller than the dimensionality $\Theta(n^d) \times \Theta(n^d)$ of $\vec\Sigma$. 
This reduction can be significant: In one of our experiments with
$\pr_2$ on the Retailer dataset \texttt{v}$_4$, there is a reduction from 46M to
36M entries in the sparse tensor representation of $\vec\Sigma$ and $\vec
c$. Proposition~\ref{prop:alternative} in Appendix~\ref{subsec:alternative:corollary} provides
the corresponding version of Corollary~\ref{cor:Sigma:h} with respect to
$\overline{\vec\Sigma}$.

This reduction in the query complexity comes at a price: The gradient solver has a new 
data-dependent computation in the regularizer. For instance, under the functional
dependency $\textsf{city}\rightarrow\textsf{country}$ used in
Section~\ref{sec:intro-fd}, $\vec\theta_\fcountry = {(\mv I_{\fcountry} + \mv R
\mv R^\top)^{-1}\mv R} \vec\gamma_\fcity$ where $\mv R$ is a matrix that maps
between cities and countries in the input database. Computing this linear
algebra expression takes time $O(N_{\fcity}^2N_{\fcountry})$ as explained in
Section~\ref{subsec:linear:algebra}, where $N_{\fcity}$ and $N_{\fcountry}$ are
the number of cities (categories for the \textsf{city} categorical feature) and
respectively countries. Assuming these quantities are small, the reduction in
the number of aggregates vastly dominates the modest increase in the complexity
of the additional linear algebra expression. Figure~\ref{fig:lmfaobreakdown} in
Section~\ref{sec:experiments} indeed shows that using one single functional
dependency for Retailer leads to a $3.5\times$ performance speedup.


\section{The Design and Implementation of AC/DC}
\label{sec:acdc}

In this section, we present the design of AC/DC, which is our
implementation of the algorithms and optimizations for the end-to-end
computation of square loss problems presented in the previous sections. AC/DC
computes each entry in the sparse tensor representation of the problem as an
aggregate over the feature extraction join query, 
following the SQL encoding developed in previous sections, e.g., 
Examples~\ref{ex:sigma:ij:cont} and
~\ref{ex:store:city}. Two key optimizations used by AC/DC for the computation of
these aggregates are: (1) {\em Factorized computation} of 
aggregates over the feature extraction query, with low complexity (Section~\ref{sec:factorized}); and (2) {\em Massively
  shared computation} across the aggregates~(Section~\ref{sec:shared}).  
AC/DC also exploits
functional dependencies to reduce the dimensionality of the problem.
By design, AC/DC does not achieve the complexity bound from
Proposition~\ref{prop:precomputation:time}. This is because it would need
different query plans for different subsets of the aggregates. Instead,
it uses one query plan for all these aggregates. This increases the opportunity
to share computation across the aggregates, which proved much more beneficial
for the overall performance.

\begin{figure*}[t]
  \begin{center}
    \begin{tabular}{|l|}\hline {\bf aggregates} (variable order $\Delta$, varMap,
      relation ranges $R_1[x_1,y_1],\ldots,R_d[x_d,y_d]$)\\\hline
      $A = root(\Delta); \TAB \text{context} = \pi_{dep(A)} (\text{varMap});
      \TAB \text{reset}(\text{aggregates}_A); \TAB \#\aggs = \left| \aggs_A \right|$;\\ \\
      $\IF\STAB (dep(A) \neq anc(A)) \STAB \{ 
      \STAB \text{aggregates}_A = \text{cache}_A[\text{context}]; \TAB \IF\STAB
      (\text{aggregates}_A[0]  \neq \emptyset) \STAB\RETURN; \STAB \}$\\ \\
      $\FOREACH\STAB i\in [d]\STAB\DO\STAB R_i[x'_i,y'_i] = R_i[x_i,y_i]$;\\
      $\FOREACH\STAB a\in \bigcap_{i\in[d] \mbox{ such that } A\in vars(R_i)} \pi_A (R_i[x_i,y_i])
      \STAB\DO\STAB \{$\\
      $\TAB\FOREACH\STAB i\in [d] \mbox{ such that } A\in vars(R_i)\STAB\DO$\\
      $\TAB\TAB\text{find range } R_i[x'_i,y'_i]
      \subseteq R_i[x_i,y_i] \mbox{ such that } \pi_A (R_i[x'_i,y'_i]) =
      \{(A:a)\};$\\
      $\TAB\MATCH\ (A):$ \\
      {\color{red}$\TAB\TAB$
      \begin{tabular}{|ll|}\hline
        {\color{black}$\textbf{continuous feature } : \STAB$}
        & {\color{black}$\lambda_A = [\{() \mapsto 1\}, \{() \mapsto a^1\} , 
          \ldots,\{()\mapsto a^{2 \cdot degree}\} \}];$}\\
        {\color{black}$\textbf{categorical feature } : \STAB$}
        & {\color{black}$\lambda_A = [\{() \mapsto 1\}, \{a \mapsto 1\}];$}\\
        {\color{black}$\textbf{no feature } : \STAB$} 
        & {\color{black}$\lambda_A = [\{() \mapsto 1\}];$}\\\hline
      \end{tabular}
      } \\
      $\TAB\MATCH\ (\Delta):$\\
      $\TAB\TAB \textbf{leaf node } A: $\\
      $\TAB\TAB\TAB$\rfbox{$\FOREACH\STAB l \in \left[\#\aggs\right]\STAB\DO\STAB\{$
      $\STAB [i_0] = \mathcal{R}_A[l]; \TAB \aggs_A[l] \pluseq \lambda_A[i_0]; \STAB\}$}\\
      $\TAB\TAB \textbf{inner node } A(\Delta_1, \ldots, \Delta_k): $\\
      $\TAB\TAB\TAB\FOREACH\STAB j\in [k]\STAB\DO\STAB$\\ 
      $\TAB\TAB\TAB\TAB\text{\bf aggregates}(\Delta_j,\text{varMap}\times\{(A:a)\},
      \text{ranges } R_1[x'_1,y'_1], \ldots, R_d[x'_d,y'_d]);$\\
      {\color{red}$\TAB\TAB\TAB $
      \begin{tabular}{|l|}\hline
      {\color{black}
      $\IF\STAB (\forall j \in [k] : \aggs_{root(\Delta_j)}[0] \neq \emptyset)
      \STAB$}\\
        {\color{black}
        $\TAB\FOREACH\STAB l \in \left[\#\aggs\right]\;\DO\STAB \{$}\\
        {\color{black}$\TAB\TAB [i_0, i_1, \ldots, i_k] = \mathcal{R}_A[l];$}\\ 
        {\color{black}$\TAB\TAB\aggs_A[l] \pluseq 
        \lambda_A[i_0]\times\bigtimes_{j \in [k]}\aggs_{root(\Delta_j)}[i_j];\; \}$}\\\hline 
      \end{tabular}
      } \\
      $\}$\\ 
      $\IF\STAB (dep(A) \neq anc(A))\TAB \text{cache}_A[\text{context}] = 
      \aggs_A;$\\\hline
    \end{tabular}
  \end{center}
  \caption{Algorithm for factorized computation of aggregates. Each aggregate
    maps tuples over its group-by variables to scalars.
    The parameters of the initial call are the variable order
    $\Delta$ of the feature extraction query, an empty map from variables to values, and the full range of tuples for each input
    relation $R_1,\ldots,R_d$.
  } \label{fig:aggcomp}\vspace*{-1em}
\end{figure*}


\subsection{Factorized aggregate computation}
\label{sec:factorized}

Factorized aggregate computation relies on a variable order for the query $Q$ to
avoid redundant computation. In this paper, we assume that we are given a
variable order. Prior work discusses the query optimization
problem of finding good orders~\cite{BKOZ13,faq}.

{\it Variable Orders.} State-of-the-art query evaluation uses relation-at-a-time
query plans.  We use variable-at-a-time query plans, which we call variable
orders. These are partial orders on the variables in the query, capture the join
dependencies in the query, and dictate the order in which we solve each join
variable.  For each variable, we join all relations with that variable.  Our
choice is motivated by the complexity of join evaluation: Relation-at-a-time
query plans are provably suboptimal, whereas variable-at-a-time query plans can
be chosen to be optimal~\cite{skew}.

For a query $Q$, a variable $X$ \emph{depends} on a variable $Y$ if both
are in the schema of a relation in $Q$.

\begin{definition}[adapted from~\cite{OlZa15}]
  A variable order $\Delta$ for a join query $Q$ is a pair $(F, dep)$, where $F$
  is a rooted forest with one node per variable in $Q$, and $dep$ is a function
  mapping each variable $X$ to a set of variables in $F$. 
  It satisfies the following constraints:
  \begin{itemize}
  \item For each relation in $Q$, its variables lie along the same root-to-leaf
    path in $F$.
  \item For each variable $X$, $dep(X)$ is the subset of its ancestors in $F$ on which
    the variables in the subtree rooted at $X$ depend.
  \end{itemize}
\end{definition}

Without loss of generality, we use variables orders that are trees instead of
forests. We can convert a forest into a tree by adding to each relation the same
dummy join variable that takes a single value.  For a variable $X$ in the
variable order $\Delta$, $anc(X)$ is the set of all ancestor variables of $X$ in
$\Delta$.  The set of variables in $\Delta$ (schema of a relation $R$) is
denoted by $vars(\Delta)$ ($vars(R)$ respectively) and the variable at the root
of $\Delta$ is denoted by $root(\Delta)$.

\begin{ex} 
  Figure~\ref{fig:varorder} shows a variable order for the natural join of
  relations $R(A,B,C)$, $T(B,D)$, and $S(A,E)$. 
  Then, $anc(D)=\{A,B\}$ and $dep(D)=\{B\}$, i.e., $D$ 
  has ancestors $A$ and $B$, yet it only depends on $B$. 
  Given $B$, the variables $C$ and $D$ are independent of each
  other. For queries with group-by variables, we choose a variable order
  where these variables sit above the other variables~\cite{BKOZ13}. \qed
\end{ex}

Figure~\ref{fig:aggcomp} presents the AC/DC algorithm for factorized computation
of SQL aggregates over the feature extraction query $Q$. The backbone of the
algorithm without the code in boxes explores the factorized join of the input
relations $R_1,\ldots,R_d$ over a variable order $\Delta$ of $Q$. As it
traverses $\Delta$ in depth-first preorder, it assigns values to the query
variables. The assignments are kept in varMap and used to compute aggregates by
the code in boxes.

The relations are sorted following a depth-first pre-order traversal of
$\Delta$. Each call takes a range $[x_i,y_i]$ of tuples in each relation
$R_i$. Initially, these ranges span the entire relations.  Once the root
variable $A$ in $\Delta$ is assigned a value $a$ from the intersection of
possible $A$-values from the input relations, these ranges are narrowed down to
those tuples with value $a$ for $A$.

To compute an aggregate over the variable order $\Delta$ rooted at $A$, we first
initialize the aggregate to zeros. This is needed since the aggregates might
have been used earlier for different assignments of ancestor variables in
$\Delta$.  We next check whether we previously computed the aggregate for the
same assignments of variables in $dep(A)$, denoted by $\text{context}$, and
cached it in a map $\text{cache}_A$.  Caching is useful when $dep(A)$ is
strictly contained in $anc(A)$, since this means that the aggregate computed at
$A$ does not need to be recomputed for distinct assignments of variables in
$anc(A)\setminus dep(A)$.  In this case, we probe the cache using as key the
assignments in varMap of the $dep(A)$ variables:
$\text{cache}_A[\text{context}]$.  If we have already computed the aggregates
over that assignment for $dep(A)$, then we can just reuse the previously
computed aggregates and avoid recomputation.

If $A$ is a group-by variable, then we compute a map from each $A$-value $a$ to
a function of $a$ and aggregates computed at children of $A$, if any. If $A$ is
not a group-by variable, then we compute a map from the empty value $()$ to such
a function; in this latter case, we could have just computed the aggregate
instead of the map though we use the map for uniformity. In case there are
group-by variables under $A$, the computation at $A$ returns maps whose keys are
tuples over all these group-by variables in $vars(\Delta)$.

\begin{ex} \label{ex:aggpushdown} 
  Consider a feature extraction query $Q$ with the variable order $\Delta$ in Figure~\ref{fig:varorder}. 
  We first compute the assignments for $A$ as
  $Q_A = \pi_A R \bowtie \pi_A T$. For each assignment $a \in Q_A$, we
  then find assignments for variables under $A$ within the narrow ranges
  of tuples that contain $a$. 
  The assignments for $B$ in the context of $a$ are given by 
  $Q^{a}_B = \pi_B(\sigma_{A = a}R) \bowtie \pi_BS$. For each
  $b \in Q^{a}_B$, the assignments for $C$ and $D$ are given by
  $Q^{a,b}_C = \pi_C(\sigma_{A = a \land B = b}R)$ and
  $Q^{b}_D = \pi_D(\sigma_{B = b}S)$. Since $D$ depends on $B$ and not on $A$, the 
  assignments for $D$ under a given $b$ are repeated for every occurrence of $b$
  with assignments for $A$. 
  The assignments for $E$ given $a\in Q_A$ are computed as
  $Q^{a}_E = \pi_E(\sigma_{A = a}T)$.

  Consider the aggregate $\COUNT(Q)$. 
  The count at each variable $X$ is computed
  as the sum over all value assignments of $X$ of the product of the counts 
  at the children of $X$ in $\Delta$; if $X$ is a leaf in $\Delta$, 
  the product at children is considered 1.
  For our variable order,
  this computation is captured by the following factorized expression:
  \begin{align}
    \COUNT =
    \sum_{a \in Q_A} \!\! 1 \cdot \left(\sum_{b \in Q^{a}_B} \!\! 1 \cdot
    \left(\sum_{c \in Q^{a,b}_C} \!\!\!1 \cdot V_D(b) \right)\right)
    \cdot \sum_{e \in Q^{a}_E} \!\! 1
    \label{eq:count}
  \end{align}
  where $V_D(b) = \sum_{d \in Q^{b}_D}\!\! 1$ is cached the first time 
  we encounter the assignment $b$ for $B$ and reused for all subsequent
  occurrences of this assignment under assignments for $A$.
  
  Summing all $X$-values in the result of $Q$ for a variable $X$ is done
  similarly, with the difference that at the variable $X$ in $\Delta$ we compute
  the sum of the values of $X$ weighted by the product of the counts of their
  children.  For instance, the aggregate $\SUM(C*E)$ is computed over our
  variable order by the following factorized expression:
  \begin{align}
    \SUM(C \cdot E) =
    \sum_{a \in Q_A} \!\! 1 \cdot \left(\sum_{b \in Q^{a}_B} \!\! 1 \cdot
    \left(\sum_{c \in Q^{a,b}_C} \!\!\!c \cdot V_D(b) \right)\right)
    \cdot \sum_{e \in Q^{a}_E} \!\! e
    \label{eq:sum_ce}
  \end{align}

  To compute the aggregate $\SUM(C*E) \texttt{ GROUP BY } A$, we compute
  $\SUM(C*E)$ for each assignment for $A$ instead of marginalizing away $A$. The
  result is a map from $A$-values to values of $\SUM(C*E)$.
  \qed
  

\end{ex}

A good variable order may include
variables that are not explicitly used in the optimization problem. 
This is the case of join variables whose presence in the variable order 
ensures a good factorization. For instance, if we remove the variable $B$ from
the variable order in Figure~\ref{fig:varorder}, 
the variables $C,D$ are no longer independent and we cannot factorize the
computation over $C$ and $D$. AC/DC exploits the conditional 
independence enabled by $B$, but computes no aggregate over $B$ if this is not
required in the problem.

The complexity bound in Proposition~\ref{prop:precomputation:time} is achieved
by factorizing the computation of each aggregate in $\vec\Sigma$ over a variable
order that has all group-by variables for this aggregate above all other
variables. Thus, different aggregates can be computed over different variable
orders.

\subsection{Shared computation of aggregates}
\label{sec:shared}

Section~\ref{sec:factorized} explains how to factorize the computation of one
aggregate in $\vec\Sigma$, $\vec c$, and $s_Y$ over the join of database relations. In this section we show how to share the computation across aggregates.

\begin{ex}
  We consider the factorized expression of the aggregates $\SUM(C)$ and
  $\SUM(E)$ over $\Delta$:
  \begin{align}
    \SUM(C) &=
              \sum_{a \in Q_A} \!\! 1 \cdot \left(\sum_{b \in Q^{a}_B} \!\! 1
              \cdot \left(\sum_{c \in Q^{a,b}_C} \!\!\!c \cdot V_D(b)\right)\right) \cdot \sum_{e \in Q^{a}_E} \!\! 1 \label{eq:sum_c}\\
    \SUM(E) &=
              \sum_{a \in Q_A} \!\! 1 \cdot \left(\sum_{b \in Q^{a}_B} \!\! 1 \cdot
              \left(\sum_{c \in Q^{a,b}_C} \!\!\!1 \cdot V_D(b)\right)\right)
              \cdot \sum_{e \in Q^{a}_E} \!\! e \label{eq:sum_e}
  \end{align}
  We can share computation across the expressions~\eqref{eq:count} to \eqref{eq:sum_e} since they are similar.
  For instance, given an assignment $b$ for $B$, all these aggregates need $V_D(b)$. Similarly, for a given assignment $a$ for $A$, the aggregates \eqref{eq:sum_ce} and \eqref{eq:sum_e} can share the computation of
  the sum aggregate over $Q_E^{a}$. For assignments $a \in Q_A$ and
  $b \in Q^{a}_B$, expressions \eqref{eq:sum_ce} and \eqref{eq:sum_c} can share the
  computation of the sum aggregate over $Q_C^{a,b}$. \qed
\end{ex}

To share as much computation as possible between aggregates, AC/DC computes all
aggregates together over a single variable order, which significantly improves
the data locality of the aggregate computation. This approach does not follow 
Proposition~\ref{prop:precomputation:time} that assumes that each aggregate is
computed over its respective best variable order. AC/DC thus decidedly
sacrifices the goal of achieving the lowest-known complexity for individual
aggregates for the sake of sharing as much computation as possible across these
aggregates.

\begin{figure*}[t] 
  \small \centering 
  
  \begin{tikzpicture}[array/.style={rectangle split,rectangle split
      horizontal, rectangle split parts=#1, draw, anchor=center}]

    \tikzstyle{rarray}= [
    minimum height=2em,
    minimum width=2.5em, draw
    ]
    
    \node at (-3.5, 0) (A) {$A$};
    \node at (-4, -4) (D1) {$\Delta_1$} edge[-] (A);
    \node at (-3, -4) (D2) {$\Delta_k$} edge[-] (A);
    \node at (-3.5, -4) {$\cdots$};
    
    \node [rarray, color=red ,draw]at (7.5,-1) (prod){
      {\large \color{black}$\alpha \pluseq \alpha_0 \times
        \bigtimes_{j \in [k]} \alpha_{j}$}};

    \begin{scope}[start chain=1 going right,node distance=-0.15mm]
      \node [rarray,on chain=1,draw=none] at (2.68,0) {$\aggs_A = \;$};
      \node [rarray,dashed, on chain=1,draw] {$\cdots$};
      \node [rarray, on chain=1,draw] (alpha){$\alpha$};
      \node [rarray,dashed, on chain=1,draw] {$\cdots$};
    \end{scope}

    \begin{scope}[start chain=1 going right,node distance=-0.15mm]
      \node [rarray,on chain=1,draw=none] at (3,-2) {}; 
      \node [rarray, on chain=1,draw] (i0){$i_0$};
      \node [rarray, on chain=1,draw] (i1){$i_1$};
      \node [rarray, on chain=1,draw] (ic){$\cdots$};
      \node [rarray, on chain=1,draw] (ik){$i_k$};
    \end{scope}

    \begin{scope}[start chain=1 going right,node distance=-0.15mm]
      \node [rarray, on chain=1,draw=none]  at (-2,-0) (lambdaA) {$\lambda_A = \;$};
      \node [rarray, on chain=1,dashed,draw] (a1) {$\cdots$};
      \node [rarray, on chain=1,draw] (a2) {$\alpha_0$};
      \node [rarray, on chain=1,dashed,draw] (a3){$\cdots$};
    \end{scope}
       
    \begin{scope}[start chain=1 going right,node distance=-0.15mm]
      \node [rarray,on chain=1,draw=none,anchor=east] at (-2,-4) {};
      \node [rarray, on chain=1,draw=none] (delta1){$\aggs_{root(\Delta_1)} = \;$};
      \node [rarray, on chain=1,dashed,draw] (i11){$\cdots$};
      \node [rarray, on chain=1,draw] (i12){$\alpha_{1}$};
      \node [rarray, on chain=1,dashed,draw] (i13){$\cdots$};
    \end{scope}

    \begin{scope}[start chain=1 going right,node distance=-0.15mm]
      \node [rarray, on chain=1,draw=none]at (3.8,-4) {};
      \node [rarray, on chain=1,draw=none] (delta2){$\aggs_{root(\Delta_k)} = \;$};
      \node [rarray, on chain=1,dashed,draw] (ik1){$\cdots$};
      \node [rarray, on chain=1,draw] (ik2){$\alpha_{k}$};
      \node [rarray, on chain=1,dashed,draw] (ik3){$\cdots$};
    \end{scope}

    \node[color=black!80, anchor=south, yshift=-1] at (a1.north) {\scriptsize $i_0 - 1$};
    \node[color=black!80, anchor=south, yshift=-1] at (a2.north) {\scriptsize $i_0$};
    \node[color=black!80, anchor=south, yshift=-1] at (a3.north) {\scriptsize $i_0 + 1$};

    \node[color=black!80, anchor=south, yshift=-1] at (i11.north) {\scriptsize $i_1 - 1$};
    \node[color=black!80, anchor=south, yshift=-1] at (i12.north) (ii1) {\scriptsize $i_1$};
    \node[color=black!80, anchor=south, yshift=-1] at (i13.north)  {\scriptsize $i_1 + 1$};

    \node[color=black!80, anchor=south, yshift=-1] at (ik1.north) {\scriptsize $i_k - 1$};
    \node[color=black!80, anchor=south, yshift=-1] at (ik2.north) (iik) {\scriptsize $i_k$};
    \node[color=black!80, anchor=south, yshift=-1] at (ik3.north)  {\scriptsize $i_k + 1$};

    \draw[->] (alpha.south) -- (ic.north west);

    \draw[->] (i0.south) -- ($ (i0.south) + (0,-0.3) $) --  ($ (a2.south) + (0,-2.3) $)
    -- (a2.south);
    \draw[->] (i1.south) -- ($ (i1.south) + (0,-0.46) $) --  ($ (ii1.north) + (0,0.5) $)
    -- (ii1.north);
    \draw[->] (ik.south) -- ($ (ik.south) + (0,-0.46) $) --  ($ (iik.north) + (0,0.5) $)
    -- (iik.north);
   
  \end{tikzpicture}%
  \caption{Index structure provided by the aggregate register for a particular
    aggregate $\alpha$ that is computed over the variable order
    $\Delta = A(\Delta_1, \ldots, \Delta_k)$. The computation of
    $\alpha$ is expressed as the sum of the Cartesian products of its aggregate components provided by the indices $i_0,\ldots,i_k$.} 
    \label{fig:genaggreg} \vspace*{-1em}
\end{figure*}

\subsubsection*{Aggregate Decomposition and Registration.}  For a model of degree
$degree$ and a set of variables $\{A_l\}_{l\in[n]}$, we have aggregates of the
form $\SUM(\prod_{l\in[n]}A_l^{d_l})$, possibly with a group-by clause, such
that $0\leq \sum_{l\in[n]} d_l \leq 2\cdot degree$, $d_l \geq 0$, and all
categorical variables are turned into group-by variables. The reason for
$2\cdot degree$ is due to the $\vec\Sigma$ matrix used to compute the
gradient of the loss function~\eqref{eqn:gradient}, which pairs
any two features of degree up to $degree$.  Each aggregate is thus defined
uniquely by a monomial $\prod_{l\in[n]}A_l^{d_l}$; we may discard the variables
with exponent 0. For instance, the monomial for $\SUM(C*E)$ is CE while for $\SUM(C*E) \text{ GROUP BY } A$ is \textbf{A}CE.

Aggregates can be decomposed into shareable components. Consider a variable order $\Delta = A(\Delta_1, \ldots, \Delta_k)$, with root $A$ and subtrees $\Delta_1$ to $\Delta_k$.
We can decompose any aggregate $\alpha$ to be
computed over $\Delta$ into $k+1$
aggregates such that aggregate $0$ is for $A$ and aggregate $j \in [k]$ is for $root(\Delta_j)$. Then $\alpha$ is computed as the product of its $k+1$ components. Each of these aggregates is defined by the projection of the monomial of $\alpha$ onto $A$ or $vars(\Delta_j)$. 
The aggregate $j$ is then pushed down the variable order and computed over the subtree $\Delta_j$. If the projection of the
monomial is empty, then the aggregate to be pushed down is $\SUM(1)$, which
computes the size of the join defined by $\Delta_j$. If
several aggregates push the same aggregate to the subtree $\Delta_j$, this
is computed only once for all of them. 

The decomposed aggregates form a hierarchy whose structure is that of the underlying variable order $\Delta$. The aggregates at a variable $X$ are denoted by $\aggs_X$. All aggregates are to be computed at the root of $\Delta$, then fewer are computed at each of its children and so on. This structure is the same regardless of the input data and can be constructed before data processing.
We therefore construct at compile time for each variable $X$ in $\Delta$ an aggregate register $\mathcal{R}_X$ that is an array of 
all aggregates to be computed  over the subtree of $\Delta$ rooted at $X$.
This register is used as an index structure to facilitate the computation of the
actual aggregates. More precisely, an entry for an aggregate $\alpha$ in the register of $X$ is labeled by the monomial of $\alpha$ and holds an array of indices of the components of $\alpha$ located in the registers at the children of $X$ in $\Delta$ and in the local register $\Lambda_X$ of $X$. 
Figure~\ref{fig:genaggreg} depicts this construction.

The hierarchy of registers forms an index structure that is used by AC/DC to compute the aggregates. This index structure is stored as one contiguous array in memory, where the entry for an aggregate $\alpha$ in the register comes with an auxiliary array with the indices of $\alpha$'s  aggregate components. The aggregates are ordered in the register so that we increase sequential access, and thus cache locality, when updating them.

\nop{
  Figure~\ref{fig:genaggreg} depicts the index structure provided by the
  aggregate register for one aggregate $\alpha \in \aggs_A$ that is computed
  over a variable order $\Delta = A (\Delta_1, \ldots, \Delta_k)$. The array
  $\aggreg_\alpha$ has the offsets of $\alpha$'s components: The offset $i_0$ is
  for the aggregate in $\Lambda_A$ that corresponds to the projection of the
  monomial of $\alpha$ on $A$. For each $j \in [k]$, the offset $i_j$ in the
  register $\mathcal{A}_{var(\Delta_j)}$ corresponds to the projection of the
  monomial of $\alpha$ onto $schema(\Delta_j)$.
}

\nop{

}

\begin{figure*}[t] 
  \small \centering
  \hspace{-0.3cm}
  \subfigure[\label{fig:varorder}Variable Order $\Delta$.]{
    \begin{minipage}{2.7cm}
      \begin{small}
        \begin{tikzpicture}[xscale=0.5, yscale=0.65]
          \node at (-3, -1) (A) {$A$};
          \node at (-3.7, -4.5) (B) {$B$} edge[-] (A);
          \node at (-2.8, -8) (D) {$D$} edge[-] (B);
          \node at (-4.2, -8) (C) {$C$} edge[-] (B);
          \node at (-2.3, -4.5) (E) {$E$} edge[-] (A);

          \begin{small}
            \node[anchor = north] at (C.south) {$\{A,B\}$};
            \node at (-4.7, -4.5) {$\{A\}$};
            \node[anchor = north] at ($(D.south)+(0.1,0)$) {$\{B\}$};
            \node[anchor = north] at (E.south) {$\{A\}$};
            \node[anchor = east] at (-3.5, -1) {$\{\;\}$};
          \end{small}

          \node at (5,-9.2) (xxx) {~};

        \end{tikzpicture}%
      \end{small}
    \end{minipage} }\hspace{-0.8cm}%
  \subfigure[\label{fig:aggreg}Aggregate Registers.]{
    \begin{minipage}{12cm}
      \begin{tikzpicture}[array/.style={rectangle split,rectangle split
          horizontal, rectangle split parts=#1,draw, anchor=center},xscale=0.65,
        yscale=0.65]

        \node[array=19] at (-2.5, 0) (aggA) {
          \nodepart{one} 1
          \nodepart{two} A
          \nodepart{three} {\bf B}
          \nodepart{four} C
          \nodepart{five} D
          \nodepart{six} {\bf E}
          \nodepart{seven} AA
          \nodepart{eight} A{\bf B}
          \nodepart{nine} AC
          \nodepart{ten} AD
          \nodepart{eleven} A{\bf E}
          \nodepart{twelve} {\bf B}C
          \nodepart{thirteen} {\bf B}D
          \nodepart{fourteen} {\bf B}{\bf E}
          \nodepart{fifteen} CC
          \nodepart{sixteen} CD
          \nodepart{seventeen} C{\bf E}
          \nodepart{eighteen} DD
          \nodepart{nineteen} D{\bf E}
        };

        \node[anchor=east] at (aggA.one west) (talphaA) {$\mathcal{R}_A = $};

        \node[array=3] at (-9.5, -3.5) (lambdaA) {
          \nodepart{one} 1
          \nodepart{two} A
          \nodepart{three} AA
        };

        \node[anchor=east] at (lambdaA.one west) (tlambaA) {$\Lambda_A = $};

        \node[array=9] at (-2, -3.5) (aggB) {
          \nodepart{one} 1
          \nodepart{two} {\bf B}
          \nodepart{three} C
          \nodepart{four} D
          \nodepart{five} {\bf B}C
          \nodepart{six} {\bf B}D
          \nodepart{seven} CC
          \nodepart{eight} CD
          \nodepart{nine} DD
        };

        \node[anchor=east] at (aggB.one west) (talphaA) {$\mathcal{R}_B = $};
        
        \node[array=2] at (6, -3.5) (aggE) { \nodepart{one} 1 \nodepart{two}
          {\bf E} };

        \node[anchor=east] at (aggE.one west) (talphaA) {$\mathcal{R}_E = \Lambda_E = $};

        \node[array=2] at (-6, -7) (lambdaB) { \nodepart{one} 1
          \nodepart{two} {\bf B} };

        \node[anchor=east] at (lambdaB.one west) (talphaA) {$\Lambda_B = $};

        \node[array=3] at (-1, -7) (aggC) { \nodepart{one} 1 \nodepart{two} C
          \nodepart{three} CC };
        
        \node[anchor=east] at (aggC.one west) (talphaA) {$\mathcal{R}_C = \Lambda_C = $};

        \node[array=3] at (5.5, -7) (aggD) { \nodepart{one} 1 \nodepart{two} D
          \nodepart{three} DD };

        \node[anchor=east] at (aggD.one west) (talphaA) {$\mathcal{R}_D = \Lambda_D = $};

        \draw[color=black!40] (aggA.one south) -- (lambdaA.one north);
        \draw[color=black!40] (aggA.two south) -- (lambdaA.two north);
        \draw[color=black!40] (aggA.three south) -- (lambdaA.one north);
        \draw[color=black!40] (aggA.four south) -- (lambdaA.one north);
        \draw[color=black!40] (aggA.five south) -- (lambdaA.one north);
        \draw[color=black!40] (aggA.six south) -- (lambdaA.one north);
        \draw[color=black!40] (aggA.seven south) -- (lambdaA.three north);
        \draw[color=black!40] (aggA.eleven south) -- (lambdaA.two north);
        \draw[color=black!40] (aggA.twelve south) -- (lambdaA.one north);
        \draw[color=black!40] (aggA.thirteen south) -- (lambdaA.one north);
        \draw[color=black!40] (aggA.fourteen south) -- (lambdaA.one north);
        \draw[color=black!40] (aggA.fifteen south) -- (lambdaA.one north);
        \draw[color=black!40] (aggA.sixteen south) -- (lambdaA.one north);
        \draw[color=black!40] (aggA.seventeen south) -- (lambdaA.one north);
        \draw[color=black!40] (aggA.eighteen south) -- (lambdaA.one north);
        \draw[color=black!40] (aggA.nineteen south) -- (lambdaA.one north);

        \draw[color=goodgreen] (aggA.eight south) -- (lambdaA.two north);
        \draw[color=goodgreen] (aggA.nine south) -- (lambdaA.two north);
        \draw[color=goodgreen] (aggA.ten south) -- (lambdaA.two north);

        \draw[color=black!40, dashed] (aggA.one south) -- (aggB.one north);
        \draw[color=black!40, dashed] (aggA.two south) -- (aggB.one north);
        \draw[color=black!40, dashed] (aggA.three south) -- (aggB.two north);
        \draw[color=black!40, dashed] (aggA.four south) -- (aggB.three north);
        \draw[color=black!40, dashed] (aggA.five south) -- (aggB.four north);
        \draw[color=black!40, dashed] (aggA.six south) -- (aggB.one north);
        \draw[color=black!40, dashed] (aggA.seven south) -- (aggB.one north);
        \draw[color=black!40, dashed] (aggA.eleven south) -- (aggB.one north);
        \draw[color=black!40, dashed] (aggA.twelve south) -- (aggB.five north);
        \draw[color=black!40, dashed] (aggA.thirteen south) -- (aggB.six north);
        \draw[color=black!40, dashed] (aggA.fourteen south) -- (aggB.two north);
        \draw[color=black!40, dashed] (aggA.fifteen south) -- (aggB.seven north);
        \draw[color=black!40, dashed] (aggA.sixteen south) -- (aggB.eight north);
        \draw[color=black!40, dashed] (aggA.seventeen south) -- (aggB.three north);
        \draw[color=black!40, dashed] (aggA.eighteen south) -- (aggB.nine north);
        \draw[color=black!40, dashed] (aggA.nineteen south) -- (aggB.four north);

        \draw[color=blue, dashed] (aggA.eight south) -- (aggB.two north);
        \draw[color=blue, dashed] (aggA.nine south) -- (aggB.three north);
        \draw[color=blue, dashed] (aggA.ten south) -- (aggB.four north);

        \draw[color=black!40,semithick,dotted] (aggA.one south) -- (aggE.one north);
        \draw[color=black!40,semithick,dotted] (aggA.two south) -- (aggE.one north);
        \draw[color=black!40,semithick,dotted] (aggA.three south) -- (aggE.one north);
        \draw[color=black!40,semithick,dotted] (aggA.four south) -- (aggE.one north);
        \draw[color=black!40,semithick,dotted] (aggA.five south) -- (aggE.one north);
        \draw[color=black!40,semithick,dotted] (aggA.six south) -- (aggE.two north);
        \draw[color=black!40,semithick,dotted] (aggA.seven south) -- (aggE.one north);
        \draw[color=black!40,semithick,dotted] (aggA.eleven south) -- (aggE.two north);
        \draw[color=black!40,semithick,dotted] (aggA.twelve south) -- (aggE.one north);
        \draw[color=black!40,semithick,dotted] (aggA.thirteen south) -- (aggE.one north);
        \draw[color=black!40,semithick,dotted] (aggA.fourteen south) -- (aggE.two north);
        \draw[color=black!40,semithick,dotted] (aggA.fifteen south) -- (aggE.one north);
        \draw[color=black!40,semithick,dotted] (aggA.sixteen south) -- (aggE.one north);
        \draw[color=black!40,semithick,dotted] (aggA.seventeen south) --
        (aggE.two north);
        \draw[color=black!40,semithick,dotted] (aggA.eighteen south) -- (aggE.one north);
        \draw[color=black!40,semithick,dotted] (aggA.nineteen south) -- (aggE.two north);

        \draw[color=red,semithick,dotted] (aggA.eight south) -- (aggE.one north);
        \draw[color=red,semithick,dotted] (aggA.nine south) -- (aggE.one north);
        \draw[color=red,semithick,dotted] (aggA.ten south) -- (aggE.one north);
        
        \draw[color = goodgreen] (aggB.one south) -- (lambdaB.one north);
        \draw[color = goodgreen] (aggB.two south) -- (lambdaB.two north);
        \draw[color = goodgreen] (aggB.three south) -- (lambdaB.one north);
        \draw[color = goodgreen] (aggB.four south) -- (lambdaB.one north);
        \draw[color = goodgreen] (aggB.five south) -- (lambdaB.two north);
        \draw[color = goodgreen] (aggB.six south) -- (lambdaB.two north);
        \draw[color = goodgreen] (aggB.seven south) -- (lambdaB.one north);
        \draw[color = goodgreen] (aggB.eight south) -- (lambdaB.one north);
        \draw[color = goodgreen] (aggB.nine south) -- (lambdaB.one north);

        \draw[color = blue, dashed] (aggB.one south) -- (aggC.one north);
        \draw[color = blue, dashed] (aggB.two south) -- (aggC.one north);
        \draw[color = blue, dashed] (aggB.three south) -- (aggC.two north);
        \draw[color = blue,dashed] (aggB.four south) -- (aggC.one north);
        \draw[color = blue, dashed] (aggB.five south) -- (aggC.two north);
        \draw[color = blue, dashed] (aggB.six south) -- (aggC.one north);
        \draw[color = blue, dashed] (aggB.seven south) -- (aggC.three north);
        \draw[color = blue, dashed] (aggB.eight south) -- (aggC.two north);
        \draw[color = blue, dashed] (aggB.nine south) -- (aggC.one north);
        
        \draw[color=red,semithick,dotted] (aggB.one south) -- (aggD.one north);
        \draw[color=red,semithick,dotted] (aggB.two south) -- (aggD.one north);
        \draw[color=red,semithick,dotted] (aggB.three south) -- (aggD.one north);
        \draw[color=red,semithick,dotted] (aggB.four south) -- (aggD.two north);
        \draw[color=red,semithick,dotted] (aggB.five south) -- (aggD.one north);
        \draw[color=red,semithick,dotted] (aggB.six south) -- (aggD.two north);
        \draw[color=red,semithick,dotted] (aggB.seven south) -- (aggD.one north);
        \draw[color=red,semithick,dotted] (aggB.eight south) -- (aggD.two north);
        \draw[color=red,semithick,dotted] (aggB.nine south) -- (aggD.three north);

        
        \node at (5,-8.2) (xxx) {~};
      \end{tikzpicture}%
    \end{minipage} }
  \caption{ (a) Variable order $\Delta$ for the natural join of the relations
    R(A,B,C), S(B,D), and T(A,E), each variable $X$ is annotated by the set that
    $dep(X)$ maps to; (b) Aggregate registers for the aggregates needed to
    compute a linear regression model with degree 1 over $\Delta$. Categorical
    variables are shown in bold.  }
  \vspace*{-1em}
  \label{fig:aggregateRegister}
\end{figure*}

\begin{ex} \label{ex:aggreg} Let us compute a regression
  model of degree $1$ over a dataset defined by the join of
  the relations $R(A,B,C), S(B,D)$, and $T(A,E)$.  We assume that $B$ and $E$
  are categorical features, and all other variables are
  continuous. The quantities ($\vec \Sigma$,$\vec c$,$s_Y$) require the
  computation of the following aggregates: $\SUM(1)$, $\SUM(X)$ for each variable $X$, and $\SUM(X * Y)$ for each pair of variables $X$ and $Y$.
  
  Figure~\ref{fig:varorder} depicts a variable order $\Delta$ for the 
  natural join of three relations, and Figure~\ref{fig:aggreg} illustrates 
  the aggregate register that assigns a list of aggregates to each variable 
  in $\Delta$. 
  The aggregates are identified by their respective monomials (the names in the register entries). The categorical
  variables are shown in bold. 
  Since they are treated as group-by variables, we do not need aggregates whose
  monomials include categorical variables with exponents higher than 1. Any such
  aggregate is equivalent to the aggregate whose
  monomial includes the categorical variable with degree 1 only.

  The register $\mathcal{R}_A$ for the root $A$ of $\Delta$  
  has all aggregates needed to compute the model. 
  The register $\mathcal{R}_B$ has all aggregates from $\mathcal{R}_A$ 
  defined over the variables in the subtree of $\Delta$ rooted at $B$. 
  The variables $C$, $D$, and $E$ are leaf nodes in 
  $\Delta$, so the monomials for the aggregates in the registers $\mathcal{R}_C$, 
  $\mathcal{R}_D$, and $\mathcal{R}_E$ are the respective variables only.
  We use two additional registers $\Lambda_A$ and $\Lambda_B$, 
  which hold the aggregates corresponding to projections of the monomials 
  of the aggregates in $\mathcal{R}_A$, and respectively $\mathcal{R}_B$, onto 
  $A$, respectively $B$. For a leaf node $X$, the registers $\Lambda_X$ and $
  \mathcal{R}_X$ are the same.
  
  A path between two register entries in Figure~\ref{fig:aggreg} indicates that
  the aggregate in the register above uses the result of the aggregate in the
  register below. For instance, each aggregate in $\mathcal{R}_B$ is computed by
  the product of one aggregate from $\Lambda_B$, $\mathcal{R}_C$, and
  $\mathcal{R}_D$. The fan-in of a register entry thus denotes the amount of
  sharing of its aggregate: All aggregates from registers above with incoming
  edges to this aggregate share its computation.  For instance, the aggregates
  with monomials \texttt{AB}, \texttt{AC}, and \texttt{AD} from $\mathcal{R}_A$
  share the computation of the aggregate with monomial \texttt{A} from
  $\Lambda_A$ as well as the count aggregate from $\mathcal{R}_E$.  Their
  computation uses a sequential pass over the register $\mathcal{R}_B$. This
  improves performance and access locality as $\mathcal{R}_B$ can be stored in
  cache and accessed to compute all these aggregates. \qed
\end{ex}

\subsubsection*{Aggregate Computation.}  Once the aggregate registers are in
place, we can ingest the input database and compute the aggregates over the join
of the database relations following the factorized structure given by a variable
order. The algorithm in Figure~\ref{fig:aggcomp} does precisely
this. Section~\ref{sec:factorized} explained the factorized computation of a
single aggregate over the join. We explain here the case of several aggregates
organized into the aggregate registers. This is stated by the pseudocode in the
red boxes.

Each aggregate is uniformly stored as a map from tuples over their categorical
variables to payloads that represent the sums over the projection of its
monomial on all continuous variables.  If the aggregate has no categorical
variables, the key is the empty tuple.

For each possible $A$-value $a$, we first compute the array $\lambda_A$ that consists of the projections of the monomials of the aggregates onto $A$. If $A$ is categorical, then we only need to compute the 0 and 1 powers of $a$. If $A$ is continuous, we need to compute all powers of $A$ from 0 to $2\cdot degree$. If $A$ is not a feature used in the model, then we only compute a trivial count aggregate.

We update the value of each aggregate $\alpha$ using the index structure depicted in Figure~\ref{fig:genaggreg} as we traverse the variable order bottom up.  Assume we are at a variable $A$ in the variable order. In case $A$ is a leaf, the update is only a specific value in the local register $\lambda_A$.
In case the variable $A$ has children in the variable order, the aggregate is updated with the Cartesian product of all its component aggregates, i.e., one value from $\lambda_A$ and one aggregate for each child of $A$. The update value  can be expressed in SQL as follows. Assume the aggregate $\alpha$ has group-by variables $C$, which are partitioned across $A$ and its $k$ children. Assume also that $\alpha$'s components are $\alpha_0$ and $(\alpha_j)_{j\in[k]}$. Recall that all aggregates are maps, which we may represent as relations with columns for keys and one column $P$ for payload. Then, the update to $\alpha$ is:
\begin{align*}
  \texttt{SELECT } \mathcal{C}, (\alpha_0.P * \ldots
    * \alpha_k.P) \texttt{ AS } P
  \texttt{ FROM } \alpha_0, \ldots, \alpha_k;
\end{align*}

\subsubsection*{Further Considerations.} The auxiliary arrays that provide the
precomputed indices of aggregate components within registers speed up the
computation of the aggregates. Nevertheless, they still represent one extra
level of indirection since each update to an aggregate would first need to fetch
the indices and then use them to access the aggregate components in registers
that may not be necessarily in the cache. We have been experimenting with an
aggressive aggregate compilation approach that resolves all these indices at
compile time and generates the specific code for each aggregate update. In
experiments with linear regression, this compilation leads to a 4$\times$
performance improvements. However, the downside is that the AC/DC code gets much
larger and the C++ compiler needs much more time to compile it. For
higher-degree models, it can get into situations where the C++ compiler
crashes. We are currently working on a hybrid approach that partially resolves
the indices while maintaining a reasonable code size.

\nop{For performance reasons, our implementation
separates the aggregates that are purely continuous from those with
categorical variables. This allows us to store all continuous aggregates
in a preallocated array of fixed size. To improve cache locality, we store
the relations for all aggregates with categorical variables in one dynamic
array, which stores generic tuples that map keys to payloads. Each aggregate
relation is kept as a consecutive subsection of the dynamic array and we store
the offsets to first and last tuples in the array by extending the aggregate
register for this aggregate. To support for efficient upserting in the dynamic
array, we keep the tuples of each aggregate relation in sorted order. The upsert
procedure then merges the existing aggregates with the result of the Cartesian
product, using a modification of the merge function of the merge sort
algorithm, where duplicate keys are combined into a single tuple with the same
key and the payload is the summation of the payloads of the original tuples.}

\subsubsection*{Point Evaluation, Gradient Computation and FD Optimization}

For the computation of the point evaluation and gradient computation we use the
optimizations we introduced in Section~\ref{sec:gradientcomputation}. Recall
that two entries in $\vec\Sigma$ can have the identical representation, which
implies that a single aggregate can be used in distinct products over different
components of $g$. In order to avoid keep track of which aggregates correspond
to which entries in $\Sigma$, we construct for each aggregate a list of index
pairs $(i,j)$ for each $\vec\sigma_{ij} \in \vec\Sigma$ that require this
aggregate. AC/DC then uses the index list and the aggregate computed at the root
of the variable order to compute the queries for point evaluation and gradient
computation that were presented in
Section~\ref{sec:gradientcomputation}. Consider the matrix vector product
$\mv p = \vec\Sigma g(\vec\theta)$, which is needed for gradient
computation. Let $A$ be the root of the variable order $\Delta$. We compute
$\mv p$ by iterating over all aggregate maps $\alpha \in \aggs_{A}$, and for
each index pair $(i,j)$ that is assigned to $\alpha$, we add to the $i$'th
component of $\mv p$ the product of $\alpha$ and $g_j(\vec\theta)$. If
$i \neq j$, we also add to $j$'s component of $\mv p$ with the product of
$\alpha$ and $g_i(\vec\theta)$.

For the FD optimization, it is required to construct the $\mv R_c$ matrices that
were introduced in Section~\ref{sec:fd-def}. In AC/DC, we represent these
matrices as maps that group the values for functionally determining variables by
the values that they determine. The maps are sparse representations of $\mv R_c$
matrices, and they are populated during the computation of the factorized
aggregates over the variable order. We choose this representation because it
allows for the efficient computation of the matrix $I + \mv R_c^\top \mv R_c$,
which is the basic building block the matrix $\mv B_{T,q}$
from~\eqref{eqn:B:T:h}. The reparameterization of the regularizer requires the
computation of the inverse of $\mv B_{T,q}$. Therefore, we store the matrix
$\mv B_{T,q}$ as a sparse matrix in the format used by the Eigen linear algebra
library~\cite{eigen}, and then use Eigen's Sparse Cholesky Decomposition to compute the
inverse of $\mv B_{T,q}$.

\section{Experiments}
\label{sec:experiments}

We report on the performance of learning regression and factorization machine
models over three real datasets used in retail and advertisement
applications. We benchmark AC/DC against state-of-the-art competitors. AC/DC can
compute the models up to $1,031\times$ faster, while the accuracy of AC/DC's
models is always at least as good as the competitor's models. For all
experiments, we assume that the model specification is given as input, and all
systems compute the same model. We do not consider the orthogonal problem of
finding the best model for a given analytics task.

\subsection{Experimental setup}

All experiments were performed on an Intel(R) Core(TM) i7-4770
3.40GHz/64bit/32GB with Ubuntu 18.04, g++7.4, and eight cores. We report
wall-clock times by running each system once and then reporting the average of
four subsequent runs with warm cache. We do not report the times to load the
database into memory for the join. All relations are sorted by their join
variables.

\subsubsection*{Competitors} We benchmark AC/DC against six competitors:
MADlib~\cite{MADlib:2012} 1.16, libFM~\cite{libfm} 1.4.2,
TensorFlow~\cite{tensorflow} 1.13.1, R~\cite{R-project} 3.4.4,
scikit-learn~\cite{scikit2011} 0.20, and Python
Statsmodels~\cite{P-StatsModels}. We evaluate the performance of learning the
models over a training dataset, and then compute the root-mean-squared-error
(RMSE) over a separate test dataset to compare the accuracy.

MADlib, R, scikit-learn, and Python StatsModels use {\em ols} (ordinary least
squares) to compute the closed-form solution of regression models, and
TensorFlow uses the LinearRegressor estimator with {\em ftrl}
optimization~\cite{ftlr}, which is based on the conventional SGD optimization
algorithm.  TensorFlow was compiled from source to enable specialized
optimizations that are native to our machine, including AVX optimizations.  We
use PostgreSQL 10.9 to compute the feature extraction query for libFM and
TensorFlow. LibFM supports factorization machines.

AC/DC uses the gradient descent algorithm with the adaptive learning rate from
Algorithm~\ref{algo:bgd}. For $\lr$ and $\pr$ models, we run the optimization
algorithm until the RMSE over the training dataset changes by less than
$10^{-15}$ in three consecutive iterations. For $\fama$ models, we use 300
iterations.

The aggregate computation in AC/DC is parallelized on eight cores. Tensorflow
also uses multiple threads by default. MADlib and libFM only use a single
thread.

The competitors come with various limitations that affect their scalability.
MADlib requires the explicit one-hot encoding of the input relations, for which
we use its predefined functions.

\nop{
}

LibFM requires as input a zero-suppressed encoding of the one-hot encoded join
result. Computing this representation is an expensive intermediate step between
exporting the query result from the database system and importing the data into
libFM. We learn the $\fama$ models using the MCMC optimization algorithm with a
fixed number of runs (300); its SGD implementation requires a fixed learning
rate and does not converge.

TensorFlow uses a user-defined iterator interface to load a batch of tuples from
the training dataset at a time.  This iterator defines a mapping from input
tuples to features and is called directly by the learning algorithm. To avoid
the explicit one-hot encoding of the features, TensorFlow encodes categorical
features using a hash function. Learning over batches requires a random
shuffling of the input data, which in TensorFlow amounts to loading the entire
dataset into memory. This failed for our experiments due to the large sizes of
the datasets.  We therefore shuffle the data in PostgreSQL instead and provide
the shuffled input to TensorFlow. We benchmark TensorFlow for $\lr$ only as it
does not provide functionality to create all pairwise interaction terms for
$\pr$ and $\fama$. The optimal batch size for our experiments is 100,000
tuples. Smaller batch sizes require loading too many batches, very large batches
cannot fit into memory. Since TensorFlow uses a fixed number of iterations, we
report the times to optimize with one epoch over the training dataset.  This
means that the algorithm learns over each data point in the training dataset
once. Our experiments show that it is often necessary to optimize with several
epochs to learn a good model.

R, scikit-learn, and Python Statsmodels fail to compute our models due to design
limitations. R limits the number of values in their data frames to $2^{31}-1$,
which is insufficient to represent out datasets. Scikit-learn and Python
Statsmodels both run out of memory for all considered models.


\begin{table}[t]
\centering
\begin{tabular}{|l|r|r|r|r|}\hline
  & \multicolumn{1}{|c|}{ Retailer }
  & \multicolumn{1}{|c|}{ Favorita }
  & \multicolumn{1}{|c|}{ Yelp } \\\hline
  Relations                              & 4             & 5             & 5                                           \\
  Variables                             & 21            & 14            & 26                                          \\
  Categorical Variables                 & 4             & 10            & 6                                           \\\hline
  Tuples in Database                     & 87M           & 125M          & 8.7M                                        \\
  Size of  Database                      & 1.5GB         & 2.5GB         & 0.2GB                                       \\\hline
  Tuples in Join Result                  & 86M           & 125M          & 360M                                        \\
  Size of  Join Result                   & 18GB          & 7GB           & 40GB                                        \\\hline
  \#values in Listing Representation     & 2.302G        & 1.735G        & 2.835G                                      \\
  \#values in  Factorized  Representation & 166M          & 372M          & 71.9M                                       \\
  Compression ( Factorized/Listing  )              & 13.91$\times$ & 4.66$\times$ & 39.43$\times$ \\\hline
\end{tabular}

\caption{Key Characteristics of the three used datasets.}
\label{table:datasetstats}\vspace*{-1em}
\end{table}

\subsubsection*{Datasets} We experimented with three real-world datasets: (1)
Retailer~\cite{SOC:SIGMOD:16} is used by a large retailer for forecasting user
demand and sales; (2) Favorita~\cite{favorita} is a public dataset used for
retail forecasting; and (3) Yelp is based on the Yelp Dataset
Challenge~\cite{yelpdataset} and used to predict ratings by a user for a
business. The structure and size of these datasets is common in retail and
advertising, where data is easily generated by sales transactions or click
streams. The feature extraction query for each dataset is the natural join of
the input relations.  For each dataset, we consider a subset of the
variables. Table~\ref{table:datasetstats} presents key characteristics for each
dataset. It shows that the join result can be orders of magnitude larger than
the input database.

Retailer has a star schema with one fact table {\sf Inventory}, which keeps
track of the number of inventory units for products (sku) in a store (locn) at a
given date, and three dimension tables: (1) \textsf{Location} keeps additional
information for each store, including its size in sqft and distances to three
competitors; (2) \textsf{Items} provides identifiers for the item category,
subcategory, category cluster, as well as the price for each {\sf sku}; and (3)
\textsf{Weather} keeps information about the weather conditions for each store
at a given date (including maximum temperature, and whether it rained).  We
design two versions of our dataset. The version \textsf{v$_1$} includes all
variables but {\sf sku}, {\sf date}, and {\sf locn} as features, and has no
functional dependencies. Version \textsf{v$_2$} extends \textsf{v$_1$} with the
categorical variable \textsf{sku}, and exploits the functional dependency {\sf
sku}$\to$\{{\sf category, subcategory, categoryCluster}\}. We learned $\lr$,
$\pr_2$, and $\fama^8_2$ models that predict the amount of inventory units based
on all other features. The test data constitutes the inventory in the last month
in the dataset (approx. 2.2\% of the data). This simulates the realistic usecase
where the ML model predicts future inventory.

  Favorita has a star schema with one fact table and 4 dimension tables. The
fact table {\sf Sales} stores the number of units sold for items for a given
date and store, and an indicator whether or not the unit was on promotion at
this time. \textsf{Items} keeps additional information about the skus, including
the item class and whether it is perishable.  \textsf{Stores} provides
additional information about the stores, such as the city and state they are
located in, and the type of store. \textsf{Transactions} gives the number of
transactions at each store on a given date. \textsf{Oil} keeps the oil price for
each date.  Our models predict the number of units sold. We designed two
variants for this dataset.  Version \textsf{v$_1$} learns the model over all
variables except \textsf{sku} and \textsf{date}, and version \textsf{v$_2$}
extends \textsf{v$_1$} with \textsf{sku}. We exploit the functional dependency
{\sf store}$\to$\{{\sf city, state, storetype}\} in both variants. The test data
constitutes the sales for the last month in the dataset (approx. 1.2\% of the
data).

Yelp has a star schema with four relations: \textsf{Review} stores, for each
review, the rating given by user, the review date, and the number of compliments
it received (e.g, useful, funny, cool); \textsf{Business} keeps the location
(city, state, coordinates), the average rating, and the total count of reviews
for each business; \textsf{User} keeps aggregated statistics for each user,
including the number of reviews they wrote, the number of compliments they
received, and the average rating they gave to businesses; \textsf{Attribute}
keeps attributes (e.g. as ``open late'') for each business. One user can review
many businesses and a business can have many attributes. The result of the
feature extraction query is much larger than the input relations.  Our models
predict the rating that users give to businesses. The models are learned over
all variables except the join keys, and exploit the functional dependency {\sf
  city}$\to${\sf state}. The test dataset is a random selection of approximately
2\% of the reviews.

\nop{
}

\subsection{Summary of findings} 

Our findings on the performance comparison between AC/DC and the three
competitors are given in Table~\ref{table:retailer}. AC/DC is the fastest system
in our experiments. It can compute the models over the input database orders of
magnitude faster than its competitors whenever they do not exceed memory
limitation, 24-hour timeout, or internal design limitations.

AC/DC learns models with at least as good accuracy as the competitors. In
particular, AC/DC's $\lr$ models have comparable accuracy to MADlib's closed
form solution (whenever MADlib does not time out), and consistently better
accuracy than the models learned by TensorFlow (trained for one epoch). With
additional features, AC/DC can learn more accurate models while the competitors
either timeout or fail due to internal design limitations.

\begin{table*}[t]
\centering
\begin{scriptsize}
\begin{tabular}{|ll||r|r|r|r|r|}\hline
  \multicolumn{2}{|l||}{}       & Retailer \texttt{v$_1$} & Retailer  \texttt{v$_2$} & Favorita  \texttt{v$_1$} & Favorita \texttt{v$_2$} & Yelp                              \\\hline\hline
  \multicolumn{2}{|l||}{Join Computation (PSQL)}
                                & 447.76        & 447.76        & 255.16       & 255.16       & 195.26        \\
  \multicolumn{2}{|l||}{Factorized Computation of Count over Join}
                                & 19.90         & 19.90         & 36.45        & 36.45        & 21.07         \\\hline\hline
  \multicolumn{7}{|c|}{\bf Linear Regression}                                                                                                                               \\\hline
   \multicolumn{2}{|l||}{Features (continuous+categorical)}
                              & 16 + 49                 & 16+3,661                 & 4 + 482                  & 4 + 4,482               & 21 + 1,068                        \\
  \multicolumn{2}{|l||}{Number of Entries in Sparse Tensor}
                              & 1,149                   & 69,777                   & 42,504                   & 455,889                 & 46,401                            \\\hline
  MADLib (ols)                & Encode                  & 0.19                     & 8.46                     & 532.52                  & 545.42         & 61.79            \\
                              & Learn                   & 1,124.84                 & $>$86,400.00             & 13,951.44               & $>$86,400.00   & 44,307.88        \\\hline
 TensorFlow (ftrl)            & Join+Shuffle+Export     & 2,266.76                 & 2,266.76                 & 1,417.87                & 1,417.87       & 1,110.41         \\
(1 epoch, batch size 100K)    & Learn                   & 3,420.91                 & 3,408.34                 & 3,649.73                & 3,648.52       & 5,763.88         \\\hline
{\bf AC/DC}                   & Aggregate               & 40.09                    & 118.47                   & 115.02                  & 1,022.44       & 42.89            \\
                              & Converge                & 0.11                     & 285.92                   & 0.94                    & 22.20          & 0.14             \\\hline
Speedup of {\bf AC/DC} over   & MADlib                  & 27.99   $\times$         & $>213.67\times$          & $124.90\times$          & $>83.23\times$ & 1,031.13$\times$ \\
                              & TensorFlow              & $141.48\times$           & $14.03\times$            & $43.70\times$           & $4.85\times$   & 159.76$\times$   \\ \hline\hline  
  \multicolumn{7}{|c|}{\bf Polynomial Regression degree $2$}                                                                                                                \\\hline
  \multicolumn{2}{|l||}{Features (continuous+categorical)}
                              & 121+980                 & 121+65,996               & 7+42,016                 & 7+451,403               & 211+41,559                        \\
  \multicolumn{2}{|l||}{Number of Entries in Sparse Tensor}
                              & 72,790                  & 2,517,600                & 498,641                  & 6,616,551               & 6,478,164                         \\\hline
  MADlib (ols)                & Encode                  & 0.19                     & 8.46                     & 532.52                  & 545.42         & 61.79            \\
                              & Learn                   & $>$86,400.00             & $>$86,400.00             & $>$86,400.00            & $>$86,400.00   & $>$86,400.00     \\\hline 
  {\bf AC/DC}                 & Aggregate               & 122.88                   & 334.35                   & 324.99                  & 7,549.99       & 3,650.47         \\
                              & Converge                & 21.92                    & 621.69                   & 45.34                   & 1,063.88       & 203.64           \\\hline
  Speedup of {\bf AC/DC} over & MADlib                  & $>596.69\times$          & $>90.38\times$           & $>234.74\times$         & $>10.09\times$ & $>22.43\times$   \\ \hline\hline       
  \multicolumn{7}{|c|}{\bf Factorization Machine degree $2$ rank $8$}                                                                                                       \\\hline
  \multicolumn{2}{|l||}{Features (continuous+categorical)}
                              & 107+980                 & 107+65,996               & 5+42,016                 & 5+451,403               & 192+41,559                        \\
  \multicolumn{2}{|l||}{Number of Entries in Sparse Tensor}
                              & 70,515                  & 2,465,443                & 497,786                  & 6,607,696               & 6,454,053                         \\\hline
  libFM (MCMC)                & Join+Ex/Import+Encode   & 3,368.06                 & 3,368.06                 & 3,214.79                & 3,214.79       & 2,719.48         \\
   (300 iterations)           & Learn                   & $>$86,400.00             & $>$86,400.00             & $>$86,400.00            & $>$86,400.00   & 67,829.59        \\\hline 
  {\bf AC/DC}                 & Aggregate               & 124.17                   &     324.492                     & 351.68                  & 7,856.88        & 3,633.93         \\ 
   (300 iterations)           & Converge                & 0.67                     &     42.74                     & 2.60                    & 25.38          & 265.94           \\\hline
  Speedup of {\bf AC/DC} over & libFM                   & $>719.06\times$          & $>244.45\times$                & $>252.95\times$               & $11.37\times$      & $18.09 \times$   \\\hline
\end{tabular}

\end{scriptsize}
\caption{Time performance (seconds) for learning $\lr$, $\pr$, and $\fama$
  models over Retailer, Favorita, and Yelp. The timeout is set to 24 hours (86,400
  seconds). R and MADlib do not support $\fama$ models. TensorFlow does not
  support $\pr$ and $\fama$ models. }
\label{table:retailer}
\end{table*}

The performance gap between competitors and AC/DC is primarily due to the following optimizations supported by AC/DC: 
\begin{enumerate}
\item It avoids the materialization of the join and the export-import step
  between database systems and statistical packages, which may take longer than
  the end-to-end learning of $\lr$ models in AC/DC. AC/DC performs the join
  together with the aggregates using one execution plan;
\item It factorizes the computation of the sparse tensor and the underlying
  join. The compression factor brought by join factorization is 13.9$\times$ for
  Retailer, $4.7\times$ for Favorita, and 21$\times$ for Yelp;
\item It massively shares the computation of many aggregates representing entries in the sparse tensor. For instance, there are up to 2.5M such sum aggregates for $\pr_2$ on Retailer \textsf{v}$_2$ and they take 150K$\times$ less time than computing the count aggregate 2.5M times, where the count takes 19.9 seconds as reported in Table~\ref{table:retailer};
\item It decouples the computation of the aggregates on the input data from the parameter
convergence step and thus avoids scanning the join result for each iteration; 
\item It avoids the upfront one-hot encoding that comes with higher asymptotic
  complexity and prohibitively large covariance matrices by only computing
  distinct, non-zero entries in the sparse tensor. For $\pr_2$ on Retailer
  \textsf{v}$_2$, this leads to $868\times$ less aggregates to compute;
\item It exploits the functional dependencies in the input data to reduce the
number of features of the model, which leads to an improvement
factor of up to 2.3$\times$.
\end{enumerate}

\subsection{Further details}

\subsubsection*{Categorical features} As we move from Retailer \textsf{v$_1$} to
\textsf{v$_2$}, we increase the number of categorical features by
approx. $75\times$ for $\lr$ (from 49 to 3.7K) and $67\times$ for $\pr_2$ and
$\fama^8_2$ (from 980 to 66K). This translates to a same-order increase in the
number of aggregates: $62\times$ ($56\times$) more distinct non-zero aggregates
in \textsf{v$_2$} vs \textsf{v$_1$} for $\lr$ (resp.\@ $\pr_2$ and $\fama^8_2$).
This increase only led to a decrease in performance of AC/DC of $10\times$ for
$\lr$ and $6.6\times$ for $\pr_2$. This sub-linear behavior is partly explained
by the ability of AC/DC to process many aggregates much faster in bulk than
individually: it takes 19.9 seconds for one count aggregate, but only 334
seconds to compute all 2.5M entries in the sparse tensors for $\pr_2$ on
\textsf{v$_2$}!

For MADlib, the performance decrease is at least $78\times$ for $\lr$ when
moving from \textsf{v$_1$} to \textsf{v$_2$} and it times out after 24 hours for
\textsf{v$_2$}. MADlib also times out for all $\pr_2$ experiments. The
performance of TensorFlow is largely invariant to the increase in the number of
categorical features, since its internal mapping from tuples in the training
dataset to the sparse representation of the features vector remains of similar
size. Nevertheless, our system is consistently faster (often by orders of
magnitude) than computing only {\em a single epoch} in TensorFlow. In addition,
AC/DC computes the end-to-end models faster than PSQL takes to materialize and
shuffle the design matrix that is the input to TensorFlow.

\nop{

}

\subsubsection*{One-hot encoding vs.\@ sparse encoding with group-by aggregates.}
One-hot encoding categorical features leads to a large number of zero and/or
redundant entries in the $\vec\Sigma$ matrix. For instance, for $\pr_2$ on
Retailer \textsf{v$_2$}, the number of features is $m$=66,117, and then the
upper half of $\vec\Sigma$ would have $m(m+1)/2 \approx 2.1\times 10^9$ entries.
Most of these are either zero or repeating, as exemplified in Section~\ref{sec:evaluation}. In contrast, AC/DC's sparse
representation only considers 2.5M non-zero and distinct aggregates.  The
number of aggregates is thus reduced by a factor of $868$!

MADlib and libFM require the data be one-hot encoded {\em before}
learning. For MADlib, the static one-hot encoding took up to 545 seconds on
Favorita. In addition to the one-hot encoding, libFM requires the input data
to be represented in a zero-supressed sparse format. This data transformation
took up to one hour in our experiments. TensorFlow one-hot encodes on the fly
using hash functions during the learning phase, which cannot be reported
separately.

When running over one-hot encoded input data, AC/DC exceeds the available
memory for all models but the linear regression model for Retailer
\textsf{v$_1$}.  

\begin{figure}
  \includegraphics[width=.49\textwidth]{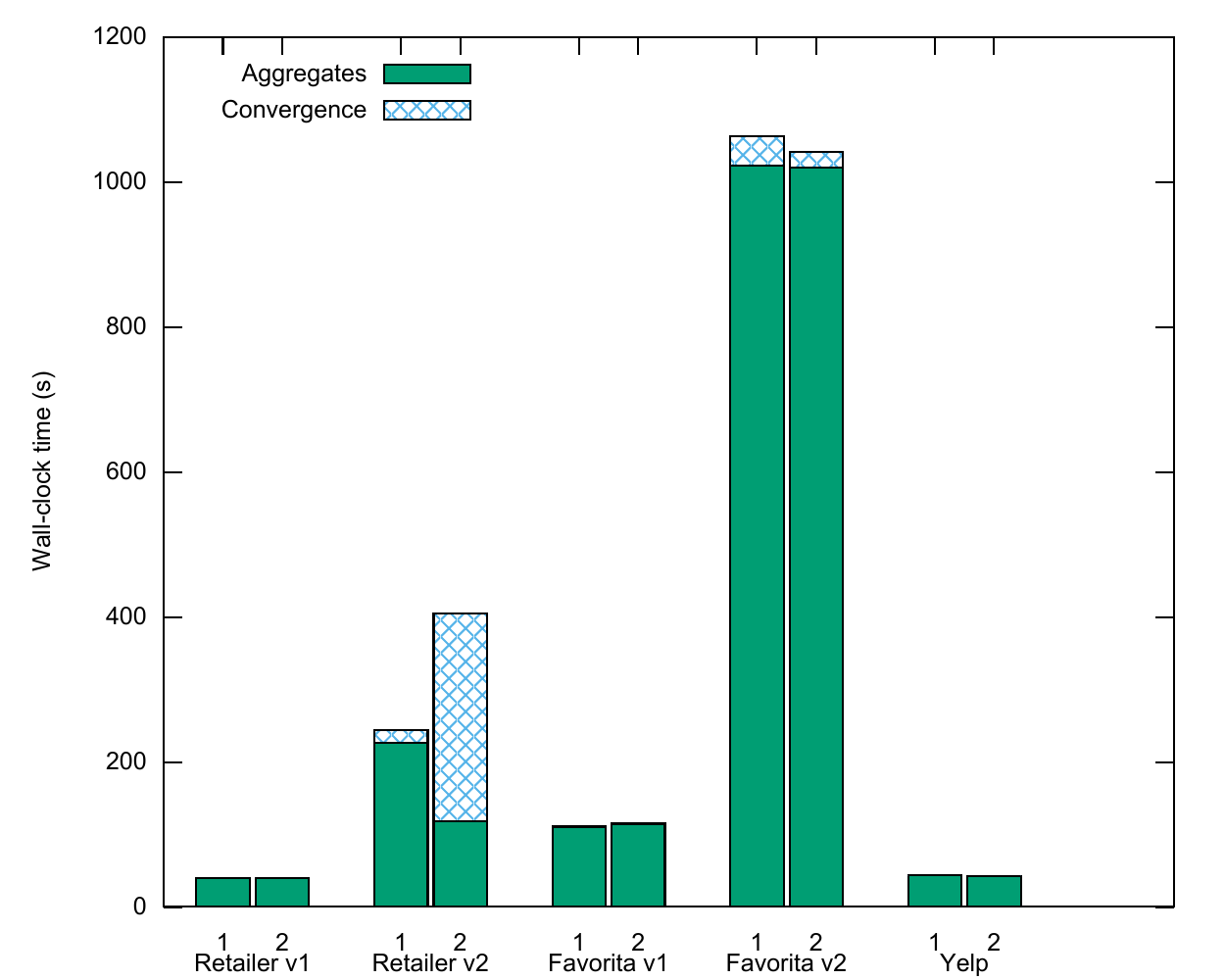}
  \includegraphics[width=.49\textwidth]{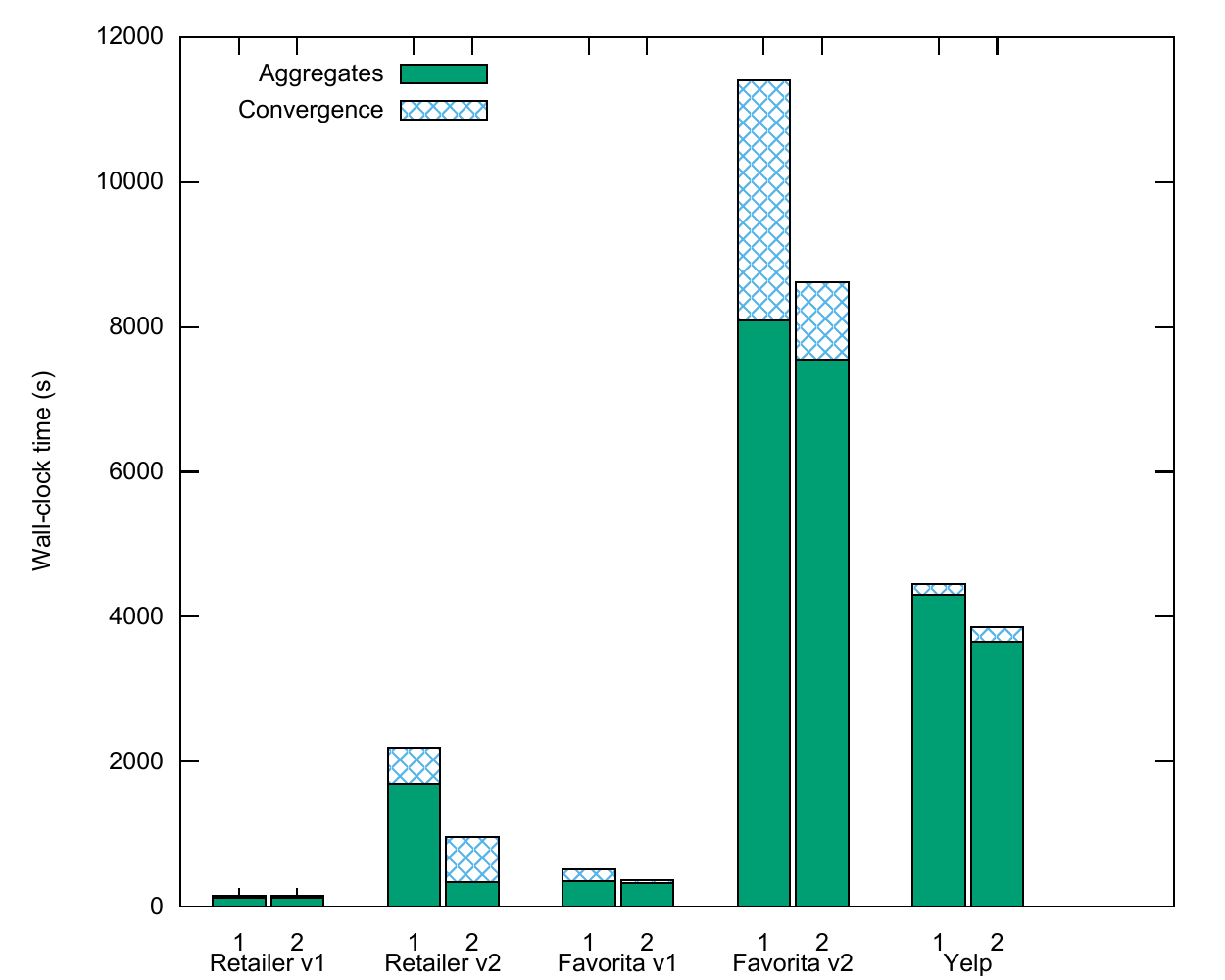}
  \caption{Breakdown of AC/DC performance for
    $\lr$ (left) and $\pr_2$ (right): (1) not using FDs; (2) using FDs.}
  \label{fig:lmfaobreakdown}
  \vspace*{-1em}
\end{figure}

\subsubsection*{Effect of functional dependencies.}
Figure~\ref{fig:lmfaobreakdown} shows the performance breakdown for AC/DC with
and without exploiting FDs. All other systems do not exploit FDs.

The FDs have a twofold effect on AC/DC: they can reduce the number of features
and aggregates, which leads to better performance of the aggregation step; yet
it requires a more expensive convergence step due to the more complex
regularizer. For $\lr$ over Retailer v2, the aggregate step becomes $2\times$
faster, while the convergence step increases $15\times$, offsetting the effect
of the faster aggregate computation. The FDs for Favorita and Yelp have a
relatively smaller effect on performance for $\lr$ (there are only 54 stores in
Favorita, so the reduction in the number of aggregates is small).

For $\pr_2$ models, the FD brings an improvement by a factor of $2.3\times$ for
Retailer.  This is due to a 18\% decrease in the number of categorical features,
which leads to a 38\% decrease in the number of group-by aggregates.  For
Favorita and Yelp, the performance improvement is 1.3$\times$ and respectively
1.2$\times$. For $\fama^8_2$ models, the effect of FDs is comparable to that of
$\pr_2$ models.

\subsubsection*{Accuracy.} The RMSE of the $\lr$ models for Retailer
$\mathsf{v}_1$ and Favorita $\mathsf{v}_1$ in AC/DC is within 1\% of that for the
closed form solutions computed in MADlib. By extending the models with the
categorical variable \textsf{sku} (version $\mathsf{v}_2$ for both datasets),
the RMSE decreases by 21\% for Retailer and by 6\% for Favorita. MADlib fails
to learn these more accurate models, because it times out after 24 hours.

For TensorFlow, the models are trained with a single epoch. The resulting models
have a consistently higher RMSE than the corresponding model computed in
AC/DC. In particular, for Retailer $\mathsf{v}_2$, the RMSE of the TensorFlow
model is 31\% higher. TensorFlow requires more epochs to achieve the
same accuracy as AC/DC.

Extending the model with pairwise interactions can also improve the model
accuracy. For Favorita $\mathsf{v}_2$, for instance, the RMSE of the $\pr_2$
model is $3\%$ lower than the RMSE of the $\lr$ model.

The $\fama$ models learned by AC/DC do not reach the same accuracy as the
$\pr_2$ models, and would therefore require more than 300 iterations to converge to an
accurate model. This is due to the non-linear structure of factorization
machines.  


\section{Related work} 
\label{sec:relatedwork}

\nop{
It has been recently acknowledged that database theory can effectively contribute to the arms race for in-database analytics~\cite{DBTHEORY:DAGSTUHL:17}. Recent works highlight the potential of applying key database theory tools to this growing research of practical interest, e.g., the relational framework for classifier engineering~\cite{Benny:PODS:2017} and in-database factorized learning of regression models  with low data complexity~\cite{SOC:SIGMOD:16}.}

Our work follows closely Chaudhuri's manifesto on SQL-aware data mining systems from two decades ago~\cite{Chaudhuri:DMDB:1998} in two key aspects. First, the goal of our work is not to invent new machine learning models or data analysis techniques, but identify common data-centric steps across a broad class of learning algorithms and investigate their theoretical and systems challenges. We show that such steps can be encoded as SQL group-by aggregate queries, which are amenable to shared batch computation. Second, our approach performs data analysis not only over materialized relations but more importantly over feature extraction queries, whose results need not be materialized. This enables the interaction between the aggregates encoding the data-centric steps and the underlying queries (this is called ad-hoc mining in Chaudhuri's terminology).

A reevaluation of Chaudhuri's manifesto in today's context brings forth two important technical changes. 
The first game-changer is represented by the recent development on query processing. This includes a new breed of worst-case optimal join algorithms, which support listing representation~\cite{NPRR12,LFTJ} and factorized representation~\cite{OlZa15} of query results, and extensions to aggregate computation~\cite{BKOZ13,faq,OS:SIGREC:2016}. These algorithms exploit developments on (fractional) hypertree decompositions of relational queries~\cite{Gottlob99,GM06,Marx:2010}. These algorithms overshadow the traditional query plans in both asymptotic complexity~\cite{skew} and practical performance~\cite{BKOZ13,LB:Experiments:2015}. 
The second change is in the workload. Whereas SQL-aware data mining systems were mostly concerned with association rules, decision trees, and clustering, current workloads feature a broader spectrum of increasingly more sophisticated machine learning (ML) models, including polynomial regression models, factorization machines, principal component analysis, generalized linear models, generalized low-rank models, sum-product networks, and convolutional networks. In this article, we introduce a unified approach to learning polynomial regression models, factorization machines, and principal component analysis over non-materialized feature extraction queries with the lowest known complexity and best performance to date. 
There is also a more profound orthogonal change: There is more data readily available in all aspects of our society and there is more appetite in industry to monetize it by turning it into knowledge.

The current landscape for analytics solutions over multi-relational data can be categorized depending on the degree of integration of the data system, which hosts the data and supports data access via query primitives, with the ML library of models and learning algorithms. 

By far the most common solutions provide no integration of the two systems,
which are distinct tools on the technology stack: The data system exports the
training dataset as one relation, commonly presented as a CSV file, and then the
ML system imports it into its own format and learns the desired model.Such
solutions are structure-agnostic as they disregard the rich structural
information of the materialized training dataset, including the input database
schema, database dependencies, and the structure of the feature extraction
queries.  Prime examples are the pairing of open-source data systems such as
Spark~\cite{Spark:NSDI:2012} or MySQL/PostgreSQL with ML systems such as
R~\cite{R-project}, Python StatsModels~\cite{P-StatsModels}, Python
Scikit~\cite{scikit2011}, MLpack~\cite{mlpack2018},
TensorFlow~\cite{tensorflow}, SystemML \cite{HBTRTR15,BTRSTBV14},
MLLib~\cite{MLlib:JMLR:2016}, and DeepDist~\cite{Neumann15}. The advantage of
this approach is that the two systems can be developed independently, with
virtually any ML model readily available for use.

Two disadvantages of such common solutions are the expensive data export/import
at the interface between the two systems and the materialization of the training
dataset as a result of a feature extraction query over multi-relational data.
The feature extraction query is computed inside the data system, its result
exported and imported into the data format of the ML system, where the model is
learned.  Furthermore, the materialized training dataset may be much larger than
the input data (cf.\@ Table~\ref{table:datasetstats}). This is exacerbated by
the stark asymmetry between the two systems: Whereas data systems tend to scale
to large datasets, this is not the case for ML libraries. Yet, such solutions
expect by design that the ML libraries work on even larger inputs than the data
systems!  A further disadvantage is that these solutions inherit the limitations
of both underlying systems.  For instance, the R data frame can host at most
$2^{31}$ values, which makes it impossible to learn models over large datasets,
even if data systems can process them. Database systems can only handle up to a
few thousand columns per relation, which is usually smaller than the number of
features of the model.

The second class of systems features a loose integration, even though they remain structure-agnostic: The ML code migrates inside the space of the data system process, with each ML task being implemented by a distinct user-defined aggregate function (UDAF). Prime examples of this class are MADlib~\cite{MADlib:2012} and GLADE PF-OLA~\cite{Rusu:2015}. MADlib casts analytics as UDAFs that can be used in SQL queries and executed inside PostgreSQL. GLADE PF-OLA casts analytics as a special form of UDAFs called Generalized Linear Aggregates that can be executed using 
the GLADE distributed engine~\cite{Glade:SIGMOD:2012}. These UDAFs remain black boxes for the underlying query engine, which has to compute the feature extraction query and delegate the UDAF computation on top of the query result to the MADLib's and GLADE PF-OLA's specialized code. The advantage of this approach is that the expensive export/import step is avoided. The disadvantage is that each ML task has to be migrated inside the data system space, which comes with design and implementation overhead. A further step towards integration is exemplified by Bismarck~\cite{Bismarck:SIGMOD:2012}, which provides a unified programming architecture for many ML tasks instead of one UDAF per task, with possible code reuse across UDAFs.

The third class of systems features a tight integration and are structure-aware: There is one execution strategy for both the feature extraction query and the subsequent learning task, with components of the latter possibly pushed past the joins in the former. Prime examples are Morpheus~\cite{KuNaPa15}, Hamlet~\cite{Kumar:SIGMOD:16}, and our prior system F~\cite{SOC:SIGMOD:16} that support generalized linear models, Na\"ive Bayes classification, and respectively linear regression models with continuous features. This class also contains the recent efforts on in-database linear algebra~\cite{Kumar:PVLDB:2017} and on scaling linear algebra using existing distributed database systems~\cite{LAoverDBMS:SIGREC:2018} and the declarative language BUDS~\cite{BUDS:SIGMOD:2017}, whose compiler can perform deep optimizations of the user's program. Our approach AC/DC generalizes F to non-linear models, categorical features, and model reparameterization under functional dependencies. A key aspect that sets apart AC/DC and its predecessor F from prior work is that they employ execution strategies for the mixed workload of queries and learning with complexity that may be asymptotically lower than that of query materialization alone. In particular, all machine learning approaches that require as input the materialization of the result of the feature extraction query  are asymptotically suboptimal. This complexity gap translates into a performance gap, cf.\@ Section~\ref{sec:experiments}.

Figure~\ref{fig:high-level-diagram-intro} sums up the difference between the first two classes that fall under structure-agnostic learning and the third class that broadly represents structure-aware learning. The inspiration for our work lies with factorized computation of aggregates over joins~\cite{BKOZ13,faq}, which avoids the materialization of joins, and with the LogicBlox system~\cite{Ron2013,LB15}, which has a unified system architecture and declarative programming language for hybrid database and optimization workloads.

Beyond the above classification, there are further directions of research looking at ML through database glasses: ML-aware query languages, the effect of dependencies on model training, sparse data representations, and implementations of gradient descent solvers.

Analytical tasks can be expressed to a varying degree within query languages possibly extended with new constructs.
Very recent works investigate query languages for matrices~\cite{MatrixQL:ICDT:2018} and a relational framework for classifier engineering~\cite{Benny:PODS:2017}. They follow works on query languages with data mining capabilities~\cite{Boulicaut:2005,Ordonez:TKDE:2016}, also called descriptive or backward-looking analytics, and on in-database data mining solutions, such as frequent itemsets~\cite{Pei:2001} and association rule mining~\cite{Agrawal:1996}. Our rewriting of ML code into aggregates falls into this line of work as well. The additional fixpoint computation needed on top of the aggregate computation for convergence of the model parameters, which is intrinsic to gradient descent approaches, can be expressed as recursive queries~\cite{abiteboul1995foundations}.

Functional dependencies (FDs) can be used to avoid key-foreign key joins and reduce the number of features in Na\"ive Bayes classification and feature selection~\cite{Kumar:SIGMOD:16}. In this article we consider the effect of FDs on the reparameterization of regression models, where a non-trivial development is on the effect of FDs on the model's non-linear regularization function, cf.\@ Section~\ref{SEC:FDS}. Our factorized learning approach exploits the join dependencies present in the training dataset, as defined by the feature extraction query. This follows prior work on factorized databases~\cite{BKOZ13,OlZa15}.

State-of-the-art machine learning systems use a sparse representation of the input data to avoid redundancy introduced by one-hot encoding ~\cite{libfm,liblinear}. In our setting, however, such systems require an additional data transformation step after the result of the feature extraction query is exported. This additional step is time consuming and makes the use of such systems inefficient in many practical applications. In statistics and machine learning, there is a rich literature on learning with sparse and/or multilinear
structures~\cite{Hastie-Tibshirani-Wainwright}. Such methods complement our framework and it would be of interest to leverage and adapt them to our setting.

Finally, there is a large collection of gradient-based methods proposed in the optimization literature. The description of our approach assumes batch gradient descent (BGD), though our insights are applicable to other methods including Quasi-Newton algorithms. The main rationale for our choice is simplicity and good statistical properties. When combined with backtracking line search (as we do in this article) or second-order gradient estimation (as in Quasi-Newton methods), BGD is guaranteed to converge to a minimum with linear asymptotic
convergence rate. A na\"ive computation of the gradient requires a full pass over the data, which can be inefficient in large-scale analytics. A popular alternative is stochastic gradient descent (SGD), which estimates the gradient with a randomly selected mini-batch of training samples. The convergence of SGD, however, is noisy, requires careful setting of hyperparameters, and does not achieve the linear asymptotic convergence rate of BGD~\cite{Bottou12}. In our setting, the entire BGD execution can be arbitrarily faster than one SGD iteration over the result of the feature extraction query. The reason is orthogonal to properties of the two gradient descent methods: The complexity of computing the sufficient statistics needed for convergence of model parameters can be asymptotically lower than the complexity of computing the training dataset.

\section{Open Problems}
\label{sec:conclusion}

Our in-database learning framework raises open questions on statistics,
algorithm design, and optimization. We next sketch a few representative questions.

One research direction is to further extend the class of statistical models that can be 
trained efficiently by exploiting the structure of the underlying relational database. Our
formulation~\eqref{eqn:generic:J} captures a common class of regression
models (such as $\pr$ and $\fama$), classification models (such as logistic
and SVM), and unsupervised learning techniques (such as principal component analysis) which is done by changing the loss function $\calLL$. It remains open
how to extend our formulation to capture latent variable models.

The aggregates defining $\vec\Sigma$, $\mv c$, point evaluation, and
gradient computation are ``multi-output'' queries. They
deserve a systematic investigation, from formulation to evaluation and
complexity analysis. In practice, one often reserves a fragment of the training
data for model validation. It is an interesting question to
incorporate this data partitioning requirement into our framework.

Understanding how to adapt further optimization algorithms, such as coordinate descent or stochastic gradient, to our structure-aware framework is an important research direction. Furthermore, our FD-aware optimization is specific to the $\ell_2$-norm in the penalty term. We would also like to understand the effect of other norms, e.g., $\ell_1$, on model reparameterization under FDs.

Finally, we conjecture that the cost function may be easier to optimize with respect to the reduced set of parameters that are not functionally determined:
As redundant variables are eliminated or optimized out, the cost function's Hessian with respect to reduced parameters becomes less ill-conditioned, resulting in faster convergence behavior for gradient-based optimization techniques. The impact of FD-based dimensionality reduction, from both computational and statistical standpoints, have not been extensively studied for learning (nonlinear) models with categorical variables, which are precisely the kind discussed in our framework.

\section*{Acknowledgements}
This project has received funding from the European Union's Horizon 2020 research and innovation programme under grant agreement No 682588.
XN is supported in part by grants NSF CAREER DMS-1351362, NSF CNS-1409303 and the Margaret and Herman Sokol Faculty Award.

\newpage

\bibliographystyle{abbrv}
\bibliography{bibtex}

\newpage

\begin{appendix}

\section{Matrix Calculus}
\label{app:matrix-calculus}

We introduce matrix inversion formulas and identities regarding tensors and the various products introduced in Section~\ref{sec:preliminaries}.

We use the following matrix inversion formulas~\cite{MR997457}.

\bprop
\label{prop:Woodbury}
We have
\begin{equation}
   (\mv{B+UCV})^{-1}=\mv B^{-1} - \mv B^{-1}\mv U(\mv C^{-1}+\mv V\mv B^{-1}\mv U)^{-1}\mv V\mv B^{-1}. 
   \label{eqn:Woodbury:main}
\end{equation}
whenever all dimensions match up and inverses on the right hand side exist. In particular, the following holds
when $\mv C = (1)$, $\mv U = \mv 1$, $\mv V = \mv 1^\top$, and $\mv J$ is the all-$1$ matrix:
\begin{equation}
   (\mv{B+J})^{-1}=\mv B^{-1} - \mv B^{-1}\mv 1(1+\mv 1^\top \mv B^{-1}\mv 1)^{-1}\mv 1^\top \mv B^{-1}. 
   \label{eqn:Woodbury:1}
\end{equation}
Another special case is
\begin{equation} (\mv A + \mv U^\top\mv U)^{-1} = \mv A^{-1} - \mv A^{-1} \mv U^\top(\mv I +
   \mv U \mv A^{-1} \mv U^\top)^{-1}\mv U \mv A^{-1}.
   \label{eqn:Woodbury}
\end{equation}
An even more special case is the Sherman-Morrison formula, where $\mv U^\top$ is just
a vector $\mv u$. The matrix $\mv A + \mv u \mv u^\top$ is typically called a
rank-$1$ update of $\mv A$:
\begin{equation} (\mv A + \mv u \mv u^\top)^{-1} = \mv A^{-1} - \frac{\mv A^{-1} 
   \mv u 
   \mv u^\top \mv A^{-1}}{1 + \mv u^\top \mv A^{-1} \mv u}.
   \label{eqn:sherman:morrison}
\end{equation}
\eprop

Next, we discuss some identities involving tensor, Khatri-Rao, and Hadamard products.

\bprop\label{prop:basic:frobenius}
We have (if the dimensionalities match up correctly):
\begin{align}
   (\mv A \mv B \otimes \mv C\mv D) &= (\mv A \otimes \mv C) (\mv B \otimes \mv D) \label{eqn:tensor:factor}\\
   (\mv A \otimes \mv B)^\top &= (\mv A^\top \otimes \mv B^\top) \label{eqn:tensor:transpose}\\
   \inner{\mv x, \mv B\mv y} &= \inner{\mv B^\top \mv x, \mv y}\label{eqn:inner:transpose}\\
   (\mv A \otimes \mv B)^{-1} &= (\mv A^{-1} \otimes \mv B^{-1}) & \text{ if
   both are square matrices}\label{eqn:otimes:inverse}\\
   \inner{\mv A \otimes \mv B, \mv R\mv X \otimes \mv S\mv Y} &= \inner{\mv R^\top
   \mv A \otimes \mv S^\top \mv B, \mv X \otimes \mv
   Y}.
\label{eqn:otimes:double:factor}
\end{align}
If $\mv x$ is a standard $n$-dimensional unit vector, $\mv A$ and $\mv B$ are two matrices 
with $n$ columns each, and $\mv a$ and $\mv b$ are two $n$-dimensional vectors, then
\begin{eqnarray}
   (\mv A \otimes \mv B)(\mv x \otimes \mv x) &=& (\mv A \star \mv B) \mv x \label{eqn:tensor:unit}\\
   \inner{\mv a \otimes \mv b, \mv x \otimes \mv x} &=& \inner{\mv a \circ \mv b, \mv x}.
 \label{eqn:tensor:unit:unit}
\end{eqnarray}
Let $\mv x$ be a standard $n$-dimensional unit vector, $\mv A_1, \dots, \mv A_k$
be $k$ matrices with $n$ columns each. Then,
\begin{equation}
   (\bigotimes_{i=1}^k \mv A_i)(\mv x^{\otimes k})
   = (\Bigstar_{i=1}^k \mv A_i)\mv x. \label{eqn:tensor:k:unit}
\end{equation}
\eprop
We note in passing that the first five identities are very useful in our dimension reduction
techniques by exploiting functional dependencies, while 
\eqref{eqn:tensor:unit}, \eqref{eqn:tensor:unit:unit}, and \eqref{eqn:tensor:k:unit}
are instrumental in achieving computational reduction in our handling 
of categorical features.

\begin{proof} The
identities~\eqref{eqn:tensor:factor},~\eqref{eqn:tensor:transpose},~\eqref{eqn:inner:transpose},
and~\eqref{eqn:otimes:inverse} can be found in the Matrix
Cookbook~\cite{IMM2012-03274}.
Identity~\eqref{eqn:otimes:double:factor} follows from~\eqref{eqn:tensor:factor}
and~\eqref{eqn:tensor:transpose}. To see~\eqref{eqn:tensor:unit}, note that
\[ (\mv A \otimes \mv B)(\mv x \otimes \mv x) = \mv A \mv x \otimes \mv B \mv x
   = (\mv A \star \mv B) \mv x, 
\]
where the last equality follows due to the following reasoning. Suppose $x_j=1$
for some $j$, then $\mv A \mv x = \mv a_j$ and $\mv B\mv x = \mv b_j$, where
$\mv a_j$ and $\mv b_j$ are the $j$th columns of $\mv A$ and $\mv B$,
respectively. Thus, $$\mv{Ax} \otimes \mv {Bx} = \mv a_j \otimes \mv b_j
= (\mv A \star \mv B)_j = (\mv A\star \mv B)\mv x.$$
Identities~\eqref{eqn:tensor:unit:unit} and~\eqref{eqn:tensor:k:unit} are proved 
similarly, where~\eqref{eqn:tensor:k:unit} is a trivial generalization
of~\eqref{eqn:tensor:unit}.
\end{proof}


\section{Tensor computation and {\sf FAQ} queries}
\label{appendix:faq-tensor}

Quite often we need to compute a product of the form $(\mv A \otimes \mv B) \mv
C$, where $\mv A, \mv B$, and $\mv C$ are tensors, provided that their dimensionalities
match up. For example, suppose $\mv A$ is an $m\times n$ matrix, 
$\mv B$ a $p \times q$ matrix, and $\mv C$ a $nq \times 1$
matrix (i.e. a vector). The result is a $mp \times 1$ tensor. 
The brute-force
way of computing $(\mv A \otimes \mv B)\mv C$ is to compute $\mv A \otimes \mv
B$ first, taking $\Theta(mnpq)$-time, and then multiply the result with $\mv C$,
for an overall runtime of $\Theta (mnpq)$. The brute-force algorithm is a horribly
inefficient algorithm. 

The better way to compute $(\mv A \otimes \mv B) \mv
C$ is to view this as an $\faq$-expression~\cite{faq} (a sum-product form): we think of $\mv A$ as a
function $\psi_A(x,y)$, $\mv B$ as a function $\psi_B(z,t)$, and $\mv C$ as 
a function $\psi_C(y,t)$. What we want to compute is the function
\begin{equation}
\varphi(x,z) = \sum_{y} \sum_{t} \psi_A(x,y)\psi_B(z,t)\psi_C(y,t).
\label{eq:faq-example}
\end{equation}
This is a $4$-cycle $\faq$ query:  
\begin{figure}[h!]
\centering{
\begin{tikzpicture}
   \node at (0,0) (x) {$x (m)$};
   \node at (0,2) (y) {$y (n)$};
   \node at (2,2) (t) {$t (q)$};
   \node at (2,0) (z) {$z (p)$};
   \draw (x) -- (y) node[left, align=center, midway] {$\psi_A(x,y)$};
   \draw (y) -- (t) node[above, align=center, midway] {$\psi_C(y,t)$};
   \draw (z) -- (t) node[right, align=center, midway] {$\psi_B(z,t)$};
   \draw[dashed] (x) -- (z) node[below, align=center, midway] {$\varphi(x,z)$};
\end{tikzpicture}
}
\end{figure}

We can pick between the following two evaluation strategies:
\bi
 \item Eliminate $t$ first, i.e., compute $\varphi_1(y,z) := \sum_{t} \psi_B(z,t) \psi_C(y,t)$
in time $O(npq)$; then, eliminate $y$, i.e., compute $\varphi(x,y) = \sum_y \varphi_1(y,z) \psi_A(x,y)$ in time 
$O(mnp)$. The overall runtime is thus $O(np(m+q))$.
 \item Eliminate $y$ first and then $t$. The
overall runtime is $O(mq(n+p))$.
\ei
This is not surprising, since the problem is just matrix chain
multiplication. In the language  of $\faq$ evaluation, we want to pick the best
tree decomposition and then compute a variable elimination order out of
it~\cite{faq}.
We shall see later that
a special case of the above that occurs often is when $\mv B = \mv I$, the
identity matrix. In that case, $\psi_B(z,t)$ is the same as the atom $z=t$,
and thus it serves as a change of variables:
\[ \varphi(x,z) 
= \sum_{y} \sum_{t} \psi_A(x,y)\psi_B(z,t)\psi_C(y,t)
= \sum_{y} \psi_A(x,y) \psi_C(y,z). 
\]
In other words, we only have to marginalize out one variable instead of two.
This situation arises, for example, in \eqref{eqn:main:derivative}
and \eqref{eqn:main:gamma:to:theta}.

Appendix~\ref{app:subsec:faqw} overviews the $\InsideOut$ algorithm for $\faq$ queries and its complexity analysis.

\section{Widths for {\sf FAQ} Queries and the {\sf InsideOut} Algorithm}
\label{app:subsec:faqw}


\subsubsection*{Background: Fractional edge cover number and output size bounds}
In what follows, we consider a conjunctive query $Q$ over a relational database instance $I$.
We use $N$ to denote the size of the largest input relation in $Q$.
We also use $Q(I)$ to denote the output and $|Q(I)|$ to denote its size.
We use the query $Q$ and its hypergraph $\calH$ interchangeably.
\bdefn[Fractional edge cover number $\rho^*$]
Let $\calH=(\calV,\calE)$ be a hypergraph (of some query $Q$). Let $B\subseteq\calV$ be any subset
of vertices. 
A {\em fractional edge cover} of $B$ using edges in $\calH$ is a feasible
solution $\vec\lambda =(\lambda_S)_{S\in\calE}$ to the following linear
program:
\begin{eqnarray*}
   \min && \sum_{S\in\calE} \lambda_S\\
   \text{s.t.}&& \sum_{S : v \in S} \lambda_S \geq 1, \ \ \forall v \in B\\
   && \lambda_S \geq 0, \ \ \forall S\in \calE.
\end{eqnarray*}
The optimal objective value of the above linear program is called
the {\em fractional edge cover number} of $B$ in $\calH$ and is denoted by $\rho^*_\calH(B)$.
When $\calH$ is clear from the context, we drop the subscript $\calH$ and use $\rho^*(B)$.

Given a conjunctive query $Q$, the fractional edge cover number of $Q$ is $\rho^*_\calH(\calV)$
where $\calH=(\calV,\calE)$ is the hypergraph of $Q$.
\edefn

\bthm[AGM-bound~\cite{AGM08,GM06}]
Given a full conjunctive query $Q$ over a relational database instance $I$,
the output size is bounded by
\[|Q(I)| \leq N^{\rho^*},\]
where $\rho^*$ is the fractional edge cover number of $Q$.
\label{thm:agm-upperbound}
\ethm

\bthm[AGM-bound is tight~\cite{AGM08,GM06}]
Given a full conjunctive query $Q$ and a non-negative number $N$,
there exists a database instance $I$ whose relation sizes are upper-bounded by $N$ and satisfies
\[|Q(I)| =\Theta(N^{\rho^*}).\]
\label{thm:agm-lowerbound}
\ethm

\emph{Worst-case optimal join algorithms}~\cite{LFTJ,NPRR12,skew,anrr} can be used to answer any full conjunctive query $Q$
in time
\begin{equation}
O(|\calV|\cdot|\calE|\cdot N^{\rho^*}\cdot \log N).
\label{eqn:runtime:lftj}
\end{equation}

\subsubsection*{Background: Tree decompositions, acyclicity, and width parameters}
\bdefn[Tree decomposition]
\label{defn:TD}
Let $\calH = (\calV, \calE)$ be a hypergraph.
A {\em tree decomposition} of $\calH$ is a pair $(T, \chi)$
where $T = (V(T), E(T))$ is a tree and $\chi : V(T) \to 2^{\calV}$ assigns to
each node of the tree $T$ a subset of vertices of $\calH$.
The sets $\chi(t)$, $t\in V(T)$, are called the {\em bags} of the 
tree decomposition.  There are two properties the bags must satisfy
\bi
\item[(a)] For any hyperedge $F \in \calE$, there is a bag $\chi(t)$, $t\in
V(T)$, such that $F\subseteq \chi(t)$.
\item[(b)] For any vertex $v \in \calV$, the set 
$\{ t \suchthat t \in V(T), v \in \chi(t) \}$ is not empty and forms a 
connected subtree of $T$.
\ei
\edefn

\bdefn[acyclicity]\label{defn:alpha-acyclic-td}
A hypergraph $\calH = (\calV, \calE)$ is {\em acyclic} iff
there exists a tree decomposition 
$(T, \chi)$ in which every bag $\chi(t)$ is a hyperedge of $\calH$.
\edefn

When $\calH$ represents a join query, the tree $T$ in the above 
definition
is also called the {\em join tree} of the query. 
A query is acyclic if and only if its hypergraph is acyclic.

For non-acyclic queries, we often need a measure of how ``close'' a query is to being acyclic. To that end, we use \emph{width} notions of a query.

\bdefn[$g$-width of a hypergraph: a generic width notion~\cite{adler:dissertation}]
\label{defn:g-width}
Let $\calH=(\calV,\calE)$ be a hypergraph, and
$g : 2^\calV \to \mathbb R^+$ be a function that assigns a non-negative
real number to each subset of $\calV$.
The {\em $g$-width} of a tree decomposition $(T, \chi)$ of $\calH$ is 
$\max_{t\in V(T)} g(\chi(t))$.
The {\em $g$-width of $\calH$} is the {\em minimum} $g$-width
over all tree decompositions of $\calH$.
(Note that the $g$-width of a hypergraph is a {\em Minimax} function.)
\edefn

\bdefn[{\em Treewidth} and {\em fractional hypertree width} are special cases of {\em $g$-width}]
Let $s$ be the following function:
$s(B) = |B|-1$, $\forall V \subseteq \calV$.
Then the {\em treewidth} of a hypergraph $\calH$, denoted by
$\tw(\calH)$, is exactly its $s$-width, and
the {\em fractional hypertree width} of a hypergraph $\calH$,
denoted by $\fhtw(\calH)$, is the $\rho^*$-width of $\calH$.
\edefn

From the above definitions, $\fhtw(\calH)\geq 1$ for any hypergraph $\calH$.
Moreover, $\fhtw(\calH)=1$ if and only if $\calH$ is acyclic.

\subsubsection*{Background: Vertex/variable orderings and their equivalence to tree decompositions}
Besides tree decompositions, there is another way to define acyclicity and width notions of a hypergraph, which is \emph{orderings} of the hypergraph vertices.
And just like we refer to queries and hypergraphs interchangeably, we also refer to query variables and hypergraph vertices interchangeably.

In what follows, we use $n$ to denote the number of vertices of the given hypergraph $\calH$.

\bdefn[Vertex ordering of a hypergraph]
A {\em vertex ordering} of a hypergraph $\calH=(\calV,\calE)$ is simply
a listing $\sigma = (v_1,\dots,v_n)$ of all vertices in $\calV$.
\edefn

\bdefn[Elimination sets $U_j^\sigma$ of a vertex ordering $\sigma$]
Given a hypergraph $\calH=(\calV,\calE)$ and a vertex ordering $\sigma = (v_1,\dots,v_n)$,
we define sets $U_1^\sigma, \ldots, U_n^\sigma\subseteq \calV$, called the \emph{elimination sets of $\sigma$}, as follows:
Let $\partial(v_n)$ be the set of hyperedges of $\calH$ that contain $v_n$.
We define $U_n^\sigma$ to be the union of all hyperedges in $\partial(v_n)$:
\[U_n^\sigma = \bigcup_{S \in \partial(v_n)} S.\]
If $n=1$, then we are done.
Otherwise, we remove vertex $v_n$ and all hyperedges in $\partial(v_n)$ from $\calH$ and add back to $\calH$
a new hyperedge $U_n^\sigma-\{v_n\}$, thus turning $\calH$ into a hypergraph with $n-1$ vertices:
\begin{eqnarray*}
   \calV &\gets& \calV -\{v_n\},\\
   \calE &\gets& \left(\calE - \partial(v_n)\right)
   \cup \bigl\{ U_n^\sigma - \{v_n\} \bigr\}.
\end{eqnarray*}
The remaining elimination sets $U_1^\sigma,\ldots,U_{n-1}^\sigma$ are defined inductively to be the elimination sets of the resulting hypergraph (whose vertices are now $\{v_1,\ldots,v_{n-1}\}$).

When $\sigma$ is clear from the context, we drop the superscript $\sigma$ and use $U_1,\ldots, U_n$.
\edefn

\bprop[Every vertex ordering has an ``equivalent'' tree decomposition~\cite{faq-arxiv}]
\label{prop:sigma-TD}
Given a hypergraph $\calH=(\calV,\calE)$, for every vertex ordering $\sigma$, there is a tree decomposition $(T,\chi)$ whose bags $\chi(t)$ are the elimination sets $U_j^\sigma$ of $\sigma$.
\eprop
By applying the GYO elimination procedure~\cite{abiteboul1995foundations} on the bags of any given tree decomposition, we can obtain an ``equivalent'' vertex ordering:
\bprop[Every tree decomposition has an ``equivalent'' vertex ordering~\cite{faq-arxiv}]
Given a hypergraph $\calH=(\calV,\calE)$, for every tree decomposition $(T,\chi)$,
there is a vertex ordering $\sigma$ such that every elimination set $U_j^\sigma$ of $\sigma $ is {\emph contained} in some bag $\chi(t)$ of the tree decomposition $(T,\chi)$.
\label{prop:TD-sigma}
\eprop

\subsubsection*{$\faq$-width of an $\faq$ query}
Just like a conjunctive query, an $\faq$ query has a query hypergraph $\calH=(\calV,\calE)$.
But unlike conjunctive queries, an $\faq$ query also specifies an order of its variables,
which is the order in which we aggregate over those variables in the given $\faq$-expression.
(For example, in expression~\eqref{eq:faq-example}, we sum over $t$ first, then over $y$, and we keep $z$ and $x$ as free variables. Hence, the $\faq$ query in \eqref{eq:faq-example} specifies the variable order $\sigma=(x,z,y,t)$.)
Such a variable order for the query can also be interpreted as a vertex order $\sigma$ for the query's hypergraph.

The $\InsideOut$ algorithm for answering $\faq$ queries is based on \emph{variable elimination}.
To eliminate variable/vertex $v_n$, we have to solve a sub-problem consisting of a smaller $\faq$ query over the variables in the elimination set $U_n^\sigma$.
This smaller query can be solved by an algorithm that is based on worst-case optimal join algorithms~\cite{LFTJ,NPRR12,skew,anrr}.
From~\eqref{eqn:runtime:lftj}, this takes time
\footnote{To achieve this runtime, we need some additional ideas that are beyond the scope of this very brief introduction to $\faq$. See~\cite{faq} for more details.}
\begin{equation}
O(|\calV|\cdot|\calE|\cdot N^{\rho^*_\calH(U_n^\sigma)}\cdot \log N).
\label{eqn:runtime:OI}
\end{equation}
After eliminating $v_n$, the remaining variables $v_{n-1},v_{n-2},\ldots,v_1$ can be eliminated similarly.
This variable elimination algorithm motivates the following width notion.

\bdefn[$\faq$-width of a given variable ordering $\sigma$]
Given an $\faq$ query $\varphi$ with a variable ordering $\sigma$, we define the $\faq$-width
of $\sigma$, denoted by $\faqw(\sigma)$, to be
\begin{equation}\label{eqn:faqw-sigma}
\faqw(\sigma) = \max_{j\in [n]} \left\{ \rho^*_\calH(U^\sigma_j)\right\}.
\end{equation}
\edefn
By the above definition, the $\faq$-width of a variable ordering $\sigma$ is the same as the fractional hypertree width of the ``equivalent'' tree decomposition that is referred to in Proposition~\ref{prop:sigma-TD}.
\bthm[Runtime of $InsideOut$~\cite{faq}]
Given an $\faq$-query $\varphi$ with a variable ordering $\sigma$,
the $InsideOut$ algorithm answers $\varphi$ in time
\begin{equation}
O\left(|\calV|^2\cdot|\calE|\cdot\left(N^{\faqw(\sigma)}+|\varphi|\right)\cdot \log N\right),
\label{eqn:runtime-IO}
\end{equation}
where $|\varphi|$ is the output size in the listing representation.
\label{thm:runtime-IO}
\ethm

Let $\varphi$ be an $\faq$ query with variable ordering $\sigma$. In many cases, there might be a different variable ordering $\sigma'$ such that if we were to permute the aggregates of $\varphi$ in the order of $\sigma'$ instead of $\sigma$, we would obtain an $\faq$-query $\varphi'$ that is ``semantically-equivalent'' to $\varphi$ (i.e. that always returns the same answer as $\varphi$ no matter what the input is). If this is the case, then we can run $\InsideOut$ on $\varphi$ using the ordering $\sigma'$ instead of $\sigma$, which can lead to a better runtime if $\faqw(\sigma')$ happens to be smaller than $\faqw(\sigma)$.
We use $\EVO(\varphi)$ to denote the set of all such ``equivalent'' orderings $\sigma'$. (For a formal definition, see~\cite{faq}.)
Therefore, it is best to consider all orderings $\sigma'$ in $\EVO(\varphi)$, pick the one with the smallest $\faqw(\sigma')$, and use it in $\InsideOut$ algorithm. This motivates the following definition.

\bdefn[$\faq$-width of an $\faq$ query]
The $\faq$-width of an $\faq$ query $\varphi$, denoted by $\faqw(\varphi)$, is the minimum one over all orderings $\sigma'$ in $\EVO(\varphi)$, i.e.
\begin{equation}
\faqw(\varphi) =
\min \left\{ \faqw(\sigma') \suchthat \sigma' \in \EVO(\varphi) \right\}.
\end{equation}
\edefn

Characterizing $\EVO(\varphi)$ for an arbitrary given $\faq$-query $\varphi$ is a technically involved problem (see~\cite{faq} for hardness background and a general solution).
However, the $\faq$ queries that we need for our machine learning tasks are of a special form that makes the problem easier.: The aggregate operator that we use in such queries is the summation operator $\sum$. We refer to those restricted $\faq$ queries as $\faqcs$ queries (see~\cite{faq}). Our $\faqcs$ queries in this work have only two types of variables:
\bi
\item Variables that we are summing over, e.g. variables $y$ and $t$ in \eqref{eq:faq-example}.
\item Free variables (i.e. Group-by variables), e.g. variables $x$ and $z$.
\ei
Given an $\faqcs$ query $\varphi$, $\EVO(\varphi)$ contains every ordering $\sigma'$ that lists
all free variables \emph{before} the non-free variables.
For example, for the $\faqcs$ query $\varphi(x, z)$ in~\eqref{eq:faq-example}, $\EVO(\varphi(x, z))$ contains all permutations of $\{x,y,z,t\}$ where $\{x,z\}$ come before $\{y, t\}$.
\bprop
For any $\faqcs$ query $\varphi$ without free variables, we have $\faqw(\varphi)=\fhtw(\calH)$, where $\calH$ is the hypergraph of $\calH$.
\eprop
\bp
In this case, $\EVO(\varphi)$ contains all $n!$ possible orderings.
By Proposition~\ref{prop:TD-sigma}, for every tree decomposition $(T,\chi)$, there is an ordering $\sigma'$ such that $\faqw(\sigma')\leq \fhtw((T,\chi))$.
By Proposition~\ref{prop:sigma-TD}, for every ordering $\sigma'$, there is a tree decomposition $(T,\chi)$ such that $\fhtw((T,\chi))=\faqw(\sigma')$. Therefore, we have
\[\min_{\sigma'\in\EVO(\varphi)} \faqw(\sigma') \quad=\quad \min_{(T,\chi)} \fhtw((T,\chi)).\]
\ep

\bprop
For any $\faqcs$ query $\varphi$ with $f\geq 1$ free variables, we have $\faqw(\varphi)\leq \fhtw(\calH)+f-1$, where $\calH$ is the hypergraph of $\calH$.
\label{prop:faqw<=fhtw+f-1}
\eprop
\bp
Find a tree decomposition $(T, \chi)$ of $\calH$ with minimal $\fhtw$, i.e. where $\fhtw((T,\chi))=\fhtw(\calH)$.
WLOG let the $f$ free variables be $v_1,\ldots, v_f$.
Construct another tree decomposition $(T,\overline\chi)$ by extending all bags $\chi(t)$ of $(T,\chi)$ with the variables $\{v_2,\ldots,v_f\}$, i.e. by defining $\overline\chi(t)=\chi(t)\cup\{v_2,\ldots,v_f\}$ for all $t$.
By Definition~\ref{defn:TD}, $(T,\overline\chi)$ \emph{is} indeed a tree decomposition.
And because $\rho^*\left(\chi(t)\cup\{v_2,\ldots,v_f\}\right) \leq \rho^*(\chi(t))+f-1$, we have
\[\fhtw((T,\overline\chi)) \leq \fhtw((T,\chi)) + f-1.\]
Moreover, since $(T, \chi)$ must have a bag $\chi(t^*)$ that contains $v_1$,
the corresponding bag $\overline\chi(t^*)$ of $(T, \overline\chi)$ contains all the free variables $\{v_1,\ldots,v_f\}$.
We designate $t^*$ as the root of $T$, and then we run GYO elimination procedure~\cite{abiteboul1995foundations}
on the bags $\overline\chi(t)$ of $(T,\overline\chi)$ to construct a vertex ordering $\sigma'$
with $\faqw(\sigma') \leq \fhtw((T,\overline\chi))$.
Moreover, if we choose to eliminate the vertices of the root $t^*$ at the end of GYO elimination (after all other vertices have already been eliminated), we can make the free variables $\{v_1,\ldots,v_f\}$ appear before all other variables in $\sigma'$, thus making sure that $\sigma'$ is indeed in $\EVO(\varphi)$ and completing the proof.
In particular, we apply GYO elimination as follows:
\bi
   \item If the tree $T$ contains only one node $t^*$:
      \bi
         \item We eliminate vertices in $\overline\chi(t^*)-\{v_1,\ldots,v_f\}$ before eliminating $\{v_1,\ldots,v_f\}$.
         \item We remove $t^*$ from $T$, thus making $T$ an empty tree.
      \ei
   \item Otherwise, we pick a \emph{leaf} node $t$ of $T$ (other than the root $t^*$). Let $t'$ be the parent of $t$ in $T$:
      \bi
         \item If $\overline\chi(t)\subseteq\overline\chi(t')$, then we remove node $t$ from $T$ along with the associated bag $\overline\chi(t)$.
         \item Otherwise, $\overline\chi(t)$ must have a vertex $u$ that is not in $\overline\chi(t')$. (Hence, by property (b) of Definition~\ref{defn:TD}, $u$ is not in $\overline\chi(t'')$ for all $t''$ in $T$ other than $t$.)
         \bi
            \item If $u$ is the only vertex in $\overline\chi(t)$, then we remove node $t$ from $T$ along with the associated bag $\overline\chi(t)$.
            \item Otherwise, we remove $u$ from $\overline\chi(t)$.
         \ei
      \ei
  \item We repeat the above steps until $T$ becomes an empty tree.
\ei
\ep

\section{Missing details from Section~\ref{SEC:ALGO}}
\label{app:sec:algo}

\bp[Proof of Theorem~\ref{thm:Sigma:h}]
We start with point evaluation:
\begin{align*}
   \frac{1}{2|D|}\sum_{(\mv x,y) \in D} (\inner{g(\vec\theta), h(\mv x)}-y)^2 
   &=
   \frac{1}{2|D|}\sum_{(\mv x,y) \in D} (\inner{g(\vec\theta), h(\mv x)}^2
   - 2 y\inner{g(\vec\theta),h(\mv x)} +y^2)\\
   &=
   \frac{1}{2|D|}   \sum_{(\mv x,y) \in D} g(\vec\theta)^\top (h(\mv x) h(\mv
   x)^\top )
   g(\vec\theta)
   - \inner{g(\vec\theta),\frac{1}{|D|}\sum_{(\mv x,y)\in D}yh(\mv x)}\\
   &\hspace*{1em}+ \frac{1}{2|D|} \sum_{(\mv x,y)\in D}y^2\\
   &=
   \frac 1 2 g(\vec\theta)^\top \left(\frac{1}{|D|}\sum_{(\mv x,y) \in D} h(\mv
x) h(\mv x)^\top \right)
   g(\vec\theta)
   - \inner{g(\vec\theta), \mv c} + \frac{s_Y}{2}\\
   &= \frac 1 2 g(\vec\theta)^\top \vec\Sigma g(\vec\theta)
   -\inner{g(\vec\theta),\mv c}+\frac{s_Y}{2}.
\end{align*}
The gradient formula follows straightforwardly from~\eqref{eqn:point:eval} and
the chain rule.
\ep

\bp[Proof of Corollary~\ref{cor:Sigma:h}]
From~\eqref{eqn:point:eval:pr} we have
\begin{align*}
J(\vec\theta)-J(\vec\theta-\alpha\mv d)
&=
   \frac 1 2 \vec\theta^\top \vec\Sigma \vec\theta
   -\frac 1 2 (\vec\theta-\alpha\mv d)^\top \vec\Sigma 
(\vec\theta-\alpha\mv d)
   - \inner{\vec\theta, \mv c} 
   + \inner{\vec\theta-\alpha\mv d, \mv c} 
   +\frac \lambda 2 \norm{\vec\theta}_2^2
   -\frac \lambda 2 \norm{\vec\theta-\alpha\mv d}_2^2\\
&=
   \frac 1 2 \vec\theta^\top \vec\Sigma \vec\theta
   -\frac 1 2 \left(\vec\theta^\top \vec\Sigma\vec\theta
-2\alpha\vec\theta^\top \vec\Sigma\mv d 
+\alpha^2 \mv d^\top \vec\Sigma \mv d\right)
   - \alpha \inner{\mv d, \mv c} 
   + \lambda \alpha \inner{\vec\theta,\mv d} 
   -\frac{\lambda \alpha^2}2 \norm{\mv d}_2^2\\
&=
\alpha\vec\theta^\top \vec\Sigma\mv d 
-\frac{\alpha^2}{2} \mv d^\top \vec\Sigma \mv d
   - \alpha \inner{\mv d, \mv c} 
   + \lambda \alpha \inner{\vec\theta,\mv d} 
   -\frac{\lambda \alpha^2}2 \norm{\mv d}_2^2.
\end{align*}
\ep

\bp[Proof of Proposition~\ref{prop:precomputation:time}]
For any event $E$, let $\delta_E$ denote the Kronecker delta, i.e. $\delta_{E}=1$
if $E$ holds, and $\delta_{E} = 0$ otherwise. 
Recall that the input query $Q$ has hypergraph $\calH= (\calV,\calE)$,
and there is an input relation $R_F$ for every hyperedge $F \in \calE$.
Recall that we can write $\vec\sigma_{ij}$ in the tensor form as shown in
Eq.~\eqref{eqn:sigma:ij:tensor}. Plugging in the definition of $h_i$ and $h_j$
from~\eqref{eqn:categorical:h}; and, let $C_{ij} = C_i \cup C_j$ and $V_{ij} =
V_i \cup V_j$,
we have
\[ \vec\sigma_{ij} = \frac{1}{|D|} \sum_{(\mv x, y) \in D} 
   \prod_{f \in V_{ij} -C_{ij}}x_{f}^{a_{i}(f)+a_j(f)} \cdot 
   \bigotimes_{f_i \in C_i} \mv x_{f_i} \otimes \bigotimes_{f_j \in C_j} \mv x_{f_j}. 
\]
As illustrated in Example~\ref{ex:store:city}, the tensor
$\bigotimes_{f \in C_i} \mv x_f \otimes \bigotimes_{f \in C_j} \mv x_f$ is very
sparse. For a fixed tuple $\mv x$, in fact, the tensor has only {\em one} $1$
entry, corresponding to the combination of values of the attributes in $C_{ij}$.
Hence, $\vec\sigma_{ij}$ is a function of the variables $C_{ij}$.
In the $\faq$-framework, the query representing $\vec\sigma_{ij}$
can be expressed as a {\sf Sum-Product} queries
with free (i.e., group-by) variables $C_{ij}$, defined by:
\begin{equation} 
   \varphi(C_{ij}) = \frac{1}{|D|}
   \sum_{x_{f'}: f' \in \calV - C_{ij}} \prod_{f \in V_{ij}-C_{ij}} 
   x_f^{a_{i}(f)+a_{j}(f)} \cdot\prod_{F\in \calE} \delta_{\pi_F(\mv x) \in R_F}.
   \label{eqn:faq}
\end{equation}
Similarly, the tensor $\mv c_j$ can be sparsely represented by an aggregate
query with group-by attributes $C_j$, which is expressed as the 
{\sf Sum-Product} query
\begin{equation}
   \varphi(C_j) = \frac{1}{|D|}
   \sum_{x_{f'}: f' \in \calV - C_j} y\cdot
    \prod_{f \in V_j-C_j} x_f^{a_{j}(f)} \cdot\prod_{F\in \calE} \delta_{\pi_F(\mv x) \in R_F}.
   \label{eqn:faq2}
\end{equation}
The overall runtimes for computing the above $\faq$-queries follow from applying the
$\InsideOut$ algorithm and Theorem~\ref{thm:runtime-IO}~\cite{faq}.
\ep

\bp[Proof of Proposition~\ref{prop:output:size}]
The fact that $\faqw(i,j)\leq \fhtw+c-1$ follows from
Proposition~\ref{prop:faqw<=fhtw+f-1}. Since $\vec\sigma_{ij}$ is a tensor of order at
most $c$, and each attribute's active domain has size at most $N$, it follows
that $|\vec\sigma_{ij}| \leq N^c$. And, $|\vec\sigma_{ij}| \leq |D|$ because the
support of the tensor $\vec\sigma_{ij}$ cannot be more than the output size.

Fix a query $Q$ with $\rho^* > \fhtw+c-1 \geq c$. 
Consider a database instance $I$
for which $|D|$ (the output size of $Q$) is $\Theta(N^{\rho^*})$. (The existence
of such database instances is guaranteed by Theorem~\ref{thm:agm-lowerbound}.)
From this~\eqref{eqn:unbounded:gap} follows trivially.
\ep

\bp[Proof of Proposition~\ref{prop:per:iteration:time}]
We first analyze the time it takes to compute expression~\eqref{eqn:point:eval},
which is dominated by the quadratic form $g(\vec\theta)^\top\vec\Sigma
g(\vec\theta)$. To compute this quadratic form, for every pair $i,j \in [m]$ we
need to compute $g_i(\vec\theta)^\top \vec\sigma_{ij}g_j(\vec\theta)$.
This product is broken up into a sum of $t_it_j$ terms when we expand $g_i$ and
$g_j$ out. Each of those terms is computed in time
$O(d_id_j|\vec\sigma_{ij}|)$.
The runtime for computing~\eqref{eqn:gradient} is analyzed similarly.
\ep

\section{Missing details from Section~\ref{SEC:FDS}}
\label{app:sec:fds}

\begin{figure}[ht!]
\centering{
   \begin{tikzpicture}
   \node at (1,2) (f1) {$f_1$};
   \node[draw,ellipse,minimum width=2cm, minimum height=1cm,dashed] at (1,0) (S1) {$S_1$};
   \draw[>=stealth,->] (f1) -- (S1);
   \draw[dashed] (0,-.5) rectangle (2,2.5) node[above,align=left] {$G_1$};
   \node at (4,2) (f2) {$f_2$};
   \node[draw,ellipse,minimum width=2cm, minimum height=1cm,dashed] at (4,0) (S2) {$S_2$};
   \draw[>=stealth,->] (f2) -- (S2);
   \draw[dashed] (3,-.5) rectangle (5,2.5) node[above,align=left] {$G_2$};
   \node at (9,2) (fk) {$f_k$};
   \node[draw,ellipse,minimum width=2cm, minimum height=1cm,dashed] at (9,0) (Sk) {$S_k$};
   \draw[>=stealth,->] (fk) -- (Sk);
   \draw[dashed] (8,-.5) rectangle (10,2.5) node[above,align=left] {$G_k$};
   \draw[dashed] (0.5,1.7) rectangle (9.5,2.3);
   \node at (6.5,2) {$F$};
   \draw[draw] (-0.5,-1) rectangle (11,3.5) node[above] {$V$};
\end{tikzpicture}
}
\caption{Groups of simple FDs. $G=G_1\cup \cdots \cup G_k$.}
\label{fig:fd:groups}
\end{figure}

In the proofs below, for each feature $w \in V$, $\mv I_w$ denote the identity matrix 
whose dimension is the size of the effective domain of $w$. This is not to be
confused with the notation $\mv I_n$ which is an order-$n$ identity matrix.

\subsection{Missing rewriting steps in the example in Section~\ref{sec:intro-fd}}
\label{app:subsec:intro-fd}

We first prove identity~\eqref{eqn:55} formally.
By setting the gradient in~\eqref{eqn:pd:10} to $0$, we have
$\vec\theta_\fcountry = {(\mv I_{\fcountry} + \mv R \mv R^\top)^{-1}\mv R} \vec\gamma_\fcity$.
Hence, it remains to show the identity
$$(\mv I_{\fcountry} + \mv R \mv R^\top)^{-1}\mv R = \mv R(\mv I_{\fcity} + \mv
R^\top \mv R)^{-1}.$$
To see this, we apply the Sherman-Morrison-Woodbury 
identity~\eqref{eqn:Woodbury} with $\mv A = \mv I_{\fcountry}$
and $\mv U = \mv R^\top$ to get
\begin{align}
(\mv I_{\fcountry} + \mv R \mv R^\top)^{-1}
     &=
     \mv I_{\fcountry} - \mv R(\mv I_{\fcity} + \mv R^\top\mv R)^{-1}\mv R^\top
     \label{eqn:78}
\end{align}
multiply both sides of~\eqref{eqn:78} on the right by $\mv R$, we obtain
\begin{align*}
(\mv I_{\fcountry} + \mv R \mv R^\top)^{-1}\mv R
  &= \mv R - \mv R(\mv I_{\fcity} + \mv R^\top\mv R)^{-1} \mv R^\top \mv R\\
  &= \mv R [\mv I_{\fcity}- (\mv I_{\fcity} + \mv R^\top\mv R)^{-1} \mv R^\top \mv R]\\
  &= \mv R [\mv I_{\fcity}- (\mv I_{\fcity} + \mv R^\top\mv R)^{-1} (\mv I_{\fcity}+\mv
  R^\top \mv R - \mv I_{\fcity})]\\
  &= \mv R [\mv I_{\fcity}- \mv I_{\fcity} + (\mv I_{\fcity} + \mv R^\top\mv R)^{-1}]\\
  &= \mv R (\mv I_{\fcity} + \mv R^\top\mv R)^{-1}.
\end{align*}
   
We next show~\eqref{eqn:56}; to do so, it is sufficient to verify that
\begin{align}
\norm{\vec\gamma_\fcity - \mv R^\top\vec\theta_\fcountry}_2^2 + \norm{\vec\theta_\fcountry}_2^2 
&=
   \inner{(\mv I_{\fcity} + \mv R^\top \mv R)^{-1}\vec\gamma_\fcity,\vec \gamma_\fcity}
   \label{eqn:81}
\end{align}
For the sake of brevity, define ${\mv B = \mv I_{\fcity} + \mv R^\top\mv R}$ so that $\vec\theta_\fcountry =  {\mv R\mv B^{-1}} \vec\gamma_\fcity$. 
We compute each term on the left hand side separately:
\begin{align*}
\vec\gamma_\fcity - \mv R^\top\vec\theta_\fcountry 
&= \vec\gamma_\fcity - \mv R^\top (\mv I_{\fcountry} + \mv R \mv R^\top)^{-1} \mv R
\vec\gamma_\fcity \\
&= [\mv I_\fcity - \mv R^\top (\mv I_{\fcountry} + \mv R \mv R^\top)^{-1} \mv R]
\vec\gamma_\fcity \\
\text{(follows from~\eqref{eqn:Woodbury})}
&= \mv B^{-1} \vec\gamma_\fcity,
\end{align*}
From here, we derive~\eqref{eqn:81} by
\begin{align*}
\norm{\vec\gamma_\fcity - \mv R^\top\vec\theta_\fcountry}_2^2 + \norm{\vec\theta_\fcountry}_2^2 
&= \norm{\mv B^{-1}\vec\gamma_\fcity}_2^2 + 
 \norm{\mv R \mv B^{-1} \vec\gamma_\fcity}^2\\
&= \inner{
    \mv B^{-1}\vec\gamma_\fcity,
    \mv B^{-1}\vec\gamma_\fcity
} + 
 \inner{
     \mv R \mv B^{-1} \vec\gamma_\fcity,
     \mv R \mv B^{-1} \vec\gamma_\fcity
 } \\
&= \inner{
    \mv B^{-1}\vec\gamma_\fcity,
    \mv B^{-1}\vec\gamma_\fcity
} + 
 \inner{
     \mv B^{-1} \vec\gamma_\fcity,
     \mv R^\top \mv R 
     \mv B^{-1} \vec\gamma_\fcity
 } \\
&= \inner{
    \mv B^{-1}\vec\gamma_\fcity,
    (\mv I_\fcity + \mv R^\top \mv R)
    \mv B^{-1}\vec\gamma_\fcity
}\\
&= \inner{ \mv B^{-1}\vec\gamma_\fcity, \vec\gamma_\fcity}.
\end{align*}

\subsection{Proof of Theorem \ref{thm:pr:d:fd}}
\label{sec:monsterproof}

\bp
We start by breaking the loss term into two parts
\[
   \inner{\vec\theta,h(\mv x)}
   = \sum_{\norm{\mv a_V}_1\leq d}\inner{\vec\theta_{\mv a},h_{\mv a}(\mv x)}\\
   =   \sum_{\substack{\norm{\mv a_V}_1\leq d\\ \norm{\mv a_G}_1 = 0}}\inner{\vec\theta_{\mv a},\mv x^{\otimes \mv a}}
      + \sum_{\substack{\norm{\mv a_V}_1\leq d\\ \norm{\mv a_G}_1 > 0}}\inner{\vec\theta_{\mv a},\mv x^{\otimes \mv a}}
\]
and rewrite the second part:
\begin{align}
   & \sum_{\substack{\norm{\mv a_V}_1\leq d\\ \norm{\mv a_G}_1 > 0}}\inner{\vec\theta_{\mv a},\mv x^{\otimes \mv a}}
   =\sum_{\substack{\norm{\mv a_V}_1\leq d\\ \norm{\mv a_G}_1 > 0}}\inner{\vec\theta_{\mv a}, \mv x_{\overline G}^{\otimes \mv a_{\overline G}} \otimes \mv x_G^{\otimes \mv a_G}}
   =\sum_{\substack{\norm{\mv a_V}_1\leq d\\ \norm{\mv a_G}_1 > 0}}\inner{\vec\theta_{\mv a}, \mv x_{\overline G}^{\otimes \mv a_{\overline G}} \otimes 
                                                                     \bigotimes_{\substack{i \in [k]\\ c \in G_i\\a_c > 0}} \mv x_c}\\
   &=\sum_{\substack{\norm{\mv a_V}_1\leq d\\ \norm{\mv a_G}_1 > 0}}\inner{\vec\theta_{\mv a}, \mv x_{\overline G}^{\otimes \mv a_{\overline G}} \otimes 
          \bigotimes_{\substack{i\in [k]\\ \norm{\mv a_{G_i}}_1>0}} \bigotimes_{\substack{c
      \in G_i\\a_c > 0}} \mv R_c \mv x_{f_i}} \\
   &= \sum_{\substack{\norm{\mv a_V}_1\leq d\\ \norm{\mv a_G}_1 > 0}}\inner{\vec\theta_{\mv a}, \mv x_{\overline G}^{\otimes \mv a_{\overline G}} \otimes 
        \bigotimes_{\substack{i\in [k]\\ \norm{\mv a_{G_i}}_1>0}} \left(\Bigstar_{\substack{c \in G_i\\a_c > 0}} \mv R_c\right) \mv x_{f_i}}\label{eqn:star:1}\\
   &=\sum_{\substack{\norm{\mv a_V}_1\leq d\\ \norm{\mv a_G}_1 > 0}}
   \inner{\vec\theta_{\mv a}, 
             \left( \bigotimes_{\substack{w\in \overline G\\ a_w>0}}  \mv I_w 
             \otimes
             \bigotimes_{\substack{i\in [k]\\ \norm{\mv a_{G_i}}_1>0}} \Bigstar_{\substack{c \in G_i\\a_c > 0}} \mv R_c\right) 
             \left( \mv x_{\overline G}^{\otimes \mv a_{\overline G}} \otimes 
             \bigotimes_{\substack{i\in [k]\\ \norm{\mv a_{G_i}}_1>0}} \mv x_{f_i} \right)
             }\label{eqn:star:2}\\
   &=\sum_{\substack{\norm{\mv a_V}_1\leq d\\ \norm{\mv a_G}_1 > 0}}
            \inner{
             \left( \bigotimes_{\substack{w\in \overline G\\ a_w>0}}  \mv I_w \otimes
             \bigotimes_{\substack{i\in [k]\\ \norm{\mv a_{G_i}}_1>0}} \Bigstar_{\substack{c \in G_i\\a_c > 0}} \mv R_c\right)^\top 
             \vec\theta_{\mv a}, 
             \mv x_{\overline G}^{\otimes \mv a_{\overline G}} \otimes 
             \bigotimes_{\substack{i\in [k]\\ \norm{\mv a_{G_i}}_1>0}} \mv x_{f_i}
             }\\
   &=\sum_{\substack{\norm{\mv a_{\overline G}}_1 = q\\ q<d}} 
     \sum_{\substack{T \subseteq [k]\\ 0<|T|\leq d-q}}
            \sum_{U \in \calU(T,q)}
            \inner{
               \bigl( \underbrace{
                  \bigotimes_{\substack{w\in \overline G\\ a_w>0}}  \mv I_w \otimes \bigotimes_{i\in T} \Bigstar_{c \in U \cap G_i} \mv R_c
               }_{\mv R_{\mv a_{\overline G}, U} \text{ defined in~\eqref{eqn:R:a:U}}}
             \bigr)^\top 
             \vec\theta_{(\mv a_{\overline G}, \mv 1_{U|G})}, 
             \mv x_{\overline G}^{\otimes \mv a_{\overline G}} \otimes 
             \bigotimes_{i \in T} \mv x_{f_i}
          }\label{eqn:tricky}\\
   &=\sum_{\substack{\norm{\mv a_{\overline G}}_1 = q\\ q<d}} 
     \sum_{\substack{T \subseteq [k]\\ 0<|T|\leq d-q}}
            \inner{
               \underbrace{
            \sum_{U \in \calU(T,q)}
             \mv R_{\mv a_{\overline G}, U}^\top
             \vec\theta_{(\mv a_{\overline G},\mv 1_{U|G})}
          }_{\vec\gamma_{(\mv a_{\overline G}, \mv 1_{F_T|F})}}, 
             \mv x_{\overline G}^{\otimes \mv a_{\overline G}} \otimes 
             \bigotimes_{i \in T} \mv x_{f_i}
             }\label{eqn:less:tricky}\\
   &=\sum_{\substack{\norm{\mv a_{\overline G}}_1 = q\\ q<d}} 
     \sum_{\substack{T \subseteq [k]\\ 0<|T|\leq d-q}}
            \inner{
               \vec\gamma_{(\mv a_{\overline G}, \mv 1_{F_T|F})}, 
             \mv x_{\overline G}^{\otimes \mv a_{\overline G}} \otimes 
             \bigotimes_{i \in T} \mv x_{f_i}
             }\\
   &=\sum_{\norm{\mv b_{\overline S}}_1 \leq d} 
            \inner{
               \vec\gamma_{\mv b_{\overline S}},
             \mv x_{\overline S}^{\otimes \mv b_{\overline S}}
             }.
\end{align}
Equality~\eqref{eqn:star:1} follows from~\eqref{eqn:tensor:k:unit}.
Equality~\eqref{eqn:star:2} follows from~\eqref{eqn:tensor:factor}.
Equality at~\eqref{eqn:tricky} is a bit loaded. What goes on there is that we
broke the sum over $\mv a_V$ for which $\norm{\mv a_V}_1 \leq d$ and $\norm{\mv
a_G}_1>0$ into a nested triple sum. First of all, in order for $\norm{\mv
a_G}_1>0$, obviously $\norm{\mv a_{\overline G}}_1 < d$ must hold, so we group by
those tuples first. The remaining mass $\norm{\mv a_G}_1$ can only be at most
$d - \norm{\mv a_{\overline G}}_1 = d-q$. Since all features in $G$ are categorical,
from the above analysis we have $\mv a_G = (a_g)_{g\in G} \in \{0,1\}^G$,
i.e., $\mv a_G$ is a characteristic vector of a subset $U \subseteq G$.
Let $T = \{ i \suchthat U_i \neq \emptyset\} \subseteq [k]$.
Then, in the second summation we group $U$ by $T$. The third summation ranges
over all choices of $U \cap G_i$, $i \in T$, for which the total mass is at 
most $d-q$. (Recall the definition of $\calU(T,q)$ in~\eqref{eqn:calU:T:h}.)

Next, in~\eqref{eqn:less:tricky} we perform the reparameterization.
Recall that $\mv 1_{F_T|F}$ is the characteristic vector of the set
$F_T = \{f_i\}_{i\in T}$ in the 
collection $F = \{f_1,\dots,f_k\}$.
The new parameter $\vec\gamma_{(\mv a_{\overline G},\mv 1_{F_T|F})}$ is indexed by
the tuple $(\mv a_{\overline G},\mv 1_{F_T|F})$ whose support is $\overline G
\cup F = \overline S$, i.e., the set of all features except for the ones
functionally determined by features in $F$. After the reparameterization, the
loss term is identical to the loss term of a $\pr^d$ model whose features are
$\overline S$. This explains the collapsed pair $(\bar g, \bar h)$ used in the
theorem. 

Next, we explore the new parameter and how it affects the penalty term.
Consider a fixed pair $\mv a_{\overline G}$ and $T \subseteq [k]$
such that $T \neq \emptyset$ and $\norm{\mv a_{\overline G}}_1+|T| \leq d$.
The last condition is implicit for the set $U$ to exist
for which $U \cap G_i \neq \emptyset$ and $\norm{\mv a_{\overline G}}_1+|U|\leq d$.
Among all choices of $U$, we single out $U = F_T$ and write
\[
   \vec\gamma_{(\mv a_{\overline G}, \mv 1_{F_T|F})}
   = \sum_{\substack{U \subseteq G\\ U \cap G_i \neq \emptyset, \forall i \in T\\\norm{\mv a_{\overline G}}_1+|U| \leq d}}
             \mv R_{\mv a_{\overline G}, U}^\top
             \vec\theta_{(\mv a_{\overline G},\mv 1_{U|G})}
   = \vec\theta_{(\mv a_{\overline G}, \mv 1_{F_T|G})} +
     \sum_{\substack{F_T \neq U \subseteq G\\ U \cap G_i \neq \emptyset, \forall i \in T\\\norm{\mv a_{\overline G}}_1+|U| \leq d}}
             \mv R_{\mv a_{\overline G}, U}^\top
             \vec\theta_{(\mv a_{\overline G},\mv 1_{U|G})}.
\]
Now we are ready to write the penalty term $\norm{\vec\theta}_2^2$ in terms of the
new parameter $\vec\gamma$ and some ``left-over'' components of $\vec\theta$.
\begin{align*}
   \norm{\vec\theta}_2^2
   &= \sum_{\norm{\mv a_V}_1\leq d} \norm{\vec\theta_{\mv a}}_2^2\\
   &= \sum_{\substack{\norm{\mv a_V}_1\leq d\\ \norm{\mv a_G}_1 = 0}}\norm{\vec\theta_{\mv a_V}}_2^2
      + \sum_{\substack{\norm{\mv a_V}_1\leq d\\ \norm{\mv a_G}_1 > 0}}\norm{\vec\theta_{\mv a_V}}_2^2\\
   &=
   \sum_{\substack{\norm{\mv a_V}_1\leq d\\ \norm{\mv a_G}_1 = 0}}\norm{\vec\theta_{\mv a_V}}_2^2
   + \sum_{\substack{\norm{\mv a_{\overline G}}_1 = q\\ q< d}} 
     \sum_{\substack{T \subseteq [k]\\ 0<|T|\leq d-q}}
     \sum_{U \in \calU(T,q)}
     \norm{\vec\theta_{(\mv a_{\overline G},\mv 1_{U|G})}}_2^2\\
   &=
   \sum_{\substack{\norm{\mv b_{\overline S}}_1\leq d\\ \norm{\mv b_F}_1 = 0}}\norm{\vec\gamma_{\mv b_{\overline S}}}_2^2
   + \sum_{\substack{\norm{\mv a_{\overline G}}_1 = q\\ q< d}} 
     \sum_{\substack{T \subseteq [k]\\ 0<|T|\leq d-q}}
     \left(
     \norm{\vec\theta_{(\mv a_{\overline G},\mv 1_{F_T|G})}}_2^2 +
     \sum_{\substack{W \in \calU(T,q)\\ W\neq F_T}}
     \norm{\vec\theta_{(\mv a_{\overline G},\mv 1_{U|G})}}_2^2
     \right)\\
   &=
   \sum_{\substack{\norm{\mv b_{\overline S}}_1\leq d\\ \norm{\mv b_F}_1 = 0}}\norm{\vec\gamma_{\mv b_{\overline S}}}_2^2
   + \sum_{\substack{\norm{\mv a_{\overline G}}_1 = q\\ q< d}} 
     \sum_{\substack{T \subseteq [k]\\ 0<|T|\leq d-q}}
     \norm{\vec\gamma_{(\mv a_{\overline G},\mv 1_{F_T|F})} -
     \sum_{\substack{U \in \calU(T,q)\\ U\neq F_T}}
             \mv R_{\mv a_{\overline G}, U}^\top
             \vec\theta_{(\mv a_{\overline G},\mv 1_{U|G})}
     }_2^2 \\
     &
   + \sum_{\substack{\norm{\mv a_{\overline G}}_1 = q\\ q< d}} 
     \sum_{\substack{T \subseteq [k]\\ 0<|T|\leq d-q}}
     \sum_{\substack{W \in \calU(T,q)\\ W\neq F_T}}
     \norm{\vec\theta_{(\mv a_{\overline G},\mv 1_{W|G})}}_2^2.
\end{align*}
Next, for every $W \in \calU(T,q) - \{F_T\}$, we optimize out  the
parameter $\vec\theta_{(\mv a_{\overline G},\mv 1_{W|G})}$ by noting that the new
loss term does not depend on these parameters. To optimize them out, we compute 
\begin{align*}
   \frac 1 2 \pd{J}{\vec\theta_{(\mv a_{\overline G},\mv 1_{W|G})}}
   &= \vec\theta_{(\mv a_{\overline G},\mv 1_{W|G})} - 
     \mv R_{\mv a_{\overline G}, W}
     \left(
        \vec\gamma_{(\mv a_{\overline G},\mv 1_{F_T|F})} -
             \sum_{\substack{U \in \calU(T, \norm{\mv a_{\overline G}}_1)\\ U\neq F_T}}
             \mv R_{\mv a_{\overline G}, U}^\top
             \vec\theta_{(\mv a_{\overline G},\mv 1_{U|G})}
     \right) \\
   &= \vec\theta_{(\mv a_{\overline G},\mv 1_{W|G})} - 
     \mv R_{\mv a_{\overline G}, W}
     \vec\theta_{(\mv a_{\overline G},\mv 1_{F_T|G})}.
\end{align*}
Setting this partial derivative to $0$, we obtain
$\vec\theta_{(\mv a_{\overline G},\mv 1_{W|G})} = 
\mv R_{\mv a_{\overline G}, W} \vec\theta_{(\mv a_{\overline G},\mv 1_{F_T|G})}$, which leads to
\begin{align*}
   \vec\theta_{(\mv a_{\overline G},\mv 1_{F_T|G})}
   &=
        \vec\gamma_{(\mv a_{\overline G},\mv 1_{F_T|F})} -
             \sum_{\substack{U \in \calU(T, \norm{\mv a_{\overline G}}_1)\\ U\neq F_T}}
             \mv R_{\mv a_{\overline G}, U}^\top
             \vec\theta_{(\mv a_{\overline G},\mv 1_{U|G})}\\
   &=
        \vec\gamma_{(\mv a_{\overline G},\mv 1_{F_T|F})} -
             \sum_{\substack{U \in \calU(T, \norm{\mv a_{\overline G}}_1)\\ U\neq F_T}}
             \mv R_{\mv a_{\overline G}, U}^\top
             \mv R_{\mv a_{\overline G}, U} \vec\theta_{(\mv a_{\overline G},\mv
             1_{F_T|G})}.
\end{align*}
Moving and grouping, we obtain
\begin{equation*}
   \left( \bigotimes_{\substack{g\in \overline G\\ a_g>0}} \mv I_g \otimes \bigotimes_{i\in T} \mv I_{f_i}
   + 
             \sum_{\substack{U \in \calU(T, \norm{\mv a_{\overline G}}_1)\\ U\neq F_T}}
             \mv R_{\mv a_{\overline G}, U}^\top
             \mv R_{\mv a_{\overline G}, U} 
     \right) 
     \vec\theta_{(\mv a_{\overline G},\mv 1_{F_T|G})}
   =
     \vec\gamma_{(\mv a_{\overline G},\mv 1_{F_T|F})}.
\end{equation*}
Or, more compactly,
\begin{align}
    \underbrace{
        \left(\sum_{W \in \calU(T, \norm{\mv a_{\overline G}}_1)}
   \mv R_{\mv a_{\overline G}, W}^\top
   \mv R_{\mv a_{\overline G}, W} \right)
   }_{\mv B_{\mv a_{\overline G},T} \text{as defined in~\eqref{eqn:B:T:h}}}
     \vec\theta_{(\mv a_{\overline G},\mv 1_{F_T|G})}
   =
     \vec\gamma_{(\mv a_{\overline G},\mv 1_{F_T|F})}.
\end{align}

Consequently, we can completely optimize out the remaining
$\vec\theta$-components, solving for them in terms of the components of
$\vec\gamma$:
\begin{align*}
   \vec\theta_{(\mv a_{\overline G},\mv 1_{F_T|G})}
   &=
   \mv B_{\mv a_{\overline G},T}^{-1} \vec\gamma_{(\mv a_{\overline G},\mv 1_{F_T|F})}\\
   \vec\theta_{(\mv a_{\overline G},\mv 1_{U|G})} 
   &= \mv R_{\mv a_{\overline G}, U} \vec\theta_{(\mv a_{\overline G},\mv 1_{F_T|G})}\\
   &= 
   \mv R_{\mv a_{\overline G}, U} 
   \mv B_{\mv a_{\overline G},T}^{-1}
   \vec\gamma_{(\mv a_{\overline G},\mv 1_{F_T|F})}
\end{align*}
Thus, for a fixed $T$ and $\overline G$,
we can simplify the total squared normed involved:
\begin{align*}
   \sum_{U \in \calU(T, \norm{\mv a_{\overline G}}_1)} 
   \norm{\vec\theta_{(\mv a_{\overline G},\mv 1_{U|G})}}_2^2
   &=\sum_{U \in \calU(T, \norm{\mv a_{\overline G}}_1)} 
   \inner{
   \mv R_{\mv a_{\overline G}, U} 
   \mv B_{\mv a_{\overline G},T}^{-1}
   \vec\gamma_{(\mv a_{\overline G},\mv 1_{F_T|F})},
   \mv R_{\mv a_{\overline G}, U} 
   \mv B_{\mv a_{\overline G},T}^{-1}
   \vec\gamma_{(\mv a_{\overline G},\mv 1_{F_T|F})}}
   \\
   &=\sum_{U \in \calU(T, \norm{\mv a_{\overline G}}_1)} 
   \inner{
   \mv R_{\mv a_{\overline G}, U}^\top 
   \mv R_{\mv a_{\overline G}, U} 
   \mv B_{\mv a_{\overline G},T}^{-1}
   \vec\gamma_{(\mv a_{\overline G},\mv 1_{F_T|F})},
   \mv B_{\mv a_{\overline G},T}^{-1}
   \vec\gamma_{(\mv a_{\overline G},\mv 1_{F_T|F})}}
   \\
   &=
   \inner{
       \left(\sum_{U \in \calU(T, \norm{\mv a_{\overline G}}_1)} 
   \mv R_{\mv a_{\overline G}, U}^\top 
   \mv R_{\mv a_{\overline G}, U} 
   \right)
   \mv B_{\mv a_{\overline G},T}^{-1}
   \vec\gamma_{(\mv a_{\overline G},\mv 1_{F_T|F})},
   \mv B_{\mv a_{\overline G},T}^{-1}
   \vec\gamma_{(\mv a_{\overline G},\mv 1_{F_T|F})}}
   \\
   &=
   \inner{
   \vec\gamma_{(\mv a_{\overline G},\mv 1_{F_T|F})},
   \mv B_{\mv a_{\overline G},T}^{-1}
   \vec\gamma_{(\mv a_{\overline G},\mv 1_{F_T|F})}}.
\end{align*}
Finally, we write $\norm{\vec\theta}_2^2$ in terms of the new parameter $\vec\gamma$ to
prove~\eqref{eqn:Omega:gamma}:
\begin{align*}
   \norm{\vec\theta}_2^2
   &= \sum_{\norm{\mv a_V}_1\leq d} \norm{\vec\theta_{\mv a}}_2^2\\
   &= \sum_{\substack{\norm{\mv a_V}_1\leq d\\ \norm{\mv a_G}_1 = 0}}\norm{\vec\theta_{\mv a_V}}_2^2
      + \sum_{\substack{\norm{\mv a_V}_1\leq d\\ \norm{\mv a_G}_1 > 0}}\norm{\vec\theta_{\mv
      a_V}}_2^2\\
   &=
   \sum_{\substack{\norm{\mv a_V}_1\leq d\\ \norm{\mv a_G}_1 = 0}}\norm{\vec\theta_{\mv a_V}}_2^2
   + \sum_{\substack{\norm{\mv a_{\overline G}}_1 = q\\ q< d}} 
     \sum_{\substack{T \subseteq [k]\\ 0<|T|\leq d-q}}
     \sum_{U \in \calU(T,q)}
     \norm{\vec\theta_{(\mv a_{\overline G},\mv 1_{U|G})}}_2^2\\
   &=
   \sum_{\substack{\norm{\mv b_{\overline S}}_1\leq d\\ \norm{\mv b_F}_1=0}}
     \norm{\vec\gamma_{\mv b_{\overline S}}}_2^2
   + \sum_{\substack{\norm{\mv a_{\overline G}}_1 = q\\ q< d}} 
     \sum_{\substack{T \subseteq [k]\\ 0<|T|\leq d-q}}
   \inner{
       \mv B_{\mv a_{\overline G},T}^{-1}
   \vec\gamma_{(\mv a_{\overline G},\mv 1_{F_T|F})},
   \vec\gamma_{(\mv a_{\overline G},\mv 1_{F_T|F})}}.
\end{align*}

\ep

\subsection{Alternative to Corollary~\ref{cor:Sigma:h}}
\label{subsec:alternative:corollary}

One big advantage of a linear model in terms of BGD is
Corollary~\ref{cor:Sigma:h}, where we do not have to redo point-evaluation for
every backtracking step. After the reparameterization exploiting FD-based
dimensionality reduction, Corollary~\ref{cor:Sigma:h} does not work as is,
because we have changed the penalty terms. However, it is easy to work out a
similar result in terms of the new parameter space; see 
The point of the following proposition is that we only need to compute 
intermediate results involving
the covariance matrix $\overline{\vec\Sigma}$ once while backtracking.
For each new value of $\alpha$, we will need to recompute the penalty's
objective $\overline \Omega(\vec\gamma-\alpha\mv d)$, which is an inexpensive
operation.  If $\lambda=0$, we can even solve for $\alpha$ directly.
\bprop With respect to the new parameters (and new objective $\overline J$
defined in~\eqref{eqn:overline:J}), the Armijo condition
$\overline J(\vec\gamma)- \overline J(\vec\gamma - \alpha \mv d) \leq \frac
\alpha 2 \norm{\mv d}_2^2$ is equivalent to
\begin{equation*}
    \alpha \left( 2 \vec\gamma^\top \overline{\vec\Sigma}\mv d
    - \alpha \mv d \overline{\vec\Sigma}\mv d 
    - 2 \inner{\mv d, \overline{\mv c}} 
    - \norm{\mv d}_2^2
    \right) +
    \lambda \overline \Omega(\vec\gamma))
    \leq \lambda \overline \Omega(\vec\gamma-\alpha\mv d),
\end{equation*}
where $\mv d = \grad \overline J(\vec\gamma)$. Furthermore, the next gradient of
$\overline J$ is also readily available: 
\[ \pd{\overline J(\vec\gamma-\alpha\mv d)}{\vec\gamma} 
   = \mv d - \alpha \overline{\vec\Sigma}\mv
d + \frac \lambda 2 \left( \pd{\Omega(\vec\gamma-\alpha\mv
d)}{\vec\gamma}-\pd{\Omega(\vec\gamma)}{\vec\gamma}\right).
\]
\label{prop:alternative}
\eprop
\bp
Let $\mv d = \grad \overline J(\vec\gamma)$. Then,
\begin{align*}
   \overline J(\vec\gamma)- \overline J(\vec\gamma - \alpha \mv d)
   &= \frac 1 2 \vec\gamma^\top \overline{\vec\Sigma}\vec\gamma 
    - \frac 1 2 (\vec\gamma - \alpha \mv d) \overline{\vec\Sigma} (\vec\gamma - \alpha \mv d)
    + \inner{\vec\gamma - \alpha \mv d, \overline{\mv c}}
    + \frac \lambda 2 ( \overline \Omega(\vec\gamma) - \overline \Omega(\vec\gamma-\alpha\mv d))\\
    &= \alpha \vec\gamma^\top \overline{\vec\Sigma}\mv d
    - \frac{\alpha^2}{2} \mv d \overline{\vec\Sigma}\mv d 
    - \alpha \inner{\mv d, \overline{\mv c}} 
    + \frac \lambda 2 ( \overline \Omega(\vec\gamma) - \overline \Omega(\vec\gamma-\alpha\mv d)).\\
\end{align*}
\ep

\subsection{Specializing Theorem~\ref{thm:pr:d:fd} to the {\sf LR} model}
\label{subsec:lr:special:case}

This section specializes Theorem~\ref{thm:pr:d:fd} to the $\lr$-model.
Let us first specialize
expressions~\eqref{eqn:calU:T:h},~\eqref{eqn:R:a:U}.
and ~\eqref{eqn:B:T:h}, 
We start with~\eqref{eqn:calU:T:h}. Since $d=1$, the only valid choice of $q$ is
$0$, and $|T|=1$. If $T = \{j\}$, then $U \in \calU(T,q)$ iff $U =\{c\}$ for
some $c \in G_j$. In other words, we can replace $\calU(T,q)$ by $G_j$ itself.
Next, consider~\eqref{eqn:R:a:U}: there is only one valid choice of $\mv
\mv a_{\overline G}$ -- the all $0$ vector -- and  $U = \{c\}$ for some $c \in
G_j$, the matrix $\mv R_{\mv a_{\overline G},U}$ is {\em exactly} $\mv R_c$.
Lastly, when $T = \{j\}$
the sum~\eqref{eqn:B:T:h} becomes $\sum_{c\in G_j}\mv R_c^\top \mv R_c$.
We have the following corollary:

\bcor
Consider a $\lr$ model with parameters $\vec\theta=(\vec\theta_w)_{w\in V}$ and 
$k$ groups of simple FDs $G_i = \{f_i\}\cup S_i$,
$i \in [k]$. Define the following reparameterization:
\[
   \vec\gamma_w = 
   \begin{cases}
      \vec\theta_w & w \in V - G,\\
      {\sum_{c\in G_i}\mv R^\top_c \vec\theta_c}& w \in F.
     \end{cases}
\]
Then, minimizing $J(\vec\theta)$ is equivalent to minimizing the
function 
$\overline J(\vec\gamma) = \frac 1 2 \vec\gamma^\top \overline{\vec\Sigma} \vec\gamma
- \inner{\vec\gamma,\overline{\mv c}} + \frac\lambda 2 \Omega(\vec\gamma),$
where
$\Omega(\vec\gamma) = 
\sum_{w \in V\setminus G}\norm{\vec\gamma_w}_2^2+\sum_{i=1}^k\inner{\mv B_i^{-1} \gamma_{f_i}, \gamma_{f_i}}
$, and matrix $B_i$ for each $i \in [k]$ is given by
\begin{equation}
   \mv B_i = \sum_{c\in G_i}\mv R^\top_c \mv R_c.\label{eqn:B:i}
\end{equation}
\label{cor:fd:lr}
\ecor

$\overline J$ is defined with respect to the FD-reduced pair of
functions $\overline{g}, \overline{h}$ and a reduced parameter
space of $\vec\gamma$. Its gradient is very simple to compute, 
where we specialize~\eqref{eqn:main:derivative}:
\begin{align}
   \frac 1 2 \pd{\Omega(\vec\gamma)}{\vec\gamma_w} &= 
   \begin{cases}
      \vec\gamma_w & w \in V-G,\\
      {\mv B_i^{-1} \vec\gamma_{f_i}}& w \in F.
   \end{cases}
\end{align}
Moreover, once a minimizer $\vec\gamma$ of $\overline J$ is obtained,
following~\eqref{eqn:main:gamma:to:theta},
we can compute a minimizer $\vec\theta$ of $J$ by setting
\[
   \vec\theta_w =
   \begin{cases}
      \vec\gamma_w & w \in V\setminus G,\\
      \mv R_w \mv B_i^{-1} \vec\gamma_{f_i} & w \in G_i, i \in [k].
   \end{cases}
\]
\subsection{Specializing Theorem~\ref{thm:pr:d:fd} to the $\pr^2$ model}
\label{subsec:pr:2:special:case}

In this section we explores Theorem~\ref{thm:pr:d:fd} for the special case of
degree-2 polynomial regression. 
This case is significant for three reasons. First, due to the explosion in the number of 
parameters, in practice one rarely runs polynomial regression of degree higher than $2$. 
In fact, $\pr^2$ may be a sufficiently rich nonlinear regression model for many real-world applications. Second, 
this is technically already a highly non-trivial application of our general theorem. \
Third, this case shares some commonality with $\fama^2_r$ model 
to be described in the next section.

As before, we first specialize
expressions~\eqref{eqn:calU:T:h},~\eqref{eqn:R:a:U}, and ~\eqref{eqn:B:T:h}.
To do so, we slightly change the indexing scheme of the model to simplify the presentation. 
In the general model, we use tuples $\mv a$ with $\norm{\mv a}_1\leq d$ to index parameters. When the
model is of degree $2$, we explicitly write down the two types of indices:
we use $\vec\theta_w$, $w \in V$ instead of $\vec\theta_{\mv a}$ with $\norm{\mv
a}_1=1$, and we use $\vec\theta_{cw}$ with $c,w \in V$ instead of
$\vec\theta_{\mv a}$ when $\norm{\mv a}_1=2$.

We start with~\eqref{eqn:calU:T:h}. Since $d=2$, two valid choices of $q$ are
$0$ and $1$. 
\bi
\item when $q=1$, $|T|=\{i\}$ for some $i \in [k]$. The set
   $\calU(\{i\},1)$ is the collection of singleton subsets of $G_i$. Hence, this
   is similar to the linear regression situation.
\item when $q=0$, $|T|$ is either $\{i\}$ or $\{i,j\}$. The set
   $\calU(\{i,j\},0)$ consists of all $2$-subsets $U$ of $G$ for which $U$
   contains one element from $G_i$ and one from $G_j$.
   The set $\calU(\{i\},0)$ contains all singletons and $2$-subsets of $G_i$.
\ei
Next, consider~\eqref{eqn:R:a:U}: there are two valid choices for the pair
$(\mv a_{\overline G}, U)$:
\bi
 \item when $\norm{\mv a_{\overline G}}_1=0$, $U \in \calU(\{i,j\},0)$
    or $U \in \calU(\{i\},0)$. In that case, we have
    \begin{align*}
       \mv R_{\emptyset,\{c,t\}} &= \mv R_c \otimes \mv R_t & (c,t) \in G_i \times G_j\\
       \mv R_{\emptyset,\{c\}} &= \mv R_c & c\in G_i \\
       \mv R_{\emptyset,\{c,t\}} &= \mv R_c \star \mv R_t & \{c,t\} \in
       \binom{G_i}{2}. \\
    \end{align*}
 \item when $\norm{\mv a_{\overline G}}_1=1$, $U \in \calU(\{i\},1)$ for some
    $i\in [k]$; and in this case we use $w \in \overline G$ to represent $\mv
    a_{\overline G}$:
    \[ \mv R_{w,\{c\}} = \mv I_w \otimes \mv R_c. \]
\ei
Finally, we write down~\eqref{eqn:B:T:h} explicitly; the valid indices for $\mv B$
are 
$\mv B_{\emptyset, \{i\}}$
$\mv B_{\emptyset, \{i,j\}}$, and
$\mv B_{w, \{i\}}$
(also recall the definition of $\mv B_i$ in~\eqref{eqn:B:i}):
\begin{align}
   \mv B_{\emptyset, \{i\}} &= 
   \sum_{c \in G_i} \mv R_c^\top \mv R_c +
   \sum_{\{c,t\} \in \binom{G_i}{2} } (\mv R_c \star \mv R_t)^\top (\mv R_c \star \mv R_t)\\
   \mv B_{\emptyset, \{i,j\}} 
   &= \sum_{c\in G_i, t\in G_j} (\mv R_c \otimes \mv R_t)^\top (\mv R_c \otimes \mv
   R_t)\nonumber\\
   &= \sum_{c\in G_i, t\in G_j} (\mv R_c^\top \mv R_c \otimes \mv R_t^\top \mv R_t)\nonumber \\
   &= \mv B_i \otimes \mv B_j
      \label{eqn:B:0:ij}\\
   \mv B_{w,\{i\}} &= 
   \sum_{c \in G_i} (\mv I_w \otimes \mv R_c)^\top (\mv I_w \otimes \mv R_c)
   = \mv I_w \otimes \mv B_i
      \label{eqn:B:w:i}
\end{align}

\bcor
Consider the $\pr^2$ model with $k$ groups of simple FDs $G_i = \{f_i\}\cup S_i$,
$i \in [k]$. Let $$\vec\theta=((\vec\theta_w)_{w\in V}, (\vec\theta_{cw})_{c,w\in
v})$$ be the original parameters, and $G = \cup_{i\in [k]}G_i$. 
Define the following reparameterization:
\begin{align}
   \vec\gamma_w &= \begin{cases}
      \vec\theta_w & w \in V \setminus G\\
      \displaystyle{\sum_{c\in G_i}\mv R^\top_c \vec\theta_c + \sum_{\{c,t\}\in \binom{G_i}{2}}
      (\mv R_c \star \mv R_t)^\top \vec\theta_{ct}} & \substack{w = f_i\\ i \in
[k].}
\end{cases}\\
\vec\gamma_{tw} &= 
\begin{cases}
   \vec\theta_{tw}, & \{t,w\}\subseteq V\setminus G\\
   \vspace{3pt}
   \displaystyle{\sum_{c\in G_i} (\mv I_w \otimes \mv R^\top_{c})
   \vec\theta_{wc}} & t=f_i, w \notin G\\
   \vspace{3pt}
   \displaystyle{\sum_{(c,c') \in G_i\times G_j} (\mv R_c \star \mv R_{c'})^\top
   \vec\theta_{cc'} }, & \{ t,w\} = \{f_i,f_j\}, \{i,j\} \in
\binom{[k]}{2}.
\end{cases}
\end{align}
Then, minimizing $J(\vec\theta)$ is equivalent to minimizing the
function 
$\overline J(\vec\gamma) = \frac 1 2 \vec\gamma^\top \overline{\vec\Sigma} \vec\gamma
- \inner{\vec\gamma,\overline{\mv c}} + \frac\lambda 2 \Omega(\vec\gamma),$
where
\begin{align*}
   \Omega(\vec\gamma) &= 
   \sum_{w \notin G}\norm{\vec\gamma_w}_2^2 + 
       \sum_{\substack{c \notin G\\ t\notin G}} \norm{\vec\gamma_{ct}}_2^2 +
       \sum_{i=1}^k \inner{\mv B_{\emptyset,\{i\}}^{-1} \vec\gamma_{f_i}, \vec\gamma_{f_i}} \\
       &\hspace*{1em}+\sum_{\substack{i\in [k]\\ w\notin G}}\inner{(\mv I_w \otimes \mv
       B_i^{-1} \vec\gamma_{wf_i}, \vec\gamma_{wf_i}}
   + \sum_{ij \in \binom{[k]}{2}} \inner{\mv B_i^{-1} \otimes \mv B_j^{-1} \vec\gamma_{f_if_j}, \gamma_{f_if_j}}.
\end{align*}
\label{cor:fd:pr2}
\ecor

The gradient of $\overline J$ is very simple to compute, by noticing that
$\overline J$ is defined with respect to the FD-reduced pair of
functions $\overline{g}, \overline{h}$ and a reduced parameter
space of $\vec\gamma$. Its gradient can be computed by 
specializing~\eqref{eqn:main:derivative}:
\begin{align}
   \frac 1 2 \pd{\Omega(\vec\gamma)}{\vec\gamma_w} &=
   \begin{cases}
      \vec\gamma_w & w \notin G\\
      {\mv B_{\emptyset,\{i\}}^{-1} \vec\gamma_{f_i}}& w = f_i
   \end{cases}\\
   \frac 1 2\pd{\Omega(\vec\gamma)}{\vec\gamma_{tw}} &=
   \begin{cases}
      \vec\gamma_{tw}& \{t,w\} \cap \{f_i\}_{i=1}^k = \emptyset\\
      \displaystyle{(\mv I_w \otimes \mv B_i^{-1}) \vec\gamma_{wf_i}}& t=f_i,w\notin G\\
      (\mv B_i^{-1} \otimes \mv B_j^{-1}) \vec\gamma_{f_if_j}  & \{ t,w\} = \{f_i,f_j\}.
   \end{cases}
\end{align}
Moreover, once a minimizer $\vec\gamma$ of $\overline J$ is obtained,
following~\eqref{eqn:main:gamma:to:theta},
we can compute a minimizer $\vec\theta$ of $J$ by setting
\begin{align*}
   \vec\theta_w &= 
   \begin{cases}
      \vec\gamma_w & w \in V\setminus G\\
      \mv R_w \mv B_{\emptyset,\{i\}}^{-1} \vec\gamma_{f_i}, & w \in G_i, i \in [k]\\
   \end{cases}\\
   \vec\theta_{ct} &= (\mv R_c \star \mv R_t) \mv B_{\emptyset,\{i\}}^{-1}
   \vec\gamma_{f_i}, \forall \{c,t\} \in \binom{G_i}{2}\\
   \vec\theta_{cw} &= 
   \begin{cases}
      \vec\gamma_{cw}, & w \in V\setminus G\\
      (\mv I_w \otimes \mv R_c\mv B_i^{-1})\vec\gamma_{wf_i}, & c \in G_i, w \notin G, i\in [k]
   \end{cases}\\
   \vec\theta_{ct} &= (\mv R_c \mv B_i^{-1} \otimes \mv R_t \mv
B_j^{-1})\vec\gamma_{f_if_j},
 (c,t) \in G_i\times G_j.
\end{align*}

\subsection{Proofs of results in Section~\ref{subsec:fama:fd}}

\bp[Proof of Theorem~\ref{thm:fd:fama2}]
We begin with a similar derivation, where ``relevant terms'' of
$\inner{g(\vec\theta), h(\mv x)}$ are the terms where $h$ contains a feature
$c \in F_i$ for some $i \in [k]$:
\begin{align*}
   & \text{relevant terms of } \inner{g(\vec\theta),h(\mv x)} \\
   &=  \sum_{\substack{c\in F_i\\ i \in [k]}}\inner{\vec\theta_c,\mv x_c}
   + \sum_{\substack{\{c,t\} \in \binom{F_i}{2}\\ i \in [k]\\\ell \in [r]}} \inner{\vec\theta_c^{(\ell)}\otimes\vec\theta_t^{(\ell)}, \mv x_c \otimes \mv x_t}
   + \sum_{ij \in \binom{[k]}{2}} \sum_{\substack{c \in F_i\\ t \in F_j\\\ell\in [r]}} \inner{\vec\theta_c^{(\ell)}\otimes\vec\theta_t^{(\ell)}, \mv x_c \otimes \mv x_t}\\
   &\hspace*{1em}+ \sum_{\substack{c \in F\\ w \notin F\\\ell\in[r]}} \inner{\vec\theta_c^{(\ell)}\otimes\vec\theta_w^{(\ell)}, \mv x_c \otimes \mv x_w}\\
   &=  \sum_{\substack{c\in F_i\\ i \in [k]}}\inner{\vec\theta_c,\mv R_c\mv x_{f_i}}
   + \sum_{\substack{\{c,t\} \in \binom{F_i}{2}\\ i \in [k]\\\ell \in [r]}} \inner{\vec\theta_c^{(\ell)}\otimes\vec\theta_t^{(\ell)}, \mv R_c \mv x_{f_i} \otimes \mv R_t \mv x_{f_i}}\\
   &
   + \sum_{ij \in \binom{[k]}{2}} \sum_{\substack{c \in F_i\\ t \in F_j\\\ell\in [r]}} \inner{\vec\theta_c^{(\ell)}\otimes\vec\theta_t^{(\ell)}, \mv R_c \mv x_{f_i} \otimes \mv R_t \mv x_{f_j}}
   + \sum_{\substack{i\in [k] \\ c \in F_i\\ w \notin F\\\ell\in[r]}} \inner{\vec\theta_c^{(\ell)}\otimes\vec\theta_w^{(\ell)}, \mv R_c \mv x_{f_i} \otimes \mv x_w}\\
   &=  \sum_{\substack{c\in F_i\\ i \in [k]}}\inner{\mv R_c^\top \vec\theta_c,\mv x_{f_i}}
   + \sum_{\substack{\{c,t\} \in \binom{F_i}{2}\\ i \in [k]\\\ell \in [r]}}
   \inner{\mv R^\top_c \vec\theta_c^{(\ell)}\otimes\mv R_t^\top \vec\theta_t^{(\ell)}, \mv x_{f_i} \otimes \mv x_{f_i}}\\
   &
   + \sum_{ij \in \binom{[k]}{2}} \sum_{\substack{c \in F_i\\ t \in F_j\\\ell\in
   [r]}} \inner{\mv R_c^\top \vec\theta_c^{(\ell)}\otimes \mv R_t^\top \vec\theta_t^{(\ell)}, \mv x_{f_i} \otimes \mv x_{f_j}}
   + \sum_{\substack{i\in [k] \\ c \in F_i\\ w \notin F\\\ell\in[r]}} \inner{\mv
   R_c^\top \vec\theta_c^{(\ell)}\otimes \vec\theta_w^{(\ell)}, \mv x_{f_i} \otimes \mv x_w}\\
   &=  \sum_{\substack{c\in F_i\\ i \in [k]}}\inner{\mv R_c^\top \vec\theta_c,\mv x_{f_i}}
   + \sum_{\substack{\{c,t\} \in \binom{F_i}{2}\\ i \in [k]\\\ell \in [r]}} \inner{\mv R^\top_c\vec\theta_c^{(\ell)}\circ \mv R_t^\top \vec\theta_t^{(\ell)}, \mv x_{f_i}}\\
   &
   + \sum_{\substack{ij \in \binom{[k]}{2}\\\ell\in[r]}} \inner{\sum_{c\in F_i}\mv R_c^\top \vec\theta_c^{(\ell)}\otimes \sum_{t\in F_j}\mv R_t^\top \vec\theta_t^{(\ell)}, \mv x_{f_i} \otimes \mv x_{f_j}}
   + \sum_{\substack{i\in [k] \\ w \notin F\\\ell\in[r]}} \inner{\sum_{c\in F_i}\mv R_c^\top \vec\theta_c^{(\ell)}\otimes\vec\theta_w^{(\ell)}, \mv x_{f_i} \otimes \mv x_w}\\
   &=  \sum_{i=1}^k \inner{\underbrace{\sum_{c\in F_i} \mv R_c^\top \vec\theta_c
   + \sum_{\ell=1}^r \sum_{\{c,t\} \in \binom{F_i}{2}} \mv R^\top_c\vec\theta_c^{(\ell)}\circ \mv R_t^\top \vec\theta_t^{(\ell)}}_{\vec\gamma_{f_i}}, \mv x_{f_i}}\\
   &
   + \sum_{\substack{ij \in \binom{[k]}{2}\\\ell\in[r]}} \inner{\underbrace{\sum_{c\in F_i}\mv R_c^\top \vec\theta_c^{(\ell)}}_{\vec\gamma_{f_i}^{(\ell)}}\otimes 
   \underbrace{\sum_{t\in F_j}\mv R_t^\top \vec\theta_t^{(\ell)}}_{\vec\gamma_{f_j}^{(\ell)}}, \mv x_{f_i} \otimes \mv x_{f_j}}
   + \sum_{\substack{i\in [k] \\ w \notin F\\\ell\in[r]}} \inner{\underbrace{\sum_{c\in F_i}\mv R_c^\top \vec\theta_c^{(\ell)}}_{\vec\gamma_{f_i}^{(\ell)}}\otimes\vec\theta_w^{(\ell)}, \mv x_{f_i} \otimes \mv x_w}\\
   &=  \sum_{i=1}^k \inner{\vec\gamma_{f_i}, \mv x_{f_i}}
   + \sum_{\substack{ij \in \binom{[k]}{2}\\\ell\in[r]}} 
     \inner{{\vec\gamma_{f_i}^{(\ell)}}\otimes {\vec\gamma_{f_j}^{(\ell)}}, \mv x_{f_i} \otimes \mv x_{f_j}}
   + \sum_{\substack{i\in [k] \\ w \notin F\\\ell\in[r]}} 
     \inner{{\vec\gamma_{f_i}^{(\ell)}}\otimes\vec\theta_w^{(\ell)}, \mv x_{f_i}
     \otimes \mv x_w}.
\end{align*}
The above derivation immediately yields the reparameterization given in the statement of the theorem,
which we reproduce here for the sake of clarity:
\begin{eqnarray*}
   \vec\gamma_w &=& {\begin{cases}
      \vec\theta_w & w\notin F\\
      \displaystyle{
         \vec\theta_{f_i} + \sum_{c\in S_i}\mv R_c^\top\vec\theta_c + \vec\beta_{f_i}} & w = f_i, i \in [k].
   \end{cases}}\\
   \vec\gamma^{(\ell)}_w &=&
   \begin{cases}
      \vec\theta^{(\ell)}_w & w \notin\{f_1,\dots,f_k\} \\ 
      \vec\theta^{(\ell)}_{f_i} + \sum_{c\in S_i}\mv R_c^\top \vec\theta^{(\ell)}_c & w=f_i, i \in [k].
   \end{cases}
\end{eqnarray*}
Note that we did not define $\vec\gamma_w$ for $w \in S_i$, $i \in [k]$.
The reason we can do so, is because we can optimize out $\vec\theta_c$ due
to the following trick we have been using (as in the proof of Theorem~\ref{thm:pr:d:fd}).
First, we rewrite all the terms in $\norm{\vec\theta}_2^2$ in terms of
$\vec\gamma$ and $\vec\theta_c$, $c \in S_i$, $i \in [k]$:
\begin{align*}
   \norm{\vec\theta}_2^2
   &= \sum_{w\notin F}\norm{\vec\theta_w}_2^2
+ \sum_{i=1}^k \sum_{t\in F_i}\norm{\vec\theta_t}_2^2
   + \sum_{\ell=1}^r \sum_{w \notin \{f_1,\dots,f_k\}} \norm{\vec\theta^{(\ell)}_w}_2^2
+ \sum_{\ell=1}^r \sum_{i=1}^k \norm{\vec\theta^{(\ell)}_{f_i}}_2^2\\
   &= \sum_{w\notin F}\norm{\vec\gamma_w}_2^2
+ \sum_{i=1}^k \sum_{t\in F_i}\norm{\vec\theta_t}_2^2
+ \sum_{\ell=1}^r \sum_{w \notin \{f_1,\dots,f_k\}} \norm{\vec\gamma^{(\ell)}_w}_2^2
+ \sum_{\ell=1}^r \sum_{i=1}^k \norm{\vec\theta^{(\ell)}_{f_i}}_2^2\\
&= \sum_{w\notin F}\norm{\vec\gamma_w}_2^2
+ \sum_{i=1}^k \norm{\vec\gamma_{f_i}-\sum_{c\in S_i}\mv R_c^\top\vec\theta_c - \vec\beta_{f_i}}_2^2
+ \sum_{i=1}^k \sum_{t\in S_i}\norm{\vec\theta_t}_2^2
+ \sum_{\ell=1}^r \sum_{w \notin \{f_1,\dots,f_k\}} \norm{\vec\gamma^{(\ell)}_w}_2^2\\
&\hspace*{1em}+ \sum_{\ell=1}^r \sum_{i=1}^k \norm{\vec\gamma^{(\ell)}_{f_i}-\sum_{c\in S_i}\mv R_c^\top \vec\gamma^{(\ell)}_c}_2^2\\
\end{align*}
Since $\vec\theta_t$,$t\in S_i$, does not depend on the loss term, we have
\begin{align}
   \frac 1 2 \pd{J}{\vec\theta_t}
   &= \displaystyle{\vec\theta_{t} - \mv R_t \left(\underbrace{\vec\gamma_{f_i}-\sum_{c\in S_i}\mv R_c^\top\vec\theta_c - \vec\beta_{f_i}}_{\vec\theta_{f_i}} \right)}
   & w \in S_i, i \in [k].
   \label{eqn:first:degree:fm:pd:2}
\end{align}
By setting \eqref{eqn:first:degree:fm:pd:2} to $0$, we have $\vec\theta_t = \mv R_t \vec\theta_{f_i}$ for all $t \in F_i$, and thus
\[ \vec \theta_{f_i} = \vec\gamma_{f_i}-\sum_{c\in S_i}\mv R_c^\top\vec\theta_c - \vec\beta_{f_i}
= \vec\gamma_{f_i}-\sum_{c\in S_i}\mv R_c^\top \mv R_c \vec\theta_{f_i} -
\vec\beta_{f_i},
\]
which implies $\vec\theta_{f_i} = \mv B_i^{-1} (\vec\gamma_{f_i} - \vec\beta_{f_i})$.
Hence, the following always holds:
\[ \vec\theta_{t} = \mv R_{t} \mv B_i^{-1} (\vec\gamma_{f_i} - \vec\beta_{f_i}), 
\qquad \qquad \forall t \in F_i, i \in [k]. 
\]
Note also that,
\begin{eqnarray*}
\sum_{t\in F_i}\norm{\vec\theta_t}_2^2
   &=& \sum_{t\in F_i} \norm{\mv R_{t} \mv B_i^{-1} (\vec\gamma_{f_i} - \vec\beta_{f_i})}_2^2\\
   &=& \sum_{t\in F_i} \inner{\mv R^\top_t \mv R_{t} \mv B_i^{-1} (\vec\gamma_{f_i} - \vec\beta_{f_i}),
   \mv B_i^{-1} (\vec\gamma_{f_i} - \vec\beta_{f_i})}\\
   &=& \inner{\left(\sum_{t\in F_i} \mv R^\top_t \mv R_{t} \right)\mv B_i^{-1} (\vec\gamma_{f_i} - \vec\beta_{f_i}),
   \mv B_i^{-1} (\vec\gamma_{f_i} - \vec\beta_{f_i})}\\
   &=& \inner{\mv B_i\mv B_i^{-1} (\vec\gamma_{f_i} - \vec\beta_{f_i}), \mv B_i^{-1} (\vec\gamma_{f_i} - \vec\beta_{f_i})}\\
   &=& \inner{(\vec\gamma_{f_i} - \vec\beta_{f_i}), \mv B_i^{-1} (\vec\gamma_{f_i} - \vec\beta_{f_i})}.
\end{eqnarray*}
Due to the fact that 
$\vec\theta^{(\ell)}_{f_i} = \vec\gamma^{(\ell)}_{f_i} - \sum_{c\in S_i}\mv R^\top_c \vec\gamma^{(\ell)}_c$, we 
can now write the penalty term in terms of the new parameter $\vec\gamma$:
\begin{align*}
   \norm{\vec\theta}_2^2
   &= \sum_{w\notin F}\norm{\vec\gamma_w}_2^2
+ \sum_{i=1}^k \sum_{t\in F_i}\norm{\vec\theta_t}_2^2
+ \sum_{\ell=1}^r \sum_{w \notin \{f_1,\dots,f_k\}} \norm{\vec\gamma^{(\ell)}_w}_2^2
+ \sum_{\ell=1}^r \sum_{i=1}^k \norm{\vec\theta^{(\ell)}_{f_i}}_2^2\\
   &= \sum_{w\notin F}\norm{\vec\gamma_w}_2^2
+ \sum_{i=1}^k \inner{(\vec\gamma_{f_i} - \vec\beta_{f_i}), \mv B_i^{-1} (\vec\gamma_{f_i} - \vec\beta_{f_i})}
+ \sum_{\ell=1}^r \sum_{w \notin \{f_1,\dots,f_k\}} \norm{\vec\gamma^{(\ell)}_w}_2^2\\
&\hspace*{1em}+ \sum_{\ell=1}^r \sum_{i=1}^k \norm{\vec\gamma^{(\ell)}_{f_i} - \sum_{c\in S_i}\mv R^\top_c \vec\gamma^{(\ell)}_c}_2^2.
\end{align*}
\ep

\bp[Proof of Proposition~\ref{prop:omega:fama2:gradient}]
The goal is to derive the gradient of $\Omega(\vec\gamma)$ w.r.t the parameters
$\vec\gamma$. Since $\vec\beta_{f_i}$ is a function of $\vec\gamma_c^{(\ell)}$, $\ell \in [r]$,
$c \in F_i$, the following is immediate:
\begin{eqnarray*}
   \frac 1 2 \pd{\norm{\vec\theta}_2^2}{\vec\gamma_w} &=& 
   \begin{cases}
      \vec\gamma_w, & w \notin F\\
      \mv B_i^{-1}(\vec\gamma_{f_i}-\vec\beta_{f_i}) & w = f_i, i \in [k].
   \end{cases}\\
   \frac 1 2 \pd{\norm{\vec\theta}_2^2}{\vec\gamma^{(\ell)}_w} &=& \vec\gamma_w, \qquad w \notin F, \ell \in [r].
\end{eqnarray*}
Next, we have to simplify $\vec\beta_{f_i}$ to facilitate fast computation:
\begin{eqnarray*} \vec\beta_{f_i} &=&
\sum_{\ell=1}^r \sum_{\{c,t\} \in \binom{F_i}{2}} \mv R^\top_c\vec\theta_c^{(\ell)}\circ \mv R_t^\top \vec\theta_t^{(\ell)}\\
   &=& 
   \sum_{\ell=1}^r \left[ \mv R^\top_{f_i} \vec\theta^{(\ell)}_{f_i} \circ \sum_{c\in S_i} \mv R^\top_c \vec\theta^{(\ell)}_c
   + \sum_{\{c,t\}\in \binom{S_i}{2}}
   \mv R_c^\top \vec\theta^{(\ell)}_c \circ \mv R_t^\top \vec\theta^{(\ell)}_t \right]\\
   &=& 
   \sum_{\ell=1}^r \left[ \vec\theta^{(\ell)}_{f_i} \circ \sum_{c\in S_i} \mv R^\top_c \vec\theta^{(\ell)}_c
   + \sum_{\{c,t\}\in \binom{S_i}{2}}
   \mv R_c^\top \vec\theta^{(\ell)}_c \circ \mv R_t^\top \vec\theta^{(\ell)}_t\right]\\
   &=& 
   \sum_{\ell=1}^r \left[ \left( \vec\gamma^{(\ell)}_{f_i} - \sum_{t \in S_i}\mv R^\top_t\vec\theta^{(\ell)}_t\right) \circ \sum_{c\in S_i} \mv R^\top_c \vec\theta^{(\ell)}_c
   + \sum_{\{c,t\}\in \binom{S_i}{2}}
   \mv R_c^\top \vec\theta^{(\ell)}_c \circ \mv R_t^\top \vec\theta^{(\ell)}_t \right]\\
   &=& 
   \sum_{\ell=1}^r \left[ \vec\gamma^{(\ell)}_{f_i} \circ \sum_{c\in S_i} \mv R^\top_c \vec\theta^{(\ell)}_c
   - \sum_{t \in S_i} \sum_{c \in S_i} \mv R^\top_t\vec\theta^{(\ell)}_t \circ \mv R^\top_c \vec\theta^{(\ell)}_c
   + \sum_{\{c,t\}\in \binom{S_i}{2}}
     \mv R_c^\top \vec\theta^{(\ell)}_c \circ \mv R_t^\top \vec\theta^{(\ell)}_t\right]\\
   &=& 
   \sum_{\ell=1}^r \left[ \vec\gamma^{(\ell)}_{f_i} \circ \sum_{c\in S_i} \mv R^\top_c \vec\theta^{(\ell)}_c
   -  \sum_{t \in S_i} \mv R^\top_t\vec\theta^{(\ell)}_t \circ \mv R^\top_t \vec\theta^{(\ell)}_t
   -  \sum_{\{c,t\}\in \binom{S_i}{2}}
   \mv R_c^\top \vec\theta^{(\ell)}_c \circ \mv R_t^\top \vec\theta^{(\ell)}_t \right]\\
   &=& 
   \sum_{\ell=1}^r \left[ \vec\gamma^{(\ell)}_{f_i} \circ \sum_{c\in S_i} \mv R^\top_c \vec\theta^{(\ell)}_c
   -  \sum_{t \in S_i} \mv R^\top_t (\vec\theta^{(\ell)}_t \circ \vec\theta^{(\ell)}_t)
   -  \sum_{\{c,t\}\in \binom{S_i}{2}}
   \mv R_c^\top \vec\theta^{(\ell)}_c \circ \mv R_t^\top \vec\theta^{(\ell)}_t\right]\\
   &=& 
   \sum_{\ell=1}^r \left[ \vec\gamma^{(\ell)}_{f_i} \circ \sum_{c\in S_i} \mv R^\top_c \vec\gamma^{(\ell)}_c
   -  \sum_{t \in S_i} \mv R^\top_t (\vec\gamma^{(\ell)}_t \circ \vec\gamma^{(\ell)}_t)
   -  \sum_{\{c,t\}\in \binom{S_i}{2}}
   \mv R_c^\top \vec\gamma^{(\ell)}_c \circ \mv R_t^\top \vec\gamma^{(\ell)}_t\right].
\end{eqnarray*}
Next, we derive the partial derivative w.r.t. $\vec\gamma_{f_i}^{(\ell)}$ for a fixed $i \in[k]$,
$\ell \in [r]$; in this computation we make use of~\eqref{eqn:mc:4} above:
\begin{eqnarray*}
   \frac 1 2 \pd{\norm{\vec\theta}_2^2}{\vec\gamma_{f_i}^{(\ell)}} &=& 
   \frac 1 2 \pd{\inner{(\vec\gamma_{f_i} - \vec\beta_{f_i}), \mv B_i^{-1} (\vec\gamma_{f_i} - \vec\beta_{f_i})}}{\vec\gamma_{f_i}^{(\ell)}} 
   + \frac 1 2 \pd{\norm{\vec\gamma^{(\ell)}_{f_i} - \sum_{c\in S_i}\mv R^\top_c \vec\gamma^{(\ell)}_c}_2^2}{\vec\gamma_{f_i}^{(\ell)}} \\
   &=& 
   \left(\sum_{c\in S_i}\DIAG(\mv R^\top_c \vec\gamma_c^{(\ell)}) \right)\mv B_i^{-1} (\vec\beta_{f_i}-\vec\gamma_{f_i})
   + \vec\gamma^{(\ell)}_{f_i} - \sum_{c\in S_i}\mv R^\top_c \vec\gamma^{(\ell)}_c\\
   &=&
   \vec\gamma^{(\ell)}_{f_i} - 
   \underbrace{\sum_{c\in S_i}\mv R^\top_c \vec\gamma^{(\ell)}_c}_{\vec\delta^{(\ell)}_i} -
   \underbrace{ \left( \sum_{c\in S_i}\mv R^\top_c \vec\gamma^{(\ell)}_c \right) }_{\vec\delta^{(\ell)}_i} 
   \circ \mv B_i^{-1} (\vec\gamma_{f_i} - \vec\beta_{f_i})\\
   &=&
   \vec\gamma^{(\ell)}_{f_i} - \vec\delta^{(\ell)}_i -
   \vec\delta^{(\ell)}_i \circ \left(\frac 1 2 \pd{\norm{\vec\theta}_2^2}{\vec\gamma_{f_i}} \right)
\end{eqnarray*}
Lastly, we move on to the partial derivative w.r.t. $\vec\gamma_{w}^{(\ell)}$ for a fixed $i \in[k]$, $w\in S_i$,
$\ell \in [r]$:
\begin{eqnarray*}
   \frac 1 2 \pd{\norm{\vec\theta}_2^2}{\vec\gamma_{w}^{(\ell)}} &=& 
   \frac 1 2 \pd{\norm{\vec\gamma_{w}^{(\ell)}}_2^2}{\vec\gamma_{w}^{(\ell)}} +   \frac 1 2 \pd{\inner{(\vec\gamma_{f_i} - \vec\beta_{f_i}), \mv B_i^{-1} (\vec\gamma_{f_i} - \vec\beta_{f_i})}}{\vec\gamma_{w}^{(\ell)}} 
   + \frac 1 2 \pd{\norm{\vec\gamma^{(\ell)}_{f_i} - \sum_{c\in S_i}\mv R^\top_c \vec\gamma^{(\ell)}_c}_2^2}{\vec\gamma_{w}^{(\ell)}} \\
   &=& 
   \vec\gamma_{w}^{(\ell)} + 
   \mv R_w \left(\sum_{c\in F_i}\DIAG(\mv R^\top_c \vec\gamma_c^{(\ell)}) \right)\mv B_i^{-1} (\vec\beta_{f_i}-\vec\gamma_{f_i})
   + \mv R_w\left( \sum_{c\in S_i}\mv R^\top_c \vec\gamma^{(\ell)}_c - \vec\gamma_{f_i}^{(\ell)}\right) \\
   &=& 
   \vec\gamma_{w}^{(\ell)} + 
   \mv R_w \left(\vec\gamma^{(\ell)}_{f_i} + \vec\delta^{(\ell)}_i \right)
   \circ \left( \frac 1 2 \pd{\norm{\vec\theta}_2^2}{\vec\gamma_{f_i}} \right)
   + \mv R_w\left( \vec\delta^{(\ell)}_i - \vec\gamma_{f_i}^{(\ell)}\right) \\
   &=& 
   \vec\gamma_{w}^{(\ell)} + 
   \mv R_w \left[ \vec\gamma^{(\ell)}_{f_i} 
   \circ \left( \frac 1 2 \pd{\norm{\vec\theta}_2^2}{\vec\gamma_{f_i}} \right)
   + \left( \vec\delta^{(\ell)}_i \circ \left( \frac 1 2 \pd{\norm{\vec\theta}_2^2}{\vec\gamma_{f_i}} \right)
   + \vec\delta^{(\ell)}_i - \vec\gamma_{f_i}^{(\ell)}\right)\right] \\
   &=& 
   \vec\gamma_{w}^{(\ell)} + 
   \mv R_w \left[ \vec\gamma^{(\ell)}_{f_i} 
   \circ \left( \frac 1 2 \pd{\norm{\vec\theta}_2^2}{\vec\gamma_{f_i}} \right)
   - \left( \frac 1 2 \pd{\norm{\vec\theta}_2^2}{\vec\gamma_{f_i}^{(\ell)}} \right)\right]. \\
\end{eqnarray*}
In particular, we were able to reuse the computation of 
$\frac 1 2 \pd{\norm{\vec\theta}_2^2}{\vec\gamma_{f_i}^{(\ell)}}$ and
$ \frac 1 2 \pd{\norm{\vec\theta}_2^2}{\vec\gamma_{f_i}}$ to compute
$\frac 1 2 \pd{\norm{\vec\theta}_2^2}{\vec\gamma_{w}^{(\ell)}}$. 
There is, however, still one complicated term $\vec\beta_{f_i}$ left to compute.
We simplify $\vec\beta_{f_i}$ to make its evaluation faster as follows.
\begin{eqnarray*}
   \vec\beta_{f_i} &=&
   \sum_{\ell=1}^r \left[ \vec\gamma^{(\ell)}_{f_i} \circ \sum_{c\in S_i} \mv R^\top_c \vec\gamma^{(\ell)}_c
   - \sum_{t \in S_i} \mv R^\top_t (\vec\gamma^{(\ell)}_t \circ \vec\gamma^{(\ell)}_t)
                        - \sum_{\{c,t\}\in \binom{S_i}{2}}
                          \mv R_c^\top \vec\gamma^{(\ell)}_c \circ \mv R_t^\top \vec\gamma^{(\ell)}_t
                  \right]\\
  &=& \sum_{\ell=1}^r \left[ 
   \vec\gamma^{(\ell)}_{f_i} \circ \sum_{c\in S_i} \mv R^\top_c \vec\gamma^{(\ell)}_c
   - \frac 1 2 \sum_{t \in S_i} \mv R^\top_t (\vec\gamma^{(\ell)}_t \circ \vec\gamma^{(\ell)}_t)
   - \frac 1 2 \sum_{c \in S_i}\sum_{t\in S_i} \mv R_c^\top \vec\gamma^{(\ell)}_c \circ \mv R_t^\top \vec\gamma^{(\ell)}_t
  \right]\\
  &=& \sum_{\ell=1}^r \left[ 
   \vec\gamma^{(\ell)}_{f_i} \circ \underbrace{\sum_{c\in S_i} \mv R^\top_c \vec\gamma^{(\ell)}_c}_{\vec\delta^{(\ell)}_{i}}
   - \frac 1 2 \sum_{t \in S_i} \mv R^\top_t (\vec\gamma^{(\ell)}_t \circ \vec\gamma^{(\ell)}_t)
   - \frac 1 2 \sum_{c \in S_i} \mv R_c^\top \vec\gamma^{(\ell)}_c \circ \sum_{t \in S_i} \mv R_t^\top \vec\gamma^{(\ell)}_t
  \right]\\
  &=& \sum_{\ell=1}^r \left[ 
   \vec\gamma^{(\ell)}_{f_i} \circ \vec\delta^{(\ell)}_i
   - \frac 1 2 \sum_{t \in S_i} \mv R^\top_t (\vec\gamma^{(\ell)}_t \circ \vec\gamma^{(\ell)}_t)
   - \frac 1 2 \vec\delta^{(\ell)}_{i} \circ \vec\delta^{(\ell)}_{i}
  \right]\\
  &=& \sum_{\ell=1}^r \left[ 
   \left(\vec\gamma^{(\ell)}_{f_i} - \frac 1 2 \vec\delta^{(\ell)}_{i} \right)\circ \vec\delta^{(\ell)}_{i}
   - \frac 1 2 \sum_{t \in S_i} \mv R^\top_t (\vec\gamma^{(\ell)}_t \circ \vec\gamma^{(\ell)}_t)
  \right].
\end{eqnarray*}
This completes the proof.
\ep

\end{appendix}

\end{document}